\documentclass[journal,twoside,web]{ieeecolor}
\usepackage{tmi}
\usepackage{cite}
\usepackage{amsmath,amssymb,amsfonts}
\usepackage{algorithmic}
\usepackage{graphicx}
\usepackage{float}
\usepackage{adjustbox}
\usepackage{caption}
\usepackage{stfloats}
\usepackage[dvipsnames]{xcolor}
\usepackage[justification=justified]{caption}
\usepackage{textcomp}
\usepackage[utf8]{inputenc}
\usepackage{longtable}
\usepackage{url}
\usepackage{xr}
\usepackage{newfloat}
\DeclareFloatingEnvironment{suppfig}
\DeclareFloatingEnvironment{supptab}
\makeatletter
\newcommand*{\addFileDependency}[1]{
  \typeout{(#1)}
  \@addtofilelist{#1}
  \IfFileExists{#1}{}{\typeout{No file #1.}}
}
\makeatother

\usepackage[belowskip=-3pt,aboveskip=3pt]{caption}
\definecolor{bondiblue}{rgb}{0.0, 0.58, 0.71}
\definecolor{brightcerulean}{rgb}{0.11, 0.62, 0.74}
\newcommand*{\revhl}{\textcolor{black}}

\usepackage{chngcntr}
\counterwithin*{subsubsection}{subsection}
\renewcommand{\thesubsubsection}{\thesubsection.\alph{subsubsection}}

\def\SB#1{\textsubscript{#1}}
\def\BibTeX{{\rm B\kern-.05em{\sc i\kern-.025em b}\kern-.08em
    T\kern-.1667em\lower.7ex\hbox{E}\kern-.125emX}}
\begin{document}
\title{Unsupervised MRI Reconstruction via Zero-Shot Learned Adversarial Transformers}
\author{Yilmaz Korkmaz, Salman UH Dar, Mahmut Yurt, Muzaffer {\"O}zbey, and Tolga \c{C}ukur \vspace{-1.5cm}
\\
\thanks{This study was supported in part by a TUBA GEBIP 2015 fellowship, and a BAGEP 2017 fellowship (Corresponding author: Tolga Çukur).}
\thanks{Y. Korkmaz, S. UH. Dar, M. Yurt, M. Özbey , and T. Çukur are with the Department of Electrical and Electronics Engineering, Bilkent University, Ankara, Turkey (e-mails: \{korkmaz, salman, mahmut, muzaffer, cukur\}@ee.bilkent.edu.tr). T. Çukur is also with the National Magnetic Resonance Research Center, Bilkent University, TR-06800 Ankara, Turkey.}
}

\maketitle

\begin{abstract}
\revhl{Supervised reconstruction models are characteristically trained on matched pairs of undersampled and fully-sampled data to capture an MRI prior, along with supervision regarding the imaging operator to enforce data consistency. To reduce supervision requirements, the recent deep image prior framework instead conjoins untrained MRI priors with the imaging operator during inference.} Yet, canonical convolutional architectures are suboptimal in capturing long-range relationships, and \revhl{priors based on randomly initialized networks may yield suboptimal performance.} To address these limitations, here we introduce a novel unsupervised MRI reconstruction method based on zero-Shot Learned Adversarial TransformERs (SLATER). SLATER embodies a deep adversarial network with cross-attention transformers to map noise and latent variables onto \revhl{coil-combined MR images. During pre-training, this unconditional network learns a high-quality MRI prior in an unsupervised generative modeling task. During inference, a zero-shot reconstruction is then performed by incorporating the imaging operator and optimizing the prior to maximize consistency to undersampled data.} Comprehensive experiments on brain MRI datasets clearly demonstrate the superior performance of SLATER against state-of-the-art unsupervised methods. 

\end{abstract}

\begin{IEEEkeywords}
adversarial, transformers, MRI, unsupervised, reconstruction, zero shot, generative. \vspace{-0.25cm}
\end{IEEEkeywords}

\section{Introduction}
\IEEEPARstart{M}{agnetic} Resonance Imaging (MRI) is a clinical powerhouse due to its excellent soft tissue contrast. Yet, low spin polarization at mainstream field strengths limits signal-to-noise ratio and necessitates prolonged exams. Extensive MRI exams incur added economic costs and lower use efficiency, and they may not be possible in uncooperative patient populations. A dire consequence is administration of exams that are in lower quality and/or diversity, which projects poorly onto diagnostic utility. Accelerated MRI addresses this fundamental limitation by performing undersampled acquisitions, and then solving an inverse problem to reconstruct images from available k-space data \revhl{\cite{Pruessmann1999,Griswold2002,Lustig2007,Shiqian2008}}.  

\revhl{Given their exceptional performance in inverse problems, supervised neural networks have been quickly adopted for MRI reconstruction \cite{DNNCS2020}. Reconstruction refers to the task of mapping undersampled acquisitions to images that are as consistent as possible with corresponding fully-sampled acquisitions. Supervised models performing this conditional mapping are trained on matched pairs of undersampled and fully-sampled data. The training process involves multiple lines of supervision including pairing of input-output data to learn an indirect prior that reduces aliasing artifacts in MR images \cite{Wang2016,ADMM-CSNET,Hammernik2017,Mardani2019b}, and enforcement of data consistency to embed the imaging operator that reflects k-space undersampling and coil-sensitivity encoding  \cite{Schlemper2017,MoDl,Quan2018c,Variatonal_end2end,Biswas2019}. Accordingly, supervised models are typically retrained for notable changes in the data distribution (e.g, different MRI contrast) or the imaging operator (e.g., different k-space sampling density) \cite{KnollGeneralization,Dar2017,Zhu2018}.}

\revhl{Several important approaches have been proposed in literature to reduce supervision requirements. A group of studies have reduced explicit supervision related to raw data, proposing models trained using unpaired sets of input-output data \cite{oh2020,Eun2020tt,chung2020progressive}, or using only undersampled data \cite{Tamir2019,Wang2020self,Cole2020,yaman2020,Huang2019self,Liu2020,aggarwal2020}. That said, these models involve implicit supervision regarding the imaging operator. As such, they are trained for a specific coil-array configuration and k-space sampling density, factors assumed to be consistent across the training and test sets \cite{Tamir2019,Wang2020self,Cole2020,yaman2020}. To remove other supervision aspects, a second group of studies have built unsupervised models by decoupling the MRI prior from the imaging operator. These models capture an MRI prior via generative networks that are either untrained \cite{Jin2019,Arora2020ismrm,Zou2021,Darestani2021} or trained to synthesize fully-sampled MR images \cite{Knoll2019inverseGANs,Konukoglu2019,Liu2020mrm}. The imaging operator is then conjoined with the MRI prior during inference on test data. This unsupervised approach excludes paired datasets for training, and promises improved generalization against deviations in the imaging operator \cite{Konukoglu2019,Liu2020mrm}. Yet, previously proposed models are commonly based on convolutional neural networks (CNNs) that suffer from limited sensitivity in capturing long-range dependencies \cite{Zhang2019,chen2021transunet,dalmaz2021resvit}.}

\begin{figure*}
\centering
\includegraphics[width=0.825\textwidth]{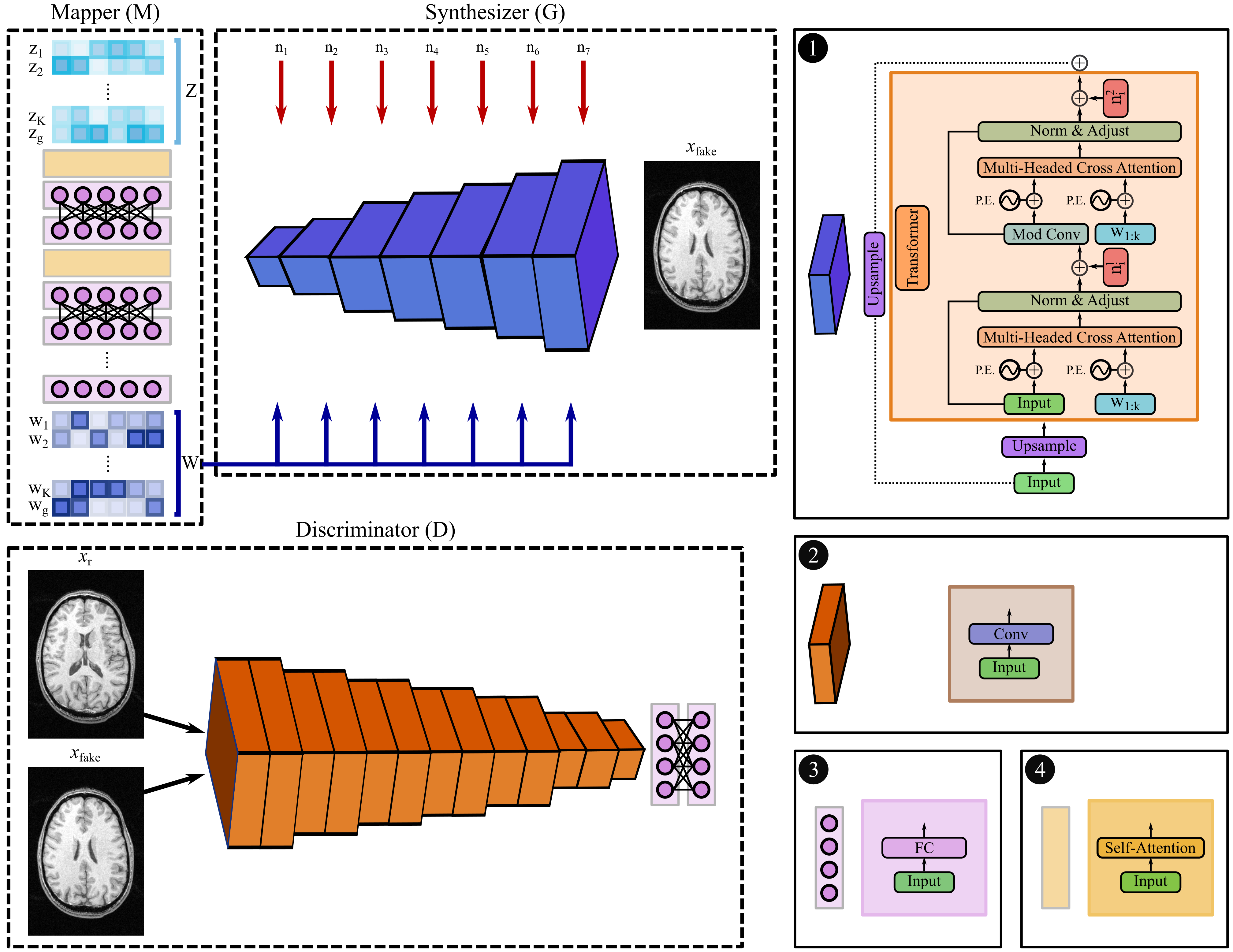}
\captionsetup{justification=justified, width=\linewidth}
\caption{SLATER is based on an unconditional adversarial architecture with a synthesizer (G), a discriminator (D), and a mapper (M). G is a multi-layer architecture where image resolution is progressively increased across layers. Within intermediate layers, a convolutional upsampling block is followed by a cross-attention transformer block (see 1). The transformer block receives global and local latent variables, noise variables and positional encoding (P.E.) for latents and image features. D enables adversarial learning while it receives as input synthetic and actual MR images. It is composed of convolutional and fully-connected (FC) blocks (see 2, 3). M projects raw latent variables (Z) onto k local (w\SB{1},...,w\SB{K}) variables and one global (w\SB{g}) latent variable. \revhl{It is composed of self-attention and FC blocks (see 3, 4, and Supp. Fig. \ref{suppfig:mapper} for details). An unsupervised generative modeling task is performed using the SLATER model to capture high-quality MRI priors.}}
\label{fig:slater}
\end{figure*}

\begin{figure*}[t]
\centering
\includegraphics[width=0.825\textwidth]{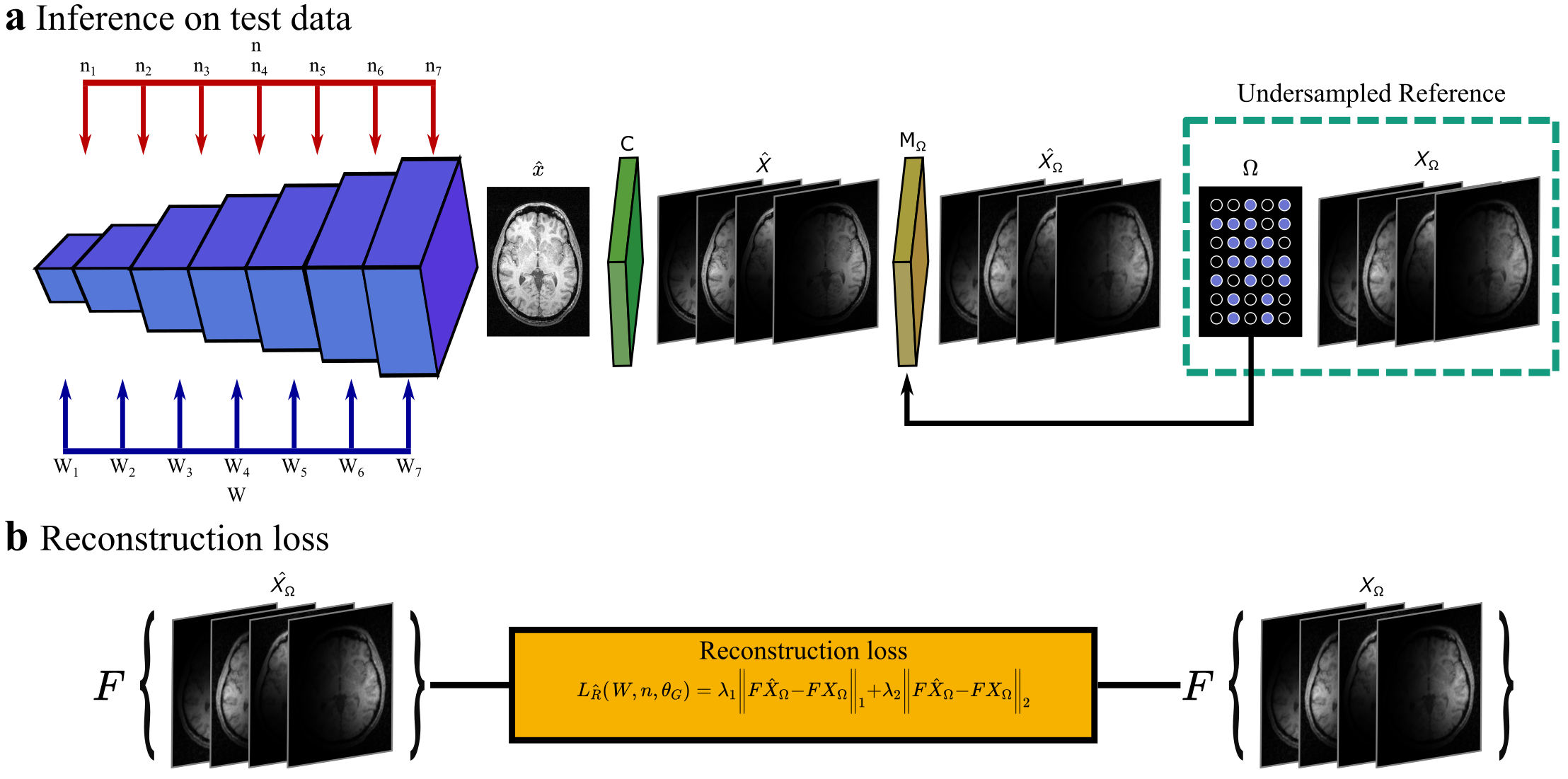}
\caption{\textbf{(a)} The generative model in SLATER is adapted to perform zero-shot reconstruction for accelerated MRI. To do this, the generator output is back-projected onto individual coils (C), and masked with the same sampling pattern as in the undersampled acquisition ($M_\Omega)$. \textbf{(b)} The network prior consisting of noise (n), latent variables (W) and weights ($\theta_G$) is then optimized to maximize consistency between the reconstructed and acquired k-space data. 
}
\label{fig:slater_opt}

\end{figure*}

\revhl{Here, we introduce a novel unsupervised MRI reconstruction based on zero-shot learned adversarial transformers (SLATER). SLATER decomposes the reconstruction process to decouple learning of the MRI prior from embedding of the imaging operator\footnote{see \cite{Yilmaz2021ismrm} for a preliminary version of this work presented at ISMRM 2021}. During a pre-training phase, an unconditional adversarial model is used to synthesize high-quality, coil-combined MR images (Fig. \ref{fig:slater}). To improve capture of long-range spatial context without introducing computational burden, we propose an architecture comprising cross-attention transformer blocks between low-dimensional latent variables and high-dimensional image features. During the inference phase, the learned MRI prior is combined with the imaging operator. This is achieved by optimizing network parameters that reflect the MRI prior to enforce a data-consistency loss on undersampled test data (Fig. \ref{fig:slater_opt}). To improve inference efficiency, a weight propagation strategy is proposed where the optimized network weights are transferred across consecutive cross-sections. }

\revhl{The proposed method performs an unsupervised generative modeling task on coil-combined images derived from fully-sampled MRI acquisitions. Adapting the generative model to the reconstruction task without any training samples, a zero-shot reconstruction then maps undersampled data to high-quality MR images during inference. The decoupled reconstruction process and model adaptation during inference contribute to improved generalization performance for SLATER. The source code for SLATER can be found at: \url{https://github.com/icon-lab/SLATER}.}

\vspace{0.1in}
\subsubsection*{\textbf{Contributions}}
\begin{itemize}

    \item \revhl{For the first time in literature, we introduce an adversarial vision transformer model for MRI reconstruction.}
    \item \revhl{Our proposed model uses cross-attention transformers to capture long-range spatial dependencies without computational burden. }

    \item \revhl{Sample-specific model adaptation and cross-sectional weight propagation strategies are introduced that respectively enhance out-of-domain generalizability and inference efficiency.}
\end{itemize}

\vspace{-0.1cm}
\section{Related Work}
\revhl{Supervised reconstruction models are current state-of-the-art in MRI with numerous successful architectures reported including basic CNNs \cite{Hyun2018,Wang2016,Yoon2018,ChulYe2018}, residual CNNs \cite{lee2018deep,Hauptmann2018}, perceptrons \cite{Kwon2017,Zhu2018}, physics-guided unrolled CNN networks \cite{Schlemper2017,Yang2016,MoDl,ADMM-CSNET,Biswas2019,KikiNet,Primal_dual,Adler2018,Wang2019}, recurrent CNNs \cite{Conv_recur,Hosseini2020b,Mardani2019b}, generative adversarial networks (GANs) \cite{Chen2021,Yu2018c,Mardani2017,rgan}, and variational networks  \cite{Hammernik2017,KnollGeneralization,Polakjointvvn2020,Variatonal_end2end}. However, supervised models are trained on paired sets of undersampled and fully-sampled data, along with supervision regarding the imaging operator in the form of a data-consistency term. Compiling large sets of paired data is non-trivial \cite{data_diff}, and supervised models often require retraining to cope with deviations in the imaging operator \cite{Zhu2018}.}

\revhl{To improve utility of deep MRI reconstruction, a common strategy has been to target explicit supervision on raw data in order to lower reliance on large, paired training datasets. Domain-transferred models are trained in a data-abundant source domain and then adopted for reconstruction in the target domain \cite{Han2018a,Dar2017}. Residual models are trained to predict residual error between ground truth and the output of a conventional MRI reconstruction \cite{Dar2021ismrm,Beker2019}. Both approaches permit training with relatively few samples, but training often requires paired datasets. Unpaired models are instead trained on input-output data collected from separate groups of subjects. This can be achieved with cycle-consistent models that learn bidirectional mappings between undersampled and fully-sampled data in order to enforce self-consistency of network inputs or outputs \cite{Quan2018c,oh2020,lei2020}. However, unpaired models can require substantially larger datasets for training compared to paired models \cite{chung2020progressive}. Lastly, several prominent approaches were proposed to train unsupervised models in the absence of ground truth. Scan-specific models learn nonlinear interpolation kernels from an auto-calibration region in undersampled acquisitions, and then recover missing k-space samples via interpolation during inference \cite{raki,sraki,loraki}. These models do not require a priori training, but their performance relies on the assumption that local dependencies between samples are largely invariant across k-space. Self-supervised models are trained via proxy loss terms substituted for the true reconstruction loss \cite{Tamir2019}. Specifically, data-consistency loss calculated on samples available in undersampled acquisitions can serve as a proxy \cite{Wang2020self,Cole2020}. An alternative is to leverage statistical estimators for mean-squared reconstruction loss \cite{aggarwal2020}. Such estimators are often analytically derived, so their use might be limited to specific types of loss functions. Recent studies have also proposed to mask a subset of k-space samples in undersampled acquisitions, where the masked subset is used to define a data-consistency loss \cite{yaman2020,Huang2019self}. Based on this loss, self-supervised models are trained to recover masked-out samples from remaining samples. A related image-domain strategy has been proposed for multi-image MRI data \cite{Liu2020,Huang2020ismrm}. In this case, models can be trained to predict a masked subset of undersampled frames in dynamic MRI from remaining frames \cite{Eldeniz2020ismrm}, or to map between differently undersampled versions of a given fully-sampled acquisition \cite{Liu2020ismrm}. Because the indirect MRI prior to reduce aliasing artifacts is learned from undersampled data, self-supervised models might show suboptimal performance at high acceleration factors. Moreover, unsupervised methods that used fixed model weights during inference might suffer from suboptimal generalization to test data \cite{Hosseini2020,aggarwal2021}.} 

\revhl{A collective attribute of the aforementioned methods to limit supervision on raw data is that they are based on conditional models that map undersampled acquisitions to MR images. Thus, they involve inherent supervision regarding the imaging operator, and they are typically retrained for varying coil-array configurations and sampling densities \cite{yaman2020}. Removing instead supervision related to the imaging operator, a fundamentally different approach decomposes the reconstruction process to decouple the MRI prior from the imaging operator. The deep image prior (DIP) approach employs an unconditional model that maps relatively low-dimensional latent variables onto images as a native MRI prior \cite{Jin2019}. The imaging operator is embedded during inference, and the prior is adapted to minimize a data-consistency loss on undersampled test data. This decoupled approach introduces flexibility in employing the same prior for various different imaging operators. Yet, DIP methods pervasively use untrained CNN architectures with randomly initialized parameters \cite{Jin2019,Arora2020ismrm,Ke2020ismrm,Zou2021,Darestani2021}. In turn, untrained priors might be suboptimal in capturing the distribution of MR images \cite{Konukoglu2019,Knoll2019inverseGANs}, and CNN models can generally suffer from limited sensitivity toward long-range spatial interactions \cite{Zhang2019,chen2021transunet,dalmaz2021resvit}.}

\vspace{-0.25cm}
\section{Theory}
\revhl{Inspired by the DIP framework, here we introduce a reconstruction method for accelerated MRI based on a deep generative model that maps noise and latent variables onto MR images. Unlike previous methods, this mapping is based on a style-generative adversarial network with transformer blocks. The network learns an MRI prior in pre-training phase, followed by a zero-shot inference phase where it is adapted to reconstruct undersampled acquisitions. We first overview the inverse problem formulation in accelerated MRI and DIP. We then describe the fundamental building blocks of SLATER.}
 
\vspace{-0.25cm} 
\subsection{Accelerated MRI Reconstruction}
In accelerated MRI, undersampled k-space acquisitions are performed to speed up scans, typically with variable-density random sampling patterns: 
\begin{equation}
\label{eq:sampling}
F_{p} C m = y_{s}
\end{equation}
$F_{p}$ is the partial Fourier operator defined by the sampling pattern, C denotes coil sensitivities, $m$ is the target MR image and $y_{s}$ are the acquired multi-coil k-space samples. The target $m$ must be computed given the available data $y_{s}$. However, the linear system in Eq.\ref{eq:sampling} is underdetermined, so MRI reconstruction is an ill-posed inverse problem. To obtain high-quality reconstruction, the solution has to be regularized with additional prior information on MR images: 
\begin{equation}
\widehat{m}=\underset{m}{\operatorname{argmin}}\|y_{s}-F_{p} C m \|_{2}^{2} + H(m)
\end{equation}
where $\widehat{m}$ is the reconstructed image, and $H(m)$ is the regularization term. The regularization term can be designed based on \revhl{sparsity priors \cite{Lustig2007,Shiqian2008}, structured low-rank priors \cite{Jin2016,Zhang2020}, or network priors \cite{Wang2016}.}

\vspace{-0.2cm}
\subsection{Deep Image Prior}
The DIP framework has recently been introduced for unsupervised learning in computer vision tasks, including super resolution, inpainting and denoising \cite{DIP}. DIP observes that local filtering operations in CNNs constrain the set of images that can be generated, so untrained CNNs can serve as native image regularizers. DIP performs random initialization of network inputs and weights without any pretraining. During inference, network parameters are optimized by enforcing consistency with the available corrupted image. Thus, DIP projects the corrupted image onto the space of CNN-generatable images to filter out corruptions such as blur or noise. DIP can be adopted for MRI reconstruction as follows: 
\begin{equation}
\label{eq:DIP}
\theta^{*}=\underset{\theta}{\operatorname{argmin}}\left\|F_{p} C d_{\theta}(z)-y_{s}\right\|_{1}
\end{equation} 
where $\theta$ are network weights, \revhl{$\theta^{*}$ are optimized network weights,} $z$ are latent variables, $d_{\theta}(z)$ is the network mapping from latents onto the reconstructed image \cite{Biswas2019,Jin2019}. Both network weights and latents are randomly initialized, and the optimization in Eq.\ref{eq:DIP} is performed over $\theta$, while $z$ is fixed. In contrast to mainstream learning-based methods, DIP inverts a random network prior to identify weights that are most consistent with the corrupted image. The reconstruction can then be expressed as: 
\begin{equation}
\widehat{m} = d_{\theta^{*}}(z)
\end{equation}

Despite its prowess in MRI reconstruction, \revhl{models with randomly initialized parameters setup a generic image prior that may not be as strongly-tuned towards the distribution of MR data as trained models \cite{Knoll2019inverseGANs}.} Furthermore, DIP is traditionally based on CNN architectures, where local kernels introduce suboptimal sensitivity in capture of long-range spatial interactions \cite{Zhang2019}.

\vspace{-0.2cm}
\subsection{Adversarial Transformer Model}
SLATER is based on an unconditional adversarial network that receives noise and latent variables to generate MR images \cite{Yilmaz2021ismrm}. Here we adopt a style-generative architecture given the success of this model family in computer vision tasks \cite{StyleGAN2}. We further propose to build the network layers with cross-attention transformer blocks as inspired by a recent study on natural image synthesis \cite{hudson2021generative}. \revhl{Self-attention transformers that compute interactions among all image pixels are prohibitive at relatively high spatial resolutions \cite{chen2021transunet,yang2021focal,liu2021Swin}. Instead, cross-attention transformers enable efficient capture of long-range context based on attentional interactions between low-dimensional latent variables and high-dimensional image features.}
Our model contains three sub-networks (Fig.\ref{fig:slater}): a synthesizer that generates MR images; a mapper that prepares the set of latent variables input to the synthesizer; and a discriminator. \revhl{The synthesizer aims to generate realistic images, while the discriminator aims to distinguish actual images from synthesized images. Thus, the discriminator aids the synthesizer in capturing an MRI prior that reflects the distribution of high-quality MR images.}  

\subsubsection{Synthesizer ($G$)} \revhl{The synthesizer contains a total of $N_L$ layers, each comprising a convolutional upsampling block to progressively increase image resolution followed by a cross-attention transformer block. Prior style-generative models typically use a global latent variable at each layer to control high-level image features related to style \cite{StyleGAN2}. In SLATER, in addition to a global latent ($w_{g}\in \mathbb{R}^{L_s}$, $L_s$: dimensionality), the synthesizer receives as input K local  ($W_{l}\in \mathbb{R}^{K\times L_s}$) latent variables at each layer. The global latent is still used to perform a spatially-uniform affine transformation of convolutional feature maps, and so it modulates high-level image features. Meanwhile, local latents are used to perform spatially-selective modulation of feature maps via cross-attention mechanisms. Each local latent variable focuses on a learned group of feature-map locations, so local latents serve to modulate relatively lower-level image features. For enabling cross-attention mechanism, a sinusoidal position encoding is used for spatial feature maps, whereas a learnable position encoding is assumed for local latent variables as their positions are unknown prior to training. Each cross-attention transformer block contains a serial cascade of the following sub-blocks: (cross-attention ($CA^1$), noise injection ($NI^1$), style-modulated convolution ($SC$), cross-attention ($CA^2$), noise injection ($NI^2$)). These sub-blocks are detailed below.}

\paragraph{\underline{\textbf{Cross-Attention ($CA^1$)}}}
\revhl{CA derives contextual representations by mediating attentional interaction between $W_{l} $ and input feature maps $X_i^{0} \in \mathbb{R}^{ h_1\times h_2 \times u}$, where $h_1$ and $h_2$ denote height and width depending on the resolution at the $i^{th}$ layer, and $u$ is the number of feature channels. 
Let $X_{i,vec}^{0} \in \mathbb{R}^{(h_1\times h_2)\times u}$ be the vectorized form of $X_i^{0}$ along the spatial dimensions.
Attention maps $\tilde{A}_{i,maps}^{1}\in \mathbb{R}^{(h_1\times h_2)\times K}$ that characterize the relation between $W_{l}$ and $X_{i,vec}^{0}$ are as follows (see Fig. \ref{fig:attmap1} and Supp. Fig. \ref{suppfig:attmap2} for sample maps):
\begin{align}
\label{eq:att_maps}
\tilde{A}_{i,maps}^{1} = smax\left( \frac{\tilde{q_{i}}^1(X_{i,vec}^{0}+PE^1_{i,X})\tilde{k_{i}}^1(W_{l}+PE_{W_l})^T} {\sqrt{u}}\right)
\end{align} 
where $smax$ is the softmax function, $\tilde{q_{i}}^1(.) \in \mathbb{R}^{(h_1\times h_2)\times u}$ are queries that receive $X_{i,vec}^{0}$ added with layer-specific position encoding variables $PE^1_{i,X} \in \mathbb{R}^{(h_1\times h_2)\times u}$ and perform a learnable linear projection, $\tilde{k_{i}}^1(.) \in \mathbb{R}^{K\times u}$
are keys that receive $W_{l}$ added with position encoding variables $PE_{W_l} \in \mathbb{R}^{K \times L_s}$ and perform a learnable linear projection. $PE^1_{i,X}$ are taken as learnable linear projections of pre-defined sinusoidal encoding variables (see Supp. Text \ref{sec:pos_enc}), while $PE_{W_l}$ are randomly initialized learnable position encoding variables. Note that $PE^1_{i,X}$ and $A_{i,maps}^{1}$ are vectorized along the spatial dimensions. Attention feature maps ($\tilde{A}_{i,feat}^{1} \in \mathbb{R}^{(h_1\times h_2)\times u}$) are then obtained by multiplying $A_{i,maps}^{1}$ with values that are learnable linear projections ($\tilde{v_{i}}^1(.) \in \mathbb{R}^{K\times u}$). 
\begin{align}
\label{eq:att_feats}
\tilde{A}_{i,feat}^{1} = \tilde{A}_{i,maps}^{1} \; \tilde{v_{i}}^1(W_{l}+PE_{W_l})
\end{align} 
Lastly, input feature maps are modulated via a single attention head that scales and shifts $X_{i,vec}^{0}$ with learnable linear projections of $\tilde{A}_{i,feat}^{1}$ to output $X^{1}_{i,vec} \in \mathbb{R}^{(h_1\times h_2)\times u}$. 
\begin{align}
\label{eq:norm_and_adjust}
X^{1}_{i,vec} = \tilde{\gamma}_1(\tilde{A}_{i,feat}^{1}) \odot  \left( \frac{X_{i,vec}^{0}-\mu(X_{i,vec}^{0}) } {\sigma(X_{i,vec}^{0})} \right) + \tilde{b}_1(\tilde{A}_{i,feat}^{1})
\end{align} 
where $\tilde{\gamma}_1(.) \in \mathbb{R}^{(h_1\times h_2)\times u}$ and $\tilde{b}_1(.) \in \mathbb{R}^{(h_1\times h_2)\times u}$ are learnable linear projections, $\mu$ denotes mean, $\sigma$ denotes variance, and $\odot$ is the Hadamard product.
The mapping through $CA^1$ can thus be summarized as:
$X_i^1 = CA^1(X_i^0)$, where $X_i^1 \in \mathbb{R}^{h_1\times h_2\times u}$ is the matrix form of $X_{i,vec}^1$.
}
\paragraph{\underline{\textbf{Noise Injection ($NI^1$)}}}
To improve control over variability in fine details of feature maps, noise variables are injected onto modulated feature maps from $CA^1$. Given input \revhl{$X_i^{1} \in \mathbb{R}^{ h_1\times h_2 \times u}$ to $NI^1$}, the output \revhl{($X_{i}^2\in \mathbb{R}^{h_1\times h_2\times u}$)} can be expressed as:
\begin{align}
X_i^{2} =\begin{bmatrix}
    X_i^{1,1} +\alpha_i^{1} n_i ^{1}\\
     \vdots \\
     X_i^{1,u} +\alpha_i^{1} n_i ^{1}\\
\end{bmatrix}
\end{align}
where \revhl{$X_i^{1,e}\in \mathbb{R}^{h_1\times h_2}$} denotes $e^{th}$ channel of $X_i^1$, \revhl{$n_i^{1}\in \mathbb{R}^{h_1\times h_2}$} is noise added to each \revhl{$e^{th}$} channel and \revhl{$\alpha_i^{1}$} is a learnable scalar. \revhl{Note that the learnable noise variables in $n_i ^{1}$ are initiated via random sampling from a standard normal distribution for each spatial location. During the course of learning, mean and standard deviation of noise variables are normalized to (0, 1) across the spatial dimensions. The final mapping through $NI^1$ is expressed as:
$X_i^2 = NI^1(X_i^1)$.
}
\paragraph{\underline{\textbf{Style Modulated Convolution ($SC$)}}} \revhl{Transformer blocks characteristically contain a feed-forward neural network (FFNN) sub-block following the attention sub-block to extract hidden features of attention-based contextual representations. For computational efficiency, here a convolutional FFNN is utilized to locally refine contextual representations while modulating feature maps to control high-level style features \cite{StyleGAN2}. Analogous to adaptive instance normalization (AdaIN) in style-transfer models, modulation is achieved via an affine transformation that controls the scale of feature maps \cite{adain}. Yet, we opted for a more compact implementation based on style-modulated convolution, with comparable complexity to convolution augmented with AdaIN and only requiring a trivial scaling operation over basic convolution \cite{StyleGAN2}. Accordingly, feature maps are scaled by modulating convolution kernels in SC, where kernel weights are multiplied via a learnable linear projection $w_{i,s}\in \mathbb{R}^{u}$ of the global latent variable $w_{g}$:}
\begin{align}
\label{eq:modulation}
\theta^{'} =\begin{bmatrix}
    w_{i,s}^{1} \theta_{i}^{1,1} & w_{i,s}^{1} \theta_{i}^{1,2} &  \dots  & w_{i,s}^{1} \theta_{i}^{1,u}\\
    w_{i,s}^{2}  \theta_{i}^{2,1} & w_{i,s}^{2}  \theta_{i}^{2,2} &  \dots  & w_{i,s}^{2}  \theta_{i}^{2,u} \\
    \vdots & \vdots  & \ddots & \vdots \\
     w_{i,s}^{u} \theta_{i}^{u,1}& w_{i,s}^{u} \theta_{i}^{u,2} &  \dots  & w_{i,s}^{u} \theta_{i}^{u,u}
\end{bmatrix}
\end{align}
where \revhl{$\theta_{i}^{e,h}\in \mathbb{R}^{r\times r}$ denotes the two-dimensional (2D) kernel for $e^{th}$ input and $h^{th}$ output channel, $w_{i,s}^{e}$ is the scaling coefficient for $e^{th}$ input channel. Note that $\theta_{i}^{'}\in \mathbb{R}^{r\times r\times u\times u}$ is a 4D tensor of modulated kernels, so the matrix expression in Eq. \ref{eq:modulation} depicts formation along the third and fourth tensor dimensions. While convolution with the modulated kernels changes the relative scaling of input feature channels for controlling style, it can also alter the overall scale of the output feature map for each channel. To restore output feature maps to unit standard deviation, the modulated kernels are normalized to unit-norm across the output channel dimension. Following the notation in Eq. \ref{eq:modulation}, this normalization can be expressed as:
\begin{align}
\theta_{i}^{''} =\begin{bmatrix}
    \frac{\theta_{i}^{'1,1}}{\sqrt{\sum_{c} (\theta_{i}^{'c,1})^2}} & \frac{\theta_{i}^{'1,2}}{\sqrt{\sum_{c} (\theta_{i}^{'c,2})^2}}&  \dots  &\frac{\theta_{i}^{'1,u}}{\sqrt{\sum_{c} (\theta_{i}^{'c,u})^2}}\\    
    \frac{\theta_{i}^{'2,1}}{\sqrt{\sum_{c} (\theta_{i}^{'c,1})^2}} & \frac{\theta_{i}^{'2,2}}{\sqrt{\sum_{c} (\theta_{i}^{'c,2})^2}}  & \dots  & \frac{\theta_{i}^{'2,u}}{\sqrt{\sum_{c} (\theta_{i}^{'c,u})^2}} \\  
    \vdots & \vdots & \ddots & \vdots \\    
    \frac{\theta_{i}^{'u,1}}{\sqrt{\sum_{c} (\theta_{i}^{'c,1})^2}}& \frac{\theta_{i}^{'u,2}}{\sqrt{\sum_{c} (\theta_{i}^{'c,2})^2}} & \dots  & \frac{\theta_{i}^{'c,u}}{\sqrt{\sum_{c} (\theta_{i}^{'c,u})^2}}
\end{bmatrix}
\end{align}
where $\theta_{i}^{''} \in \mathbb{R}^{r\times r\times u\times u}$ denotes de-modulated kernels and $c$ corresponds to channel index.
Finally, the output feature maps ($X_{i}^{3} \in \mathbb{R}^{h_1\times h_2 \times u}$) can be computed as:
\begin{align}
X_{i}^{3} =\begin{bmatrix}
    \sum_{c} X_{i}^{2,c} \circledast \theta_{i}^{''c,1} \\
     \vdots \\
     \sum_{c}  X_{i}^{2,c} \circledast \theta_{i}^{''c,u} \\
\end{bmatrix}
\end{align}
where $X_{i}^{2,c} \in \mathbb{R}^{h_1\times h_2}$ are input feature maps, $\circledast$ is convolution. The mapping through the convolution sub-block $SC^1$ is expressed as: $X_i^3 = SC(X_i^2)$.
}
\paragraph{\underline{\textbf{Cross-Attention ($CA^2$)}}}
\revhl{A second cross-attention sub-block is used following the $SC$ sub-block to further boost sensitivity of the model to global context. This maps the output of the SC sub-block to attention-modulated feature maps ($X_{i}^4 \in \mathbb{R}^{h_1\times h_2\times u} $) as:
$X_{i}^4 = CA^2(X_{i}^3)$.}
\paragraph{\underline{\textbf{Noise-Injection ($NI^2$)}}}
\revhl{Finally, noise variables are injected to control variability in the fine details of feature maps following the second CA sub-block:
$X_{i}^5 = NI^2(X_{i}^4)$.
where $X_{i}^5\in \mathbb{R}^{h_1\times h_2\times u}$ is the output of $NI^2$.
}

\subsubsection{Mapper ($M$)} 
The mapper projects independent and identically distributed random variables ($Z=z_1, z_2,..., z_K, z_{g}; z\in \mathbb{R}^{1 \times L_s}$) onto a refined set of local and global latent variables ($w_1, w_2, ..., w_K, w_{g}; w \in \mathbb{R}^{1 \times L_s}$) expected by the synthesizer \cite{StyleGAN2}. 
\revhl{M is a multi-layered architecture comprising a first stream dedicated to the global latent and a second stream of dedicated to local latents (see Supp. Fig. \ref{suppfig:mapper}). The global latent is processed with a cascade of fully-connected layers  \cite{StyleGAN2}. Local latents are instead processed with self-attention blocks to enable interactions among these variables. Self-attention blocks contain a cascade of self-attention (SA), fully-connected ($FC^1$), fully-connected ($FC^2$) sub-blocks as detailed below. 
\paragraph{\underline{\textbf{Self-Attention (SA)}}}
At the $i^{th}$ self-attention block, intermediate activations for the $K$ latent variables are concatenated as input: 
\begin{align}
Z_i^{0}= (z_{1} \; \oplus_1 \dots \oplus_1 \; z_{K})_i^0
\end{align} 
where $Z_i^{0} \in \mathbb{R}^{K \times L_s}$ and $\oplus_1$ is the concatenation operator along the first dimension.
Attention maps ($\hat{A}_{i,maps}\in \mathbb{R}^{K \times K}$) are first obtained:
\begin{align}
\hat{A}_{i,maps} = smax\left( \frac{\hat{q_i}(Z_i^{0}+PE_{Z^{0}}) \hat{k_i}(Z_i^{0}+PE_{Z^{0}})^T} {\sqrt{L_s}}\right) 
\end{align} 
where $PE_{Z^{0}} \in \mathbb{R}^{K \times L_s}$ denotes learnable position encoding variables, $\hat{q_i}(.) \in \mathbb{R}^{K \times L_s}$ and $\hat{k_i}(.) \in \mathbb{R}^{K \times L_s}$ are queries and keys respectively.
Attention feature maps ($\hat{A}_{i,feat}\in \mathbb{R}^{K \times L_s}$) are given as:
\begin{align}
\hat{A}_{i,feat} = \hat{A}_{i,maps} \; \hat{v_i}(Z_i^{0}+PE_{Z^{0}})
\end{align} 
where $\hat{v_i}(.) \in \mathbb{R}^{K \times L_s}$ are values.
$\hat{A}_{i,feat}$ is then used to scale and shift $Z_i^0$:
\begin{align}
Z_i^{1} = \hat{\gamma}(\hat{A}_{i,feat}) \odot  \left( \frac{Z_i^{0}-\mu(Z_i^{0}) } {\sigma(Z_i^{0})} \right) + \hat{b}(\hat{A}_{i,feat})
\end{align}
where $Z_i^{1} \in \mathbb{R}^{K \times L_s}$ is the output, $\hat{\gamma}(.)$ and $\hat{b}(.)$ are learnable linear projections.
Lastly, $Z_i^{1}$ is decomposed into individual latents, $Z_i^{1} \rightarrow \left\{z_{1}, \dots , z_{K}\right\}_i^1$.}
\revhl{\paragraph{\underline{\textbf{Fully Connected ($FC$)}}}
Next, each latent is separately processed with two fully-connected sub-blocks ($FC^1$ and $FC^2$):
\begin{align}
\left\{z_j\right\}_i^{2}=FC^2(FC^1(\left\{z_j\right\}_i^{1}))
\end{align}
where $\left\{z_j\right\}_i^{2} \in \mathbb{R}^{1 \times L_s}$ is the output $FC^2$ for the $j^{th}$ local latent.
}
\subsubsection{Discriminator ($D$)} 
\revhl{Adversarial models involve an interplay between the synthesizer that generates images and a separate discriminator sub-network \cite{Goodfellow2014a}. In SLATER, $D$ aims to accurately distinguish images generated by the synthesizer from actual MR images. A feed-forward architecture is employed here with a cascade of convolutional layers augmented with several fully-connected layers. The mapping through $D$ can be compactly expressed as:
\begin{align}
x_D=D_{\theta_{D}}(x)
\end{align}
where $x_D \in \mathbb{R}^{1}$ is the output of the discriminator, and $x$ is either an actual MR image ($x_r$) or an image generated by the synthesizer $G(M(Z))$.}  

\begin{figure}[H]
\centering
\includegraphics[width=0.45\textwidth]{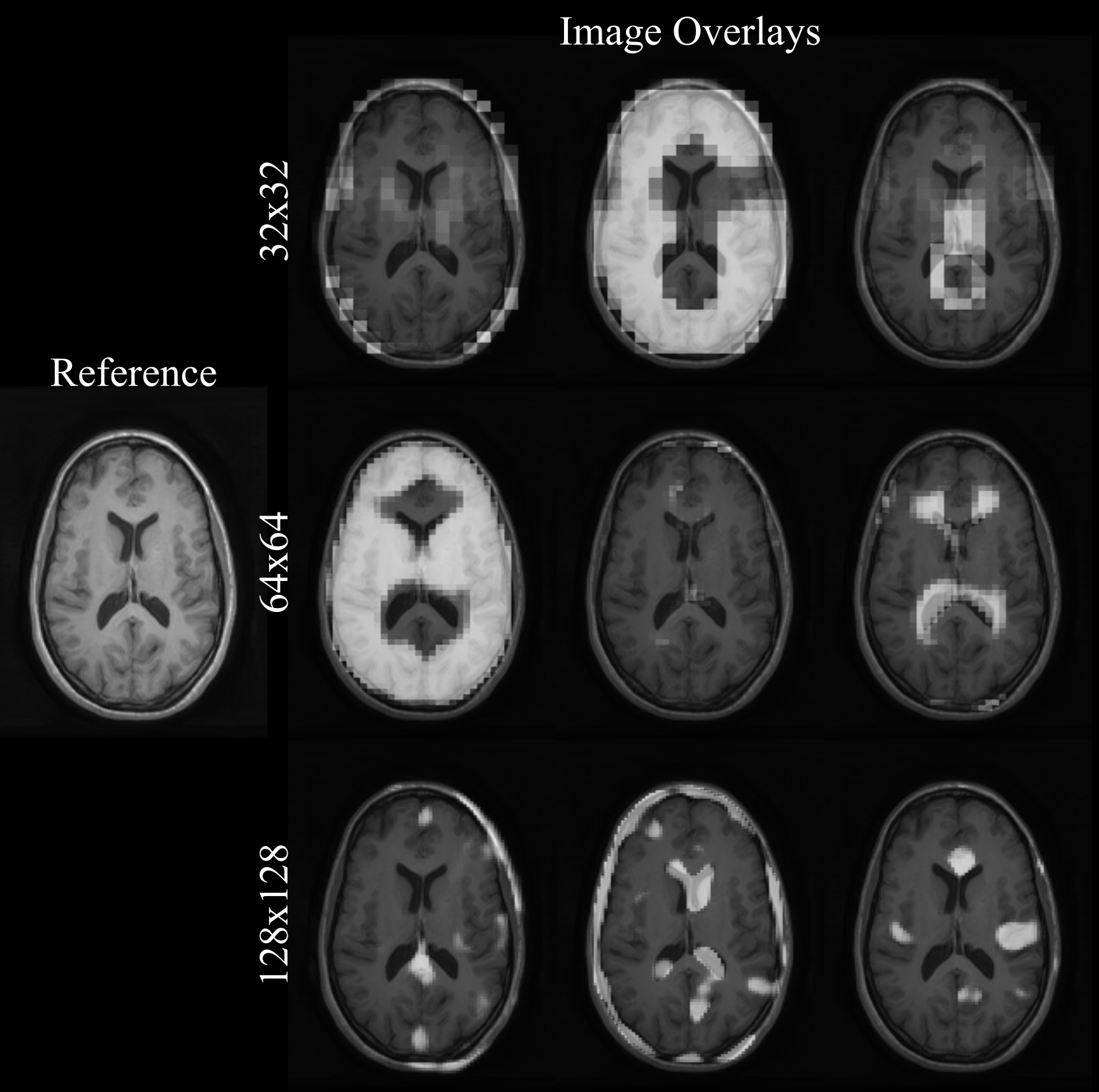}
\captionsetup{justification=justified,width=\linewidth}
\caption{Cross-attention maps in SLATER for a T\SB{1}-weighted acquisition. Sample maps from the first CA sub-block are displayed by overlay onto the respective MR image across three resolutions (i.e., at network layers 4-6). Attention maps for separate latents typically show segregated spatial distribution, and tend to group tissue clusters with similar signal intensity and texture broadly distributed across the image.}
\label{fig:attmap1}
\vspace{-0.2in}
\end{figure} 

\subsection{Self- versus Cross-Attention Transformers}
\revhl{The main components of vanilla transformers are a self-attention (SA) sub-block followed by a feed-forward neural network (FFNN) sub-block \cite{chen2021transunet}. Self-attention mechanisms serve to explicitly relate different positions within an image to compute contextual representations at each image pixel. Thus, SA is the primary component that learns long-range dependencies, while FFNN performs nonlinear transformations to extract hidden features of attention-based contextual representations. While vanilla transformers with fully-connected FFNN are common in natural language processing \cite{transformer1}, computer vision studies often introduce vision-specific modifications for computational efficiency \cite{hudson2021generative}. Cross-attention transformers in SLATER involve two key modifications: cross-attention sub-blocks that use a compact set of latents implicitly relating image pixels to learn contextual representations, and a convolutional FFNN to improve computational efficiency. Here, we overview self-attention and cross-attention transformers in terms of their ability to capture long-range spatial relationships and model complexity. For simplicity, remaining operations in transformers such as normalization or skip connections are ignored, and only a single feature channel and a single attention head are considered.}

\subsubsection{Self-attention transformer}
\revhl{Given an input feature map $X^0_{vec} \in \mathbb{R}^{(h_1 \times h_2)}$, let $a,b$ be two distant pixels whose $r \times r$ neighborhoods $\Delta a, \Delta b$ do not spatially overlap. A convolution sub-block with kernel size $r$ would perform localized processing for each pixel in its $r \times r$ neighborhood:
\begin{align}
\label{eq:dense_SC}
X^1_{vec}=Conv(X^0_{vec})
\end{align}
where $X^1_{vec} \in \mathbb{R}^{(h_1 \times h_2)}$. Intensities of $a,b$ in $X^1_{vec}$ will thus be conditionally independent given intensities of $X^0_{vec}$ in $\Delta a, \Delta b$:
\begin{align}
\mathrm{E}\{X^{1}_{vec}[a] X^{1}_{vec}[b] \mid X_{c}\} = \mathrm{E}\{X^{1}_{vec}[a] \mid X_{c}\} \mathrm{E}\{X^{1}_{vec}[b] \mid X_{c}\}
\end{align} 
where $\mathrm{E}\{.\}$ denotes expectation, $X_c = X^0_{vec}[\Delta a \cup \Delta b]$ and $\cup$ is the union operator. As a result, convolutional processing does not leverage long-range relationships in feature maps.}

\revhl{In contrast, the SA sub-block in a transformer characterizes the relation between all spatial locations in $X^0_{vec} \in \mathbb{R}^{(h_1 \times h_2)}$ via an attention map:
\begin{align}
\label{eq:att_maps_self}
A_{maps} = smax\left( q(X_{vec}^{0}+PE_{X})k(X_{vec}^{0}+PE_{X})^T\right)
\end{align}
where $PE_{X}, q, k \in \mathbb{R}^{(h_1 \times h_2)}$ are positional encoding, query, and key, and $A_{maps}\in \mathbb{R}^{(h_1 \times h_2) \times (h_1 \times h_2) }$. Attention-modulated feature map is then derived:
\begin{align}
\label{eq:att_feat_maps_self}
A_{feat} = A_{maps} v(X_{vec}^{0}+PE_{X})
\end{align}
where $A_{feat} \in \mathbb{R}^{(h_1 \times h_2)}$, and $v \in \mathbb{R}^{(h_1 \times h_2)}$ is value. Intensities of $A_{feat}$ at pixels $a,b$ are: 
\begin{align}
\label{eq:att_feats_self}
A_{feat}[a]=A_{maps}[a,1:(h_1 \times h_2)] \ v [1:(h_1 \times h_2)] \notag \\
A_{feat}[b]=A_{maps}[b,1:(h_1 \times h_2)] \ v [1:(h_1 \times h_2)] 
\end{align}
As seen here, both  $A_{feat}[a]$ and $A_{feat}[b]$ are functions of all pixels in $v$ and thereby $X_{vec}^{0}$. Next, the FFNN sub-block processes $A_{feat}$ to extract hidden representations, typically via a fully-connected architecture: 
\begin{align}
\label{eq:dense_FC_self}
X^1_{vec}=FC(A_{feat})
\end{align}
where $X^1_{vec} \in \mathbb{R}^{(h_1 \times h_2)}$. Note that $A_{feat}[a]$ and $A_{feat}[b]$ are readily dependent on all pixels. Intensities of pixels $a,b$ in output feature maps are then statistically dependent even when conditioned on $X_c = X^0_{vec}[\Delta a \cup \Delta b]$: 
\begin{align}
\mathrm{E}\{X^{1}_{vec}[a] X^{1}_{vec}[b] \mid X_c \} \neq \mathrm{E}\{X^{1}_{vec}[a] \mid X_c \} \mathrm{E}\{X^{1}_{vec}[b] \mid X_c \}
\end{align} 
Thus, vanilla transformers utilize dependencies across distant pixels to compute contextual representations. }

\revhl{The SA sub-block examines interactions among all possible pairs of pixels as described in Eqs. \ref{eq:att_maps_self} and \ref{eq:att_feats_self}, so it has a computational complexity of $O((h_1 \times h_2)^2)$. Likewise, the FFNN sub-block as described in Eq. \ref{eq:dense_FC_self} exhaustively considers inter-pixel interactions with a complexity of $O((h_1 \times h_2)^2)$. This quadratic complexity with respect to image size limits the applicability of self-attention transformers at relatively high resolutions encountered in MRI \cite{chen2021transunet}.}

\subsubsection{Cross-attention transformer}
\revhl{Instead of exhaustively modeling inter-pixel interactions in high-dimensional feature maps, the CA sub-block uses a small set of $K$ latent variables to implicitly characterize these interactions:
\begin{align}
\label{eq:att_maps_cross_comp}
A_{maps} = smax\left( q(X_{vec}^{0}+PE_{X})k(W_{l}+PE_{W_{l}})^T\right)
\end{align} 
where $A_{maps} \in \mathbb{R}^{(h_1 \times h_2) \times K}$, $W_l \in \mathbb{R}^{K \times L_{s}}$: local latent variables, $PE_{W_{l}} \in \mathbb{R}^{K \times L_{s}}$: positional encoding for $W_l$. Attention-modulated feature map can be expressed as:
\begin{align}
\label{eq:att_feat_maps_cross}
A_{feat} = A_{maps} v(W_{l}+PE_{W_{l}})
\end{align}
where $A_{feat} \in \mathbb{R}^{(h_1 \times h_2)}$, and $v \in \mathbb{R}^{K}$ is value. In turn, intensities of pixels $a,b$ are:
\begin{align}
\label{eq:att_feats_cross}
A_{feat}[a]=A_{maps}[a,1:K] \ v[1:K] \notag \\
A_{feat}[b]=A_{maps}[a,1:K] \ v[1:K]
\end{align} 
As seen above, both $A_{feat}[a]$ and $A_{feat}[b]$ are functions of all latent variables in $W_{l}$, so they are statistically dependent.}

\revhl{Next, a convolutional FFNN processes $A_{feat}$ to extract hidden representations: 
\begin{align}
\label{eq:dense_SC_self}
X^1_{vec}=Conv(A_{feat})
\end{align}
where $X^1_{vec} \in \mathbb{R}^{(h_1 \times h_2)}$. Although convolutional processing is local, the dependency introduced in the CA sub-block carries over to the output feature maps:
\begin{align}
& \mathrm{E}\{A_{feat}[a] A_{feat}[b] \mid X_c \} \neq \mathrm{E}\{A_{feat}[a] \mid X_c \} \mathrm{E}\{A_{feat}[b] \mid X_c \} \notag \\
& \mathrm{E}\{X^{1}_{vec}[a] X^{1}_{vec}[b] \mid X_c \} \neq \mathrm{E}\{X^{1}_{vec}[a] \mid X_c \} \mathrm{E}\{X^{1}_{vec}[b] \mid X_c \}
\end{align}
Therefore, cross-attention transformers can model long-range dependencies to compute contextual representations.}

\revhl{The CA sub-block examines interactions among image pixels and local latents as described in Eqs. \ref{eq:att_maps_cross_comp} and \ref{eq:att_feats_cross}, so it has a computational complexity of $O((h_1 \times h_2) \times K)$. The convolutional FFNN considers local interactions with a complexity of $O((h_1 \times h_2) \times r^2)$. Because $K<<(h_1 \times h_2)$ and  $r^2<<(h_1 \times h_2)$ typically, the cross-attention transformer achieves notably lower complexity to permit use at higher spatial resolutions.
}

\subsection{Learning Procedures}
\revhl{SLATER uses a two-stage strategy towards MRI reconstruction with a pre-training phase to learn the MRI prior, followed by a zero-shot reconstruction phase to embed the imaging operator. These two phases are detailed below.}

\subsubsection{Pre-training of the MRI prior}
\revhl{Since SLATER completely decouples the MRI prior from the imaging operator, pre-training assumes no prior knowledge on the imaging operator such as undersampling patterns or coil sensitivity encoding. Instead, the adversarial transformer model is trained to capture a prior on coil-combined, complex MR images, derived from fully-sampled acquisitions. Note that the synthesizer in this unconditional model maps noise and latent variables onto MR images, unlike conditional models with explicitly defined input-output relationships (i.e. undersampled versus fully-sampled data). Therefore, SLATER's pre-training is categorized as an unsupervised generative modeling task where the distribution of MR images is learned so that new, random samples can be drawn from the distribution \cite{Bishop2006}.}

\revhl{Adversarial models commonly involve synthesizer and discriminator sub-networks that are trained with inter-linked loss functions to improve quality of synthesized images \cite{Goodfellow2014a}. In SLATER, the synthesizer along with the mapper that provides latent variables are trained to minimize a common adversarial loss based on non-saturating logistic function: 
\begin{equation}
\label{synthesizer_loss}
L_{G,M}(\theta_G,\theta_M) = -\mathrm{E}_{p(Z)}\{log(f(D(G_{\theta_G}(M_{\theta_M}(Z))))\}
\end{equation}
where $\mathrm{E}_{p(.)}$ is expectation with probability density $p$, $f(.)$ is the sigmoid function, $\theta_G$ are parameters of the synthesizer, $\theta_M$ are parameters of the mapper. This particular loss is preferred in order to prevent the saturation problem in adversarial learning, where the discriminator starts outperforming the synthesizer by a significant margin and learning stops prematurely \cite{Goodfellow2014a}.}

\revhl{Meanwhile, the discriminator is trained to minimize an adversarial loss based on non-saturating logistic function augmented with a gradient penalty term:
\begin{align}
\label{discriminator_loss}
L_{D}(\theta_D) = -\mathrm{E}_{p(Z)}\{log(1 - f(D_{\theta_D}(G(M(Z))))\}\notag \\ 
- \mathrm{E}_{p(x_r)}\{log(f(D_{\theta_D}(x_r))\} \notag \\  
+ \frac{\eta}{2} \mathrm{E}_{p(x_r)}\{\left\|\nabla D_{\theta_D}(x_r)\right\|^{2}\}
\end{align}
where $x_r$ denotes coil-combined, complex MR images derived from actual scans, $\eta$ is the regularization parameter, and $\theta_D$ are the parameters of the discriminator. The first two terms define the adversarial loss, whereas the third term has been suggested to improve adversarial learning by enforcing limited gradients in $x_r$ according to the learned distribution \cite{pmlr-v80-mescheder18a}.}
\begin{figure}
\centering
\includegraphics[width=0.45\textwidth]{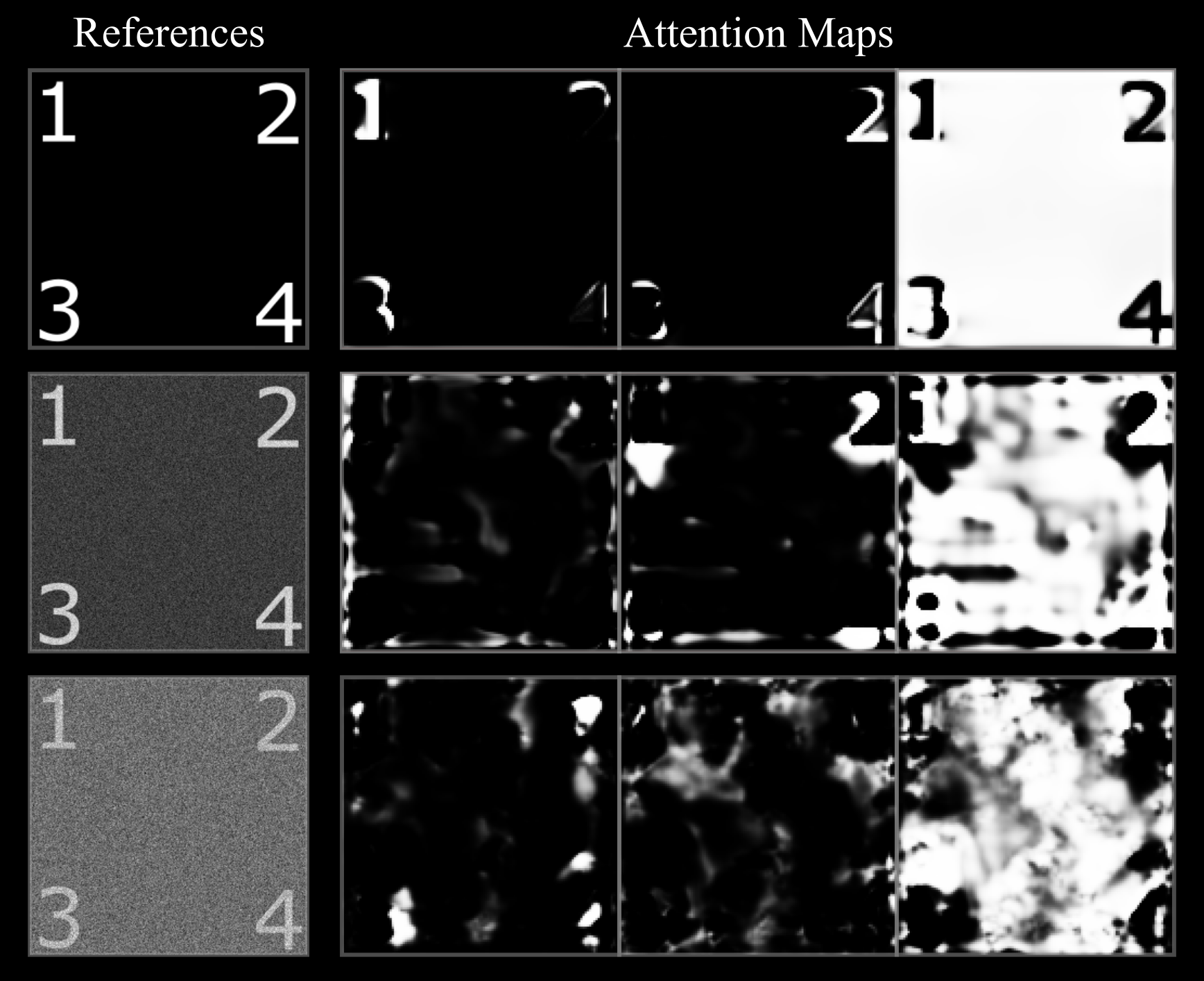}
\caption{Cross-attention maps in SLATER for a simulated phantom with varying levels of normally-distributed white noise. Sample maps from the first CA sub-block are displayed at 128x128 resolution (network layer 6). Relative to a peak signal intensity of 1, top, middle and bottom rows show results for no noise, noise variance of 0.01, and noise variance of 0.1, respectively.}
\label{fig:phantom}
\vspace{-0.1in}
\end{figure}

\subsubsection{Zero-shot reconstruction}
\revhl{The learned MRI prior does not contain any information regarding the conditional mapping from undersampled to fully-sampled data. Thus, the prior is conjoined with the imaging operator during inference to reconstruct test data. To adapt the pre-trained generative model to the reconstruction task, a data-consistency loss is employed on undersampled acquisitions to optimize synthesizer parameters (Fig. \ref{fig:slater_opt}). Note that zero-shot learning is an unsupervised task-adaptation approach, where a model trained for an initial task is later transferred to a different target task without using additional training samples \cite{XLSA18}. Analogously, SLATER adapts its adversarial model pre-trained to perform generative modeling of MR images to perform MRI reconstruction without any extra training samples. Therefore, the inference phase of SLATER is categorized as zero-shot reconstruction.} 

\revhl{During inference, we optimize all components of the synthesizer including noise ($n$), latent variables ($W$) and weights ($\theta_G$) to minimize the data-consistency loss. The synthesizer outputs a coil-combined, complex MR image, which is back-projected onto individual coils given sensitivity estimates, and then Fourier transformed to select available k-space coefficients according to the undersampling pattern \cite{rgan}. Consistency of acquired and reconstructed k-space coefficients in the test data is then computed:
\begin{align}
L_{\hat{R}}(W, n, \theta_G)=&\lambda_{1}\left\|F_{p} C G(W, n, \theta_G)-k_{x}\right\|_{1} \notag \\
&+\lambda_{2}\left\|F_{p} C G(W, n, \theta_G)-k_{x}\right\|_{2}
\label{optimizationequationstyle}
\end{align} 
where $\hat{R}$: reconstructed image, $k_{x}$: acquired k-space coefficients, $F_p$: partial Fourier operator, $G$: synthesizer. Data consistency is taken as a weighted $\ell_1-\ell_2$-norm loss, where ($\lambda_{1}, \lambda_{2}$): weightings of ($\ell_1$, $\ell_2$) loss components. This loss function is considered to offer a more balanced weighting of errors across k-space compared to $\ell_2$-norm that can be over-sensitive to lower spatial frequencies \cite{yaman2020}. During pre-training, a common set of latent variables produced by the mapper are input uniformly across all CA and SC sub-blocks, whereas sub-block specific noise is included. To improve performance during zero-shot reconstruction, latent variables for each sub-block within each layer are instead segregated for independent optimization. In Eq. \ref{optimizationequationstyle}, $W = W_{1,..,N_L}$ where $W_i$ denotes the collection of latent variables for the $i^{th}$ layer and contains the local latents for the two CA sub-blocks ($W_{l,i}^{1,2}$) and the global latent for the SC sub-block ($w_{g,i}$); $n = n_{1,..,N_L}$ where $n_i$ denotes the collection of noise components for the $i^{th}$ layer and contains noise for the two NI sub-blocks ($n_{i}^{1,2}$).}

\begin{figure*}
\centering
\includegraphics[width=1\textwidth]{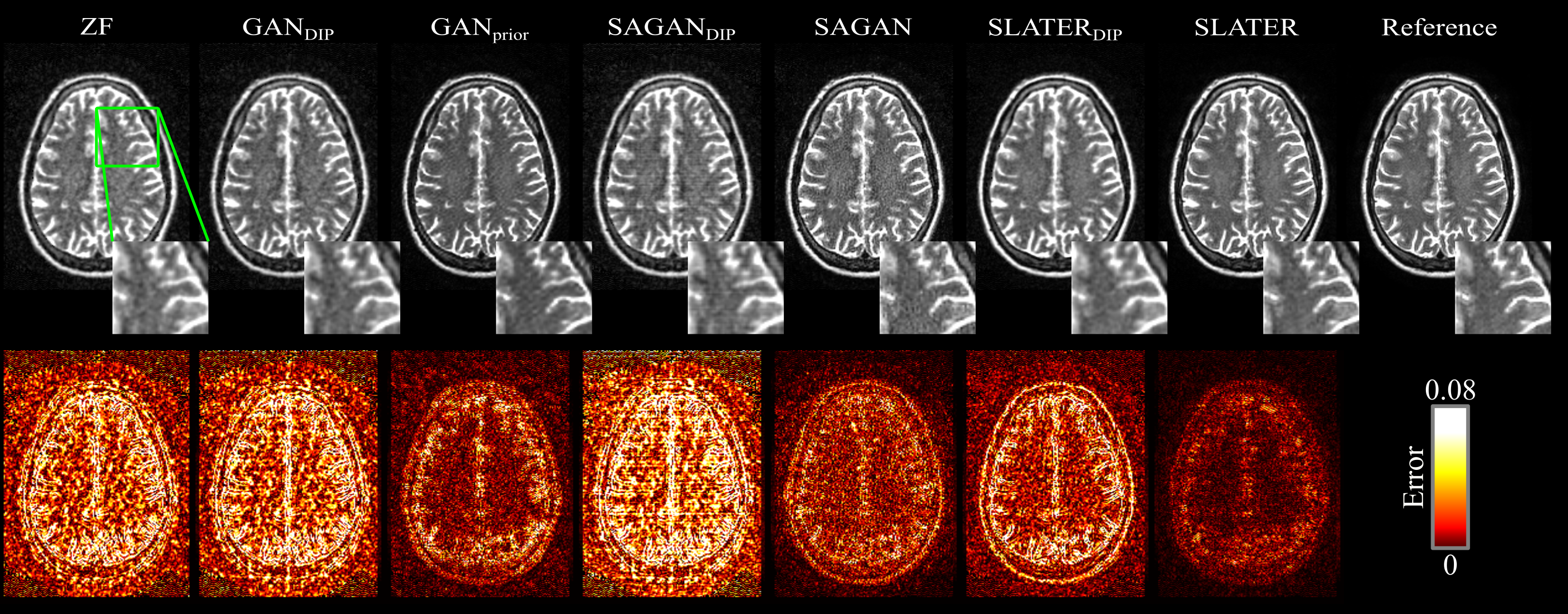}
\caption{Reconstructions of a representative T\SB{2}-weighted acquisition at R=4 are shown for the Fourier method (ZF), DIP methods (GAN\SB{DIP}, SAGAN\SB{DIP}, SLATER\SB{DIP}) and zero-shot reconstructions (GAN\SB{prior}, SAGAN, SLATER) along with the reference image. \revhl{Zoom-in display windows are added to aid visualization of performance differences.} Corresponding error maps are underneath the images for each method.}
\label{fig:T2_IXI_DIP}
\end{figure*}

\begin{table*}[]
\centering
\captionsetup{justification=justified ,width=\linewidth}
\caption{Reconstruction performance of DIP and zero-shot reconstructions for T\SB{1}- and T\SB{2}-weighted acquisitions IXI at R=4 and 8. \revhl{Performance metrics are presented as mean $\pm$ standard deviation across test subjects.} Results are listed for GAN\SB{prior}, SAGAN, SLATER, their DIP variants, and ZF along with the reference image.}
\resizebox{\textwidth}{!}{%
\begin{tabular}{|l|l|l|l|l|l|l|l|l|l|l|l|l|}
\hline
        & \multicolumn{2}{c|}{\textbf{GAN\SB{DIP}}} & \multicolumn{2}{c|}{\textbf{SAGAN\SB{DIP}}} & \multicolumn{2}{c|}{\textbf{SLATER\SB{DIP}}} & \multicolumn{2}{c|}{\textbf{GAN\SB{prior}}} & \multicolumn{2}{c|}{\textbf{SAGAN}} & \multicolumn{2}{c|}{\textbf{SLATER}} \\ \hline
 &
  \multicolumn{1}{c|}{PSNR} &
  \multicolumn{1}{c|}{SSIM(\%)} &
  \multicolumn{1}{c|}{PSNR} &
  \multicolumn{1}{c|}{SSIM(\%)} &
  \multicolumn{1}{c|}{PSNR} &
  \multicolumn{1}{c|}{SSIM(\%)} &
  \multicolumn{1}{c|}{PSNR} &
  \multicolumn{1}{c|}{SSIM(\%)} &
  \multicolumn{1}{c|}{PSNR} &
  \multicolumn{1}{c|}{SSIM(\%)} &
  \multicolumn{1}{c|}{PSNR} &
  \multicolumn{1}{c|}{SSIM(\%)} \\ \hline
T\SB{1}, R=4 &   26.7$\pm$1.2         &  87.4$\pm$1.0         & 26.6$\pm$1.1           &    86.5$\pm$1.0       
&     32.6$\pm$1.8       &      94.6$\pm$0.8
&       34.4$\pm$0.8     &   94.4$\pm$0.7
&       32.1$\pm$0.9     &      92.1$\pm$0.7
&     38.8$\pm$0.8       &    97.9$\pm$0.5       \\ \hline
T\SB{1}, R=8 &  23.5$\pm$0.9          &     83.0$\pm$1.1      &    23.4$\pm$0.9        &   82.5$\pm$1.2       
&     30.3$\pm$1.7       &      91.6$\pm$1.4
&    29.3$\pm$1.2        &      89.7$\pm$1.4    
&   28.6$\pm$0.9         &      88.3$\pm$1.2     
&     33.2$\pm$0.9       &     95.2$\pm$0.9      \\ \hline
T\SB{2}, R=4 &      30.2$\pm$0.5      &   80.8$\pm$1.2        &   29.9$\pm$0.4         &       79.4$\pm$1.4
&     32.7$\pm$0.7       &       87.7$\pm$1.7    
&      33.4$\pm$0.9      &       87.5$\pm$1.0    
&     34.9$\pm$0.6       &      91.6$\pm$1.1     
&     40.0$\pm$0.8       &      97.7$\pm$0.5     \\ \hline
T\SB{2}, R=8 &  27.6$\pm$0.4          &    76.2$\pm$1.3       &      27.3$\pm$0.4      &     74.1$\pm$1.6      
&      29.9$\pm$0.7      &        84.0$\pm$1.9   
&       31.2$\pm$0.7     &         85.3$\pm$1.0  
&      30.7$\pm$0.5      &           86.4$\pm$1.4
&      34.1$\pm$0.8      &        94.8$\pm$0.7   \\ \hline
\end{tabular}%
}
\label{tab:model_invertibility_table}
\end{table*}

\vspace{-0.1cm}
\section{Methods}
\subsection{Network Architecture}
\revhl{Architecture of SLATER's synthesizer, mapper and discriminator are described below (see also Supp. Text \ref{sec:architecture}).}

\subsubsection{Synthesizer}
\revhl{The synthesizer is a multi-layer architecture where image resolution is progressively increased. Each layer comprises a convolutional upsampling block to increase image resolution by a factor of 2 followed by a cross-attention transformer block, and a skip connection to add the upsampled input. The first layer receives a constant input randomly drawn from a standard normal distribution. As the input layer, the first layer does not contain the upsampling blocks and the first attention sub-block is replaced with an identity transformation. The last layer calculates the final synthesizer output. Attention sub-blocks are omitted in the last layer to preserve precise localization in high-resolution images with convolution operators, as contextual representations have been extracted in previous layers \cite{chen2021transunet}. Two separate channels are used to output real and imaginary parts of images. The upsampling block uses transpose convolution, and upsampling and modulated convolutions have a kernel size of 3x3. The cross-attention transformer block contains a cascade of cross-attention and convolutional sub-blocks. The fully-connected sub-block in vanilla transformers is replaced with a convolutional sub-block to improve computational efficiency and permit use at high resolutions. Since the convolutional sub-block inherently focuses on local relationships, a second cross-attention block is used to reinforce long-range interactions. Note that the imaging matrix sizes differ between the IXI and fastMRI datasets analyzed here. The number of synthesizer layers was adjusted accordingly. The resolution of the final layer was set to 256x256 for IXI with a total of seven layers, whereas the resolution of the final layer was set to 512x512 for fastMRI with a total of eight layers. During unsupervised pretraining, MR images were zero-padded to the resolution of the final layer. During inference, the output of the synthesizer was cropped to match the matrix size of the MRI acquisition.}

\subsubsection{Mapper}
\revhl{The mapper comprises two streams of multi-layer architectures for processing local and global latent variables. The local stream has a total of five layers, with the first four containing a self-attention sub-block and the last containing a fully-connected sub-block. The global stream has a total of nine layers, each containing fully-connected sub-blocks.} 

\subsubsection{Discriminator}
\revhl{The discriminator aggregates information across multiple spatial scales in a multi-layered architecture. Real and imaginary parts of synthesized and actual MR images are represented in separate channels. Each layer comprises a convolution block followed by downsampling by a factor of 2 and a skip connection to add the downsampled input. The downsampling block uses convolution with a kernel size of 3x3. The resolution of the first layer was set to 256x256 for IXI with a total of seven layers, and 512x512 for fastMRI with a total of eight layers.}

\subsection{Competing Methods}
SLATER was comparatively demonstrated against state-of-the-art techniques based on supervised and unsupervised models, \revhl{as well as a traditional method.} For each technique, hyperparameter optimization was performed via cross-validation on a three-way split of subjects. Optimization was performed for number of epochs, number of inference iterations, and weights for regularization terms based on performance on the validation set. \revhl{For supervised models, performance in the validation set was quantified as $\ell_2$-norm difference between the reconstructed and fully-sampled ground-truth images. For unsupervised models, validation performance was instead quantified as the $\ell_2$-norm difference between reconstructed and available k-space samples in undersampled acquisitions. A single set of hyperparameters yielding near-optimal results in all tasks were selected for each technique. Please see Supp. Fig. \ref{suppfig:hyper} for sample performance curves in the validation set versus number of training epochs where training was continued up to 1500 for methods that perform generative modeling of MR images. During inference on test data, strict data consistency was enforced to the network outputs. Codes were run on an eight-core Intel Xeon E5-2690v3 CPU for LORAKS, and in parallel on five nVidia 2080 Ti GPUs for all network models.}

\textbf{SLATER}: SLATER was first pretrained to map random noise and latent codes onto high-quality MR images. In the IXI dataset, the model was trained to map onto single-coil magnitude images. In fastMRI dataset, the model was instead trained to map onto coil-combined complex images, with real and imaginary channel outputs. \revhl{To do this, coil sensitivity maps were derived via ESPIRiT with default parameters \cite{Uecker2014}. Using these estimates, an optimal linear combination on multi-coil complex images was then performed \cite{Bydder2002}.} \revhl{Pre-training was performed via the Adam optimizer with $\beta_1$:0.0, $\beta_2$:0.99, $\eta=10$ and a learning rate of 0.001 as adopted from \cite{StyleGAN2}. The dimensionality of latent variables were set as $K$=16 and $L_s$=32. Network weights were randomly initialized using a standard normal distribution. Inference was performed via the RMSprop optimizer with learning rate 0.1, momentum parameter 0.9, early-stopping and learning rate schedule as adopted from \cite{StyleGAN2}. Cross-validation indicated 1000 iterations as a favorable early-stopping point for maintaining near-optimal performance and computational efficiency. For IXI, $470^{th}$ pretraining epoch, $\lambda_1=1.0, \lambda_2=0.0$ were selected. For fastMRI, $1280^{th}$ pretraining epoch, $\lambda_1=0.5, \lambda_2=0.5$ were selected. Separate SLATER models were trained to build MRI priors for each tissue contrast within each dataset. Reconstruction was then performed via inference on individual test data conditioned by the MRI prior.}

\textbf{LORAKS}: \revhl{A traditional parallel-imaging reconstruction based on low-rank modeling of local k-space neighborhoods was performed \cite{Haldar2016} via libraries in the LORAKS V2.1 toolbox \cite{lorakstoolbox}. Here, an autocalibrated reconstruction was performed where the structured low-rank matrix was formed based on limited image support assumption \cite{Haldar2016}. Accordingly, the k-space neighborhood radius and the rank of the resultant matrix were selected via cross-validation as: (2,6) for IXI, and (2,30) for fastMRI.}

\textbf{GAN\SB{sup}}:
A fully-supervised conditional generative adversarial network (GAN\SB{sup}) was trained using paired ground-truth and undersampled acquisitions. Network architecture and loss functions were adopted from \cite{rgan}. \revhl{Training was performed via the Adam optimizer with $\beta_1=0.5$, $\beta_2=0.999$, dropout regularization rate 0.5, and a learning rate of 0.0002. Training was continued over 100 epochs, with learning rate schedule from \cite{rgan}. Network weights were randomly initialized using a normal distribution with zero mean and 0.02 standard deviation. Regularization parameters for (pixel-wise, perceptual, adversarial) losses were selected as (100,100,1).} A separate GAN\SB{sup} model was trained for each contrast within each dataset and acceleration rate. 

\textbf{SSDU}:
A self-supervised version of the conditional GAN model in GAN\SB{sup} was trained on undersampled data \cite{yaman2020}. \revhl{Acquired k-space samples were split into two sets of nonoverlapping points, where 60\% of samples were used to estimate model weights and 40\% were used to define the network loss. Analogous to Eq. \ref{optimizationequationstyle}, the network loss was taken as a weighted sum of $\ell_1$-, $\ell_2$-norm differences between recovered and acquired k-space samples. For both datasets, $\lambda_1=1.0, \lambda_2=0.0$ were selected. All other procedures were identical to GAN\SB{sup}.} Separate SSDU models were trained for each contrast within each dataset and acceleration rate. 

\textbf{GAN\SB{prior}}:
Following \cite{Knoll2019inverseGANs}, unsupervised pretraining on fully-sampled MRI data was performed using an unconditional GAN. Network architecture was adopted from \cite{StyleGAN2} for fair comparison against SLATER.  
Training and inference procedures were identical to SLATER with minor modifications for enhanced performance. \revhl{The synthesizer in GAN\SB{prior} was trained to minimize the same loss as in Eq. \ref{synthesizer_loss} with an additional path length regularization parameter adopted from \cite{StyleGAN2}. The discriminator was trained to minimize Eq. \ref{discriminator_loss}. A matching number of latents to SLATER were prescribed with $L_s$=512. For IXI, $60^{th}$ pretraining epoch, $\lambda_1=1.0,\lambda_2=0.0$ were selected. For fastMRI, $180^{th}$ pretraining epoch, $\lambda_1=1.0, \lambda_2=0.0$ were selected.}
Separate GAN\SB{prior} models were trained for each contrast within each dataset.

\textbf{SAGAN}:
Zero-shot learned reconstructions were also implemented using a self-attention GAN model. The network architecture was adopted from \cite{StyleGAN2} for fair comparison. Training and inference procedures were identical to SLATER. \revhl{Optimization of network weights was not performed as it was observed to degrade reconstruction performance. Instead contrast-specific epoch selection was adopted for SAGAN since in this case it yielded enhanced performance. For T\SB{1} and T\SB{2} reconstructions, $798^{th}$, $399^{th}$ epochs in IXI, and $967^{th}$, $2661^{th}$ epoch in fastMRI, along with $\lambda_1=1.0, \lambda_2=0.0$ were selected.} Separate models were trained for each contrast within each dataset.

\textbf{GAN\SB{DIP}, SAGAN\SB{DIP}, SLATER\SB{DIP}}:
DIP reconstructions were performed via untrained GAN\SB{prior}, SAGAN and SLATER models respectively. \revhl{The inference procedures were identical to pre-trained counterparts. Network weights were randomly initialized using a normal distribution with zero mean and unit standard deviation.}

\vspace{-0.2cm}
\subsection{Datasets}
 Demonstrations were performed on single-coil brain MRI data from IXI (http://brain-development.org/ixi-dataset/) and multi-coil brain MRI data from fastMRI \cite{fastmri}. T\SB{1}-weighted and T\SB{2}-weighted acquisitions were considered. In IXI, 25 subjects were used for training, 5 for validation and 10 for testing. \revhl{Parameters for T\SB{1}-weighted scans are: repetition time (TR)=9.813 ms, echo time (TE)=4.603 ms, flip angle=$8^{\circ}$, matrix size=256x256x150, voxel size=0.94x0.94x1.2 mm$^3$; and those for T\SB{2}-weighted scans are: TR=8178 ms, TE=100 ms, flip angle=$90^{\circ}$, matrix size=256x256x130, voxel size=0.94x0.94x1.2 mm$^3$.} In fastMRI, 100 subjects were used for training, 10 for validation and 40 for testing. \revhl{Data from multiple sites are included with no common protocol. For consistency, only volumes with at least 10 cross-sections and acquired with at least 5 coils were selected.} To reduce computational complexity, GCC \cite{Zhang2013} was used to decrease the number of coils to 5. For both datasets, subject selection and splitting was done sequentially. Data were retrospectively undersampled using variable-density random patterns \cite{Lustig2007}. Undersampling masks were generated based on a 2D normal distribution with covariance adaptively adjusted to obtain the desired acceleration rates of R=[4, 8].

\vspace{-0.2cm}
\subsection{Quantitative Assessments}
To assess reconstruction quality, quantitative comparisons were performed against reference images Fourier-reconstructed from fully-sampled acquisitions. Both reconstructed and reference images were normalized to a maximum of 1 prior to measurement. Peak signal-to-noise ratio (PSNR) and structural similarity index (SSIM) were calculated between the reconstructed and reference images. In Tables, summary statistics for quantitative metrics were provided as mean $\pm$ standard deviation across test subjects. Statistical significance of differences between methods was assessed via nonparametric Wilcoxon signed-rank tests.

\begin{figure*}[tp]
\centering
\includegraphics[width=1\textwidth]{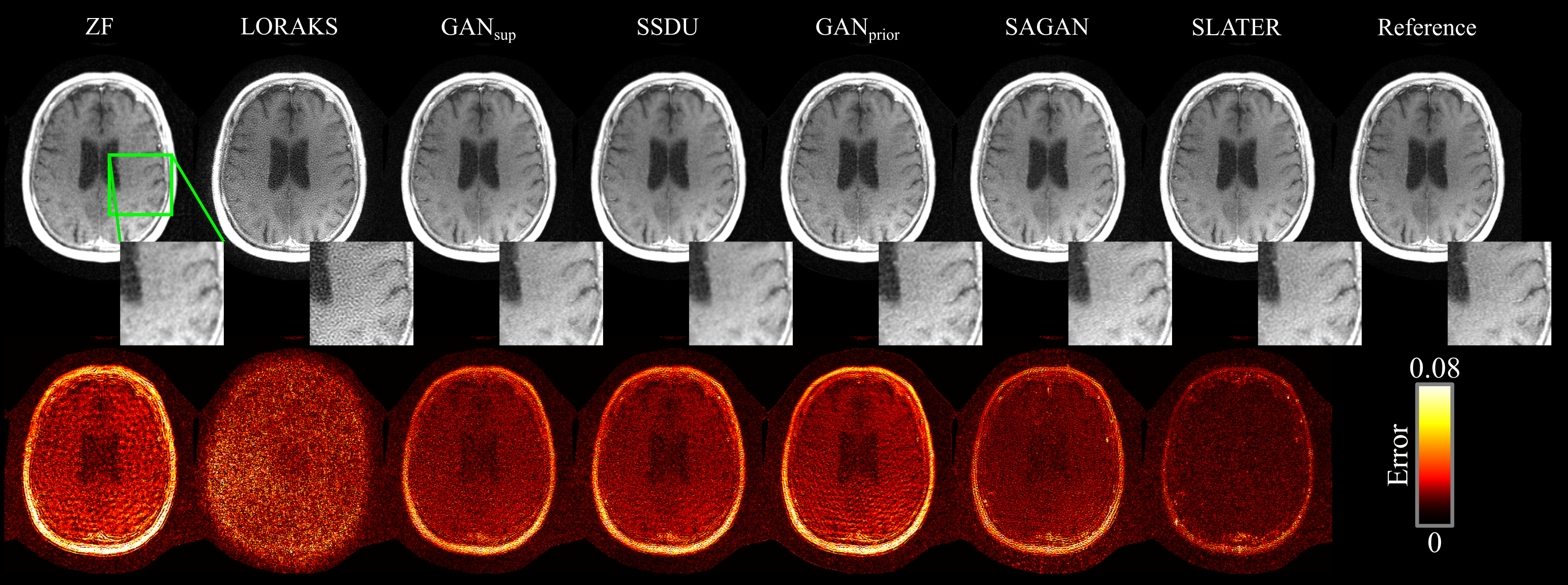}
\captionsetup{justification   = justified,singlelinecheck = false}
\caption{Within-domain reconstructions of a T\SB{1}-weighted acquisition in fastMRI at R=4. Results are shown for ZF, \revhl{LORAKS}, GAN\SB{sup}, SSDU, GAN\SB{prior}, SAGAN and SLATER along with the reference image, and error maps in the bottom row.}
\label{fig:T1_FASTMRI_IN}
\end{figure*}

\begin{table*}[]
\centering
\captionsetup{justification=justified ,width=\linewidth}
\caption{Within-domain reconstruction performance for T\SB{1}- and T\SB{2}-weighted acquisitions in fastMRI at R=4 and 8.}
\resizebox{\textwidth}{!}{%
\begin{tabular}{|l|l|l|l|l|l|l|l|l|l|l|l|l|}
\hline
 &
    \multicolumn{2}{c|}{\textbf{LORAKS}} &
  \multicolumn{2}{c|}{\textbf{GAN\SB{sup}}} &
  \multicolumn{2}{c|}{\textbf{SSDU}} &
  \multicolumn{2}{c|}{\textbf{GAN\SB{prior}}} &
  \multicolumn{2}{c|}{\textbf{SAGAN}} &
  \multicolumn{2}{c|}{\textbf{SLATER}} \\ \hline
 &
  \multicolumn{1}{c|}{PSNR} &
  \multicolumn{1}{c|}{SSIM(\%)} &
  \multicolumn{1}{c|}{PSNR} &
  \multicolumn{1}{c|}{SSIM(\%)} &
  \multicolumn{1}{c|}{PSNR} &
  \multicolumn{1}{c|}{SSIM(\%)} &
  \multicolumn{1}{c|}{PSNR} &
  \multicolumn{1}{c|}{SSIM(\%)} &
  \multicolumn{1}{c|}{PSNR} &
  \multicolumn{1}{c|}{SSIM(\%)} &
  \multicolumn{1}{c|}{PSNR} &
  \multicolumn{1}{c|}{SSIM(\%)} \\ \hline
T\SB{1}, R=4 &33.4$\pm$2.7 & 82.2$\pm$7.7&  37.2$\pm$2.6& 94.4$\pm$5.7 & 37.2$\pm$3.3 & 94.2$\pm$7.5 &  32.8$\pm$2.0& 92.5$\pm$5.2 & 36.1$\pm$2.6 &  94.1$\pm$5.1 &37.6$\pm$3.2 & 93.9$\pm$9.5 \\ \hline
T\SB{1}, R=8 &32.5$\pm$2.4 &  83.0$\pm$9.0 &34.6$\pm$2.4  & 92.0$\pm$7.0 & 33.9$\pm$2.6 & 90.6$\pm$8.5 &  29.8$\pm$1.9& 88.3$\pm$6.7 & 33.1$\pm$2.1 & 90.4$\pm$6.7 &34.2$\pm$2.4 &  90.7$\pm$7.9 \\ \hline
T\SB{2}, R=4 &34.3$\pm$1.0 &90.8$\pm$1.6 & 35.4$\pm$1.2 & 95.5$\pm$0.5 & 33.0$\pm$2.9 & 94.6$\pm$1.3 &  33.5$\pm$1.1&  91.5$\pm$1.8& 33.5$\pm$1.3 & 94.1$\pm$0.8 &36.3$\pm$1.2 &95.5$\pm$0.7 \\ \hline
T\SB{2}, R=8 & 33.1$\pm$1.0&91.7$\pm$1.1 & 32.7$\pm$1.3 & 93.0$\pm$0.8 &31.3$\pm$1.3  &91.0$\pm$1.5  & 28.0$\pm$1.4 & 85.1$\pm$2.5 & 30.9$\pm$1.3 & 91.2$\pm$1.1& 33.4$\pm$1.1&93.0$\pm$1.0 \\ \hline
\end{tabular}%
}
\label{tab:FASTMRI_IN_DOMAIN}
\end{table*}
\vspace{-0.2cm}
\section{Results}
\subsection{Cross-Attention Transformers}
\revhl{We first examined the utility of cross-attention transformers in capturing contextual representations via experiments on a simulated phantom. The phantom had numerical digits of unit intensity on four corners against a zero-intensity background (Fig. \ref{fig:phantom}). This design creates a strong dependency among digit pixels in distant corners, and separately among background pixels. Normally-distributed white noise was added to vary the degree of contextual information present in phantom images by dampening inter-pixel correlations. DIP reconstructions at R=4 were then performed using SLATER. The spatial distribution of cross-attention maps in Eq. \ref{eq:att_maps} characterize learned groupings of correlated pixels. Thus, we reasoned that the attention maps for local latent variables should span over distant albeit correlated image pixels for relatively limited noise levels, and the maps should degrade for higher noise levels due to weakened dependencies. To test this prediction, we inspected the attention maps for the phantom image in Fig. \ref{fig:phantom} (see Supp. Fig. \ref{suppfig:att_maps_for_all_phantoms} for details). As expected, attention maps clearly span across digit pixels or across background pixels for limited noise. Note that attention maps learned on brain images also manifest similar grouping of correlated albeit spatially-distant pixels (see Fig. \ref{fig:attmap1} and Supp. Fig. \ref{suppfig:attmap2} for representative maps). Yet, towards substantially higher noise levels in the simulated phantom, attention maps show a less clear grouping structure as contextual information is gradually weakened. Taken together, these results suggest that SLATER can capture long-range spatial interactions among distant image pixels.}  

\subsection{Model Invertability}
During inference on test data, SLATER inverts its generative model to identify noise, latents and network weights that are most consistent with the undersampled MRI acquisition. To evaluate model invertability, we compared SLATER against CNN-based and self-attention GAN models. Each model was used in both DIP and zero-shot reconstructions. Representative results on T\SB{1}- and T\SB{2}-weighted acquisitions from IXI at R=4 are displayed in Fig. \ref{fig:T2_IXI_DIP} and Supp. Fig. \ref{suppfig:T1_IXI_DIP} respectively. DIP reconstructions tend to suffer from visible loss of spatial resolution, and GAN\SB{prior} and SAGAN have elevated noise and artifacts. In contrast, SLATER yields low residual errors and high visual acuity. Performance metrics are listed in Table \ref{tab:model_invertibility_table}. SLATER achieves superior performance against GAN\SB{prior} and SAGAN in both DIP and zero-shot reconstructions ($p<0.05$). Compared to the second-best method, SLATER yields 4.4dB higher PSNR and 7.7\% higher SSIM in DIP, and \revhl{4.1}dB higher PSNR and \revhl{5.9}\% higher SSIM in zero-shot reconstruction. Furthermore, SLATER yields 5.1dB higher PSNR and 7.0\% higher SSIM over SLATER\SB{DIP}. These results indicate that the cross-attention transformer blocks in SLATER enhance model invertability compared to CNN architectures with or without self-attention and that the unsupervised pretraining stage in SLATER improves \revhl{reconstruction performance}. 

\vspace{-0.2cm}
\subsection{Within-Domain Reconstructions}
Next, we assessed within-domain reconstruction performance when the training and testing domains matched (e.g., T\SB{1} reconstruction based on a T\SB{1}-prior for SLATER). SLATER was compared against \revhl{LORAKS,} GAN\SB{sup}, SSDU, GAN\SB{prior} and SAGAN at R=4 and 8. Representative reconstructions are shown for IXI in Supp. Figs. \ref{suppfig:T1_IXI_IN} and \ref{suppfig:T2_IXI_IN}, and for fastMRI in Fig. \ref{fig:T1_FASTMRI_IN} and Supp. Fig. \ref{suppfig:T2_FASTMRI_IN}. SLATER yields lower residual errors and higher acuity in depicting detailed tissue structure than competing methods. Quantitative assessments are listed in Supp. Table \ref{supptab:IXI_IN_DOMAIN} and Table \ref{tab:FASTMRI_IN_DOMAIN}. SLATER achieves significantly enhanced reconstruction quality against all competing unsupervised methods ($p<0.05$) offering 1.1dB higher PSNR and 1.1\% higher SSIM compared to the second-best method. Furthermore, it offers 4.4dB higher PSNR and 5.5\% higher SSIM compared to GAN\SB{prior}. It also yields higher performance than GAN\SB{sup} in all tasks ($p<0.05$), except at R=8 in IXI where the two methods perform similarly and R=8 in fastMRI where GAN\SB{sup} yields higher SSIM. These results indicate that SLATER offers enhanced reconstruction quality for within-domain tasks compared to unsupervised baselines, while offering on par performance to a supervised baseline.

\vspace{-0.2cm}
\subsection{Across-Domain Reconstructions}
\begin{figure*}
\centering
\includegraphics[width=1\textwidth]{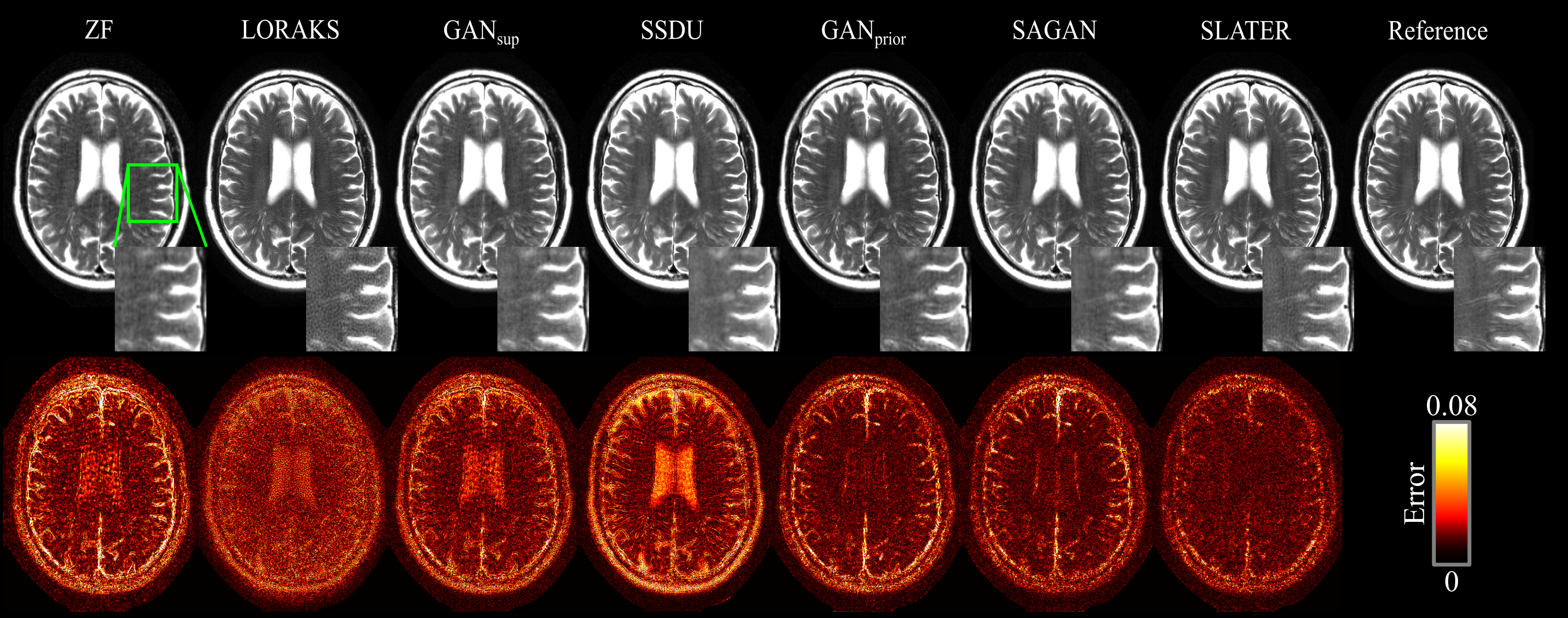}
\captionsetup{justification   = justified,singlelinecheck = false}
\caption{Across-domain reconstructions of a T\SB{2}-weighted acquisition in fastMRI at R=4. Results are shown for ZF, \revhl{LORAKS}, GAN\SB{sup}, SSDU, GAN\SB{prior}, SAGAN and SLATER along with the reference image, and error maps in the bottom row.}
\label{fig:T2_FASTMRI_CD}
\end{figure*}

We then evaluated across-domain reconstruction performance when the tissue contrast in the test domain was different than that in the domain of the trained MRI prior (e.g., T\SB{2} reconstructions based on a T\SB{1}-prior). SLATER was compared against \revhl{LORAKS,} GAN\SB{sup}, SSDU, GAN\SB{prior} and SAGAN at R=4. Representative results are shown for IXI in Supp. Figs. \ref{suppfig:T2_IXI_CD} and \ref{suppfig:T1_IXI_CD}, and for fastMRI in Fig. \ref{fig:T2_FASTMRI_CD} and Supp. Fig. \ref{suppfig:T1_FASTMRI_CD}. Quantitative assessments are listed in Table \ref{tab:ALL_CROSS_DOMAIN}. Again, SLATER yields lower residual errors and higher acuity in depicting detailed tissue structure. It also achieves superior reconstruction quality against all competing supervised and unsupervised methods ($p<0.05$), offering 2.5dB higher PSNR and 1.5\% higher SSIM compared to the second-best method. \revhl{Static models that are not adapted during inference such as GAN\SB{sup} and SSDU can yield suboptimal performance when the data distribution differs between the training and testing domains. Compared to within-domain reconstruction, we find that GAN\SB{sup} and SSDU suffer from 1.7dB PSNR, 1.7\% SSIM loss on average in across-domain reconstruction. In contrast, adaptive models such as SLATER can attain more similar performance for across-domain and within-domain cases. We find that SLATER's average performance differs less than 0.01dB PSNR and 0.11\% SSIM between the two cases. Note that SLATER reconstructions based on within-domain priors generally yield on par or better performance than those based on across-domain priors, yet there are few exceptions with higher across-domain performance that might be attributed to empirical limitations in gradient-descent optimization (see Discussion).}

\revhl{We also assessed across-domain reconstructions when the acceleration rate of the imaging operator differed between training and testing (i.e., trained at R=8 and tested at R=4). Performance measurements are listed in Supp. Table \ref{supptab:CROSS_DOMAIN_R}. SLATER achieves superior reconstruction quality against all competing supervised and unsupervised methods ($p<0.05$), offering 4.0dB PSNR, 1.5\% SSIM improvement over the second-best method. Note that pre-training for zero-shot methods including SLATER is agnostic to the imaging operator, so these methods yield equivalent performance for within-domain and across-domain reconstructions. In contrast, GAN\SB{sup} and SSDU that are explicitly trained for a specific acceleration rate suffer from 3.3dB PSNR, 2.1\% SSIM loss on average in across-domain reconstructions.} Collectively, these results demonstrate that SLATER has improved generalization capabilities compared to static supervised and unsupervised models with fixed weights during inference, while still outperforming zero-shot reconstructions based on pure CNN architectures. 

\vspace{-0.2cm}
\subsection{Ablation Experiments}
We examined the contributions of individual parameter sets in SLATER that are optimized during inference. Variant models were built by progressively introducing optimization for latent variables, noise and network weights. Supp. Table \ref{supptab:ablation} lists performance metrics for experiments on T\SB{1}- and T\SB{2}-weighted acquisitions in the IXI dataset at R=4. On average, the incurred performance gain in (PSNR, SSIM) is (4.1dB, 7.2\%) with latent optimization, (6.5dB, 11.6\%) with latent and noise optimization, and (10.8dB, 14.0\%) with latent, noise and weight optimization. These results indicate the importance of each component in SLATER.

\vspace{-0.2cm}
\subsection{Weight Propagation}
\revhl{Average inference times for competing methods are listed in Table \ref{tab:inference} (see Supp. Table \ref{supptab:training} for training times in the IXI dataset). Model adaptation to specific test samples in zero-shot reconstructions leads to prolonged inference.} In principle, neighboring cross-sections with structural correlations should have similar reconstructions. \revhl{Thus we reasoned that propagating model parameters across consecutive cross-sections should increase efficiency by accelerating the progression onto high-quality reconstructions. Accordingly, the network weights for a given cross-section at the end of its inference optimization were stored, and then used to initialize inference optimization for the next cross-section within each subject.} Latent variables and noise were not shared as they control context and fine details that could vary between cross-sections. \revhl{Table \ref{tab:inference} lists inference times with this weight propagation procedure (see Supp. Table \ref{supptab:WP_performance} for reconstruction performance). While methods that do not perform model adaptation still provide faster inference, weight propagation substantially increases the inference efficiency for SLATER to improve its practicality.} 

\begin{table*}
\centering
\caption{Across-domain reconstruction performance for T\SB{1}- and T\SB{2}-weighted acquisitions in the IXI and fastMRI datasets at R=4. In A-\textgreater{}B, A is the training domain and B is the test domain.}
\resizebox{\textwidth}{!}{%
\begin{tabular}{|l|l|l|l|l|l|l|l|l|l|l|l|l|l|}
\hline
  & \multicolumn{2}{c|}{\textbf{LORAKS}} &
  \multicolumn{2}{c|}{\textbf{GAN\SB{sup}}} &
  \multicolumn{2}{c|}{\textbf{SSDU}} &
  \multicolumn{2}{c|}{\textbf{GAN\SB{prior}}} &
  \multicolumn{2}{c|}{\textbf{SAGAN}} &
  \multicolumn{2}{c|}{\textbf{SLATER}} \\ \hline
   &
  \multicolumn{1}{c|}{PSNR} &
  \multicolumn{1}{c|}{SSIM(\%)} &   
  \multicolumn{1}{c|}{PSNR} &
  \multicolumn{1}{c|}{SSIM(\%)} &
  \multicolumn{1}{c|}{PSNR} &
  \multicolumn{1}{c|}{SSIM(\%)} &
  \multicolumn{1}{c|}{PSNR} &
  \multicolumn{1}{c|}{SSIM(\%)} &
  \multicolumn{1}{c|}{PSNR} &
  \multicolumn{1}{c|}{SSIM(\%)} &
  \multicolumn{1}{c|}{PSNR} &
  \multicolumn{1}{c|}{SSIM(\%)} \\ \hline
\textbf{IXI} T\SB{1}-\textgreater{}T\SB{2}
& 35.4$\pm$0.5 & 92.3$\pm$1.2
&36.8$\pm$0.7 & 95.3$\pm$0.6
&34.8$\pm$1.0 &90.3$\pm$1.4
& 36.4$\pm$0.7&93.2$\pm$0.6
& 32.3$\pm$0.7&86.1$\pm$1.3
& 39.9$\pm$0.8&97.7$\pm$0.5
   \\ \hline

\textbf{IXI} T\SB{2}-\textgreater{}T\SB{1}
& 30.7$\pm$1.2 & 91.7$\pm$1.0
& 35.3$\pm$0.6&96.4$\pm$0.3
& 35.7$\pm$0.4&96.3$\pm$0.3
& 32.5$\pm$1.0&91.6$\pm$1.1
& 32.8$\pm$0.7&93.1$\pm$0.6
& 38.7$\pm$0.9&97.9$\pm$0.5
   \\ \hline 

\textbf{fastMRI}  T\SB{1}-\textgreater{}T\SB{2} & 34.3$\pm$1.0 &90.8$\pm$1.6 &  33.8$\pm$1.2& 94.1$\pm$0.7 & 34.8$\pm$1.1 & 95.1$\pm$0.7 &  33.5$\pm$1.5& 94.3$\pm$0.8& 34.5$\pm$1.2  & 94.4$\pm$0.9& 36.2$\pm$1.1&94.6$\pm$1.0 \\ \hline
\textbf{fastMRI}  T\SB{2}-\textgreater{}T\SB{1}  &33.4$\pm$2.7 & 82.2$\pm$7.7& 36.9$\pm$2.4 & 93.6$\pm$5.8 & 34.3$\pm$3.3 & 92.9$\pm$9.2 & 35.9$\pm$2.1 &90.8$\pm$5.4 & 34.4$\pm$2.1 & 92.2$\pm$4.9 & 38.0$\pm$2.6&95.3$\pm$6.2\\ \hline

\end{tabular}%
}
\label{tab:ALL_CROSS_DOMAIN}
\end{table*}

\begin{table}[ht]
\centering
\caption{Average inference times in sec per cross section. Run times without and with weight propagation (WP) are listed for zero-shot reconstructions. SAGAN does not perform weight optimization, so it is unaffected by WP.}
\resizebox{\columnwidth}{!}{%
\begin{tabular}{|l|c|c|c|c|c|c|}
\hline
  & {\textbf{LORAKS}} & {\textbf{GAN\SB{sup}}} & {\textbf{SSDU}} &  {\textbf{GAN\SB{prior}}} & {\textbf{SAGAN}} & {\textbf{SLATER}} \\ \hline
No WP & 4.10 & 0.002 & 0.002 &  12.73  &  7.74   & 14.80  \\ \hline
With WP & -- & -- & -- &    2.52    &    7.74     & 2.63 \\ \hline
\end{tabular}%
}
\label{tab:inference}
\end{table}

\vspace{-0.17cm}
\section{Discussion}
Here we propose zero-shot learned adversarial transformers for unsupervised reconstruction in accelerated MRI. To our knowledge, this is the first study to introduce a transformer network for MRI reconstruction. Traditional GANs contain cascades of convolutional layers, which might limit ability to capture long-range spatial dependencies. \revhl{We instead employed cross-attention transformer blocks to efficiently capture contextual image features.} Note that self-attention among all feature map locations leads to excessive computational burden \cite{chen2021transunet}. As such, self-attention modules following convolutional blocks have only been leveraged in layers with modest resolution to prevent quadratic complexity \cite{Zhang2019,Wu2019,Lan2020,Oktay2018,Yuan2020}. \revhl{In contrast, the cross-attention mechanism between low-dimensional latent variables and image features permits use at higher resolutions \cite{hudson2021generative}.}
 
Our demonstrations clearly indicate the superiority of the proposed method over a fully-supervised GAN model (GAN\SB{sup}), a self-supervised model (SSDU), DIP based on CNN and self-attention GANs (GAN\SB{DIP}, SAGAN\SB{DIP}), and zero-shot reconstructions (GAN\SB{prior}, SAGAN). SLATER offers on par performance with GAN\SB{sup} for within-domain tasks, while it outperforms GAN\SB{sup} for across-domain tasks. SLATER also outperforms DIP by a substantial margin, which can be attributed to its MRI prior. Lastly, SLATER yields superior performance to GAN\SB{prior} that indicates the importance of transformer blocks in learning a high-fidelity prior. 

\revhl{Recent studies aiming to reduce supervision requirements related to raw data have proposed unpaired or unsupervised learning strategies for MRI reconstruction. A successful approach is to train models on unpaired samples of undersampled and fully-sampled acquisitions \cite{Quan2018c,oh2020,lei2020}. Another promising approach is to train models via self-supervision on undersampled data \cite{Cole2020,yaman2020,Huang2019self,Tamir2019}. Both approaches perform model training in conjunction with the imaging operator to map undersampled acquisitions to MR images. Trained model weights that reflect an indirect prior to reduce artifacts are then fixed during inference. Thus, reconstructions based on an across-domain prior can suffer from suboptimal generalization due to inconsistencies in the data distribution (e.g., tissue contrast) or the imaging operator (e.g., acceleration rate) between training and testing domains.} 

\revhl{SLATER instead learns an MRI prior agnostic to the imaging operator that is later adapted to test data during inference. Prior adaptation involves an inference optimization with considerable computational burden compared to static models with fixed priors. Yet, our experiments indicate that it mitigates potential performance losses during across-domain reconstructions. Ideally, a within-domain prior would be expected to initialize the inference optimization closer to a desirable local minimum, resulting in on par or higher performance than an across-domain prior. Given complex loss surfaces for deep models, however, gradient-descent optimizers can occasionally show non-monotonic behavior approaching less preferred minima despite favorable initialization \cite{optimizersoverview}. The few cases where an across-domain prior yielded relatively higher performance are likely attributed to such behavior, since we observed that the performance gap between the priors diminished with substantially prolonged optimizations in unreported experiments. That said, we opted for early stopping during inference to maintain a desirable compromise between performance and inference time while also mitigating potential risks for over-fitting \cite{DIP,Zou2021}.}

\revhl{The inference optimization performed during zero-shot reconstruction is closely related to DIP methods \cite{DIP}. Compared to DIP with randomly initialized networks, here we show that SLATER yields enhanced reconstruction quality via a pre-trained prior that is more strongly tuned towards the distribution of MR images. Yet, it remains an important research topic to examine the convergence behavior of both untrained and pre-trained models. Several recent studies on accelerated imaging report theoretical and empirical validation of convergence for iterative optimization with traditional recovery methods \cite{Zhang2021} as well as untrained generative priors \cite{Huang2021,Heckel2020}. Further theoretical and empirical research is warranted to investigate whether these results generalize to pre-trained priors.}

Among recent efforts, the closest to the presented approach are prior-adaptation methods in \cite{Konukoglu2019} that uses a variational auto-encoder to learn patch priors and in \cite{Knoll2019inverseGANs} that uses a GAN to learn image priors. Our work differs from these efforts in the following aspects: (i) Compared to \cite{Konukoglu2019} that uses a patch-based implementation, we leverage an image-based implementation that can improve performance in leveraging non-local context. (ii) \cite{Konukoglu2019} uses a fixed patch prior during inference, whereas we perform test-sample specific adaptation of the prior. (iii) Compared to \cite{Knoll2019inverseGANs} that uses a CNN-based GAN, here we use cross-attention transformers to better capture long-range spatial dependencies. (iv) SLATER includes noise and latents at each synthesizer layer to better control image features.     

\revhl{Flow-based models have also been proposed to learn priors for inverse problems \cite{Asim2020}. These models use a composition of fully-invertible flow steps that transform latent variables onto data samples, and they offer exact estimates of the maximum likelihood of data samples and improved immunity against representation errors due to bias. In comparison, adversarial models such as SLATER implicitly minimize discrepancy between generated and actual data distributions without explicit estimation of probability densities. Note, however, that flow-based models characteristically require high-dimensional latent spaces to retain a target level of expressiveness, which might result in a less favorable trade-off between computational burden and quality of generated samples \cite{flowgan}. It remains important future work to compare flow-based and adversarial methods as well as their hybrids in the context of MRI reconstruction.}

Several lines of development can be pursued for the proposed technique. First, zero-shot reconstructions can be initialized with latent estimates based on conventional parallel imaging/compressed sensing reconstructions. This can increase computational efficiency by shortening the inference procedure. Second, the reconstruction loss in SLATER can be combined with SURE-type estimates or regularization terms on network weights \cite{aggarwal2020}. Here we did not explicitly regularize the weights, latents or noise to allow higher degree of consistency to acquired data. When desired, network regularization and on-line error estimates can be introduced to reduce potential for overfitting. \revhl{Third, SLATER learns a coil-combined MRI prior and subsequently incorporates coil-sensitivity information during zero-shot reconstruction. This is achieved by back-projecting the synthesizer output onto individual coils. A powerful alternative is to build a synthesizer with a consistent prior across coils to generate multi-coil images \cite{Darestani2021}. Similar computational efficiency might be expected from both approaches that use a common MRI prior across coils. That said, effects of the decoupling strategy in SLATER against the channel-consistence prior on reconstruction quality remain to be investigated.} 

\revhl{SLATER pre-training performs unsupervised generative modeling of coil-combined MR images derived from fully-sampled acquisitions. Combined with the imaging operator during inference, the learned prior is then used to perform reconstruction of undersampled acquisitions via unsupervised model adaptation. This decoupled approach bypasses the need for paired training datasets. Moreover, the MRI prior is agnostic to the imaging operator and flexibly adapted to the test domain. Thus, a learned prior can be used to reconstruct undersampled acquisitions at varying contrasts or acceleration rates. To lower reliance on fully-sampled datasets, pre-training can be instead performed on undersampled acquisitions. While the resultant prior will not entirely reflect the distribution of high-quality MR images, the model adaptation procedures in SLATER might limit potential performance losses.}

\vspace{-0.2cm}
\section{Conclusion}
Here we introduced a novel unsupervised MRI reconstruction based on an unconditional deep adversarial network. SLATER leverages cross-attention transformers to improve capture of contextual image features. Benefits of SLATER over state-of-the-art supervised and unsupervised methods were demonstrated in brain MRI. SLATER can also be adopted for structural and dynamic MRI in other anatomies, or other imaging modalities such as CT. Reduced supervision requirements and subject-specific adaptation render SLATER a promising candidate for high-performance accelerated MRI. 

\vspace{-0.2cm}
\bibliographystyle{IEEEtran}
\bibliography{Papers}

\begin{thebibliography}{10}
\providecommand{\url}[1]{#1}
\csname url@samestyle\endcsname
\providecommand{\newblock}{\relax}
\providecommand{\bibinfo}[2]{#2}
\providecommand{\BIBentrySTDinterwordspacing}{\spaceskip=0pt\relax}
\providecommand{\BIBentryALTinterwordstretchfactor}{4}
\providecommand{\BIBentryALTinterwordspacing}{\spaceskip=\fontdimen2\font plus
\BIBentryALTinterwordstretchfactor\fontdimen3\font minus
  \fontdimen4\font\relax}
\providecommand{\BIBforeignlanguage}[2]{{%
\expandafter\ifx\csname l@#1\endcsname\relax
\typeout{** WARNING: IEEEtran.bst: No hyphenation pattern has been}%
\typeout{** loaded for the language `#1'. Using the pattern for}%
\typeout{** the default language instead.}%
\else
\language=\csname l@#1\endcsname
\fi
#2}}
\providecommand{\BIBdecl}{\relax}
\BIBdecl

\bibitem{Pruessmann1999}
K.~P. Pruessmann, M.~Weiger, M.~B. Scheidegger, and P.~Boesiger, ``{SENSE:
  sensitivity encoding for fast MRI.}'' \emph{Magnetic Resonance in Medicine},
  vol.~42, no.~5, pp. 952--62, 1999.

\bibitem{Griswold2002}
M.~A. Griswold, P.~M. Jakob, R.~M. Heidemann, M.~Nittka, V.~Jellus, J.~Wang,
  B.~Kiefer, and A.~Haase, ``{Generalized autocalibrating partially parallel
  acquisitions (GRAPPA)},'' \emph{Magnetic Resonance in Medicine}, vol.~47,
  no.~6, pp. 1202--1210, 2002.

\bibitem{Lustig2007}
M.~Lustig, D.~Donoho, and J.~M. Pauly, ``{Sparse MRI: The application of
  compressed sensing for rapid MR imaging},'' \emph{Magnetic Resonance in
  Medicine}, vol.~58, no.~6, pp. 1182--1195, 2007.

\bibitem{Shiqian2008}
S.~Ma, W.~Yin, Y.~Zhang, and A.~Chakraborty, ``An efficient algorithm for
  compressed mr imaging using total variation and wavelets,'' in \emph{IEEE
  CVPR}, 2008, pp. 1--8.

\bibitem{DNNCS2020}
C.~M. {Sandino}, J.~Y. {Cheng}, F.~{Chen}, M.~{Mardani}, J.~M. {Pauly}, and
  S.~S. {Vasanawala}, ``Compressed sensing: {From} research to clinical
  practice with deep neural networks: {Shortening} scan times for {Magnetic
  Resonance Imaging},'' \emph{IEEE Signal Processing Magazine}, vol.~37, no.~1,
  pp. 117--127, 2020.

\bibitem{Wang2016}
S.~Wang, Z.~Su, L.~Ying, X.~Peng, S.~Zhu, F.~Liang, D.~Feng, and D.~Liang,
  ``{Accelerating magnetic resonance imaging via deep learning},'' in
  \emph{{IEEE ISBI}}, 2016, pp. 514--517.

\bibitem{ADMM-CSNET}
Y.~{Yang}, J.~{Sun}, H.~{Li}, and Z.~{Xu}, ``{ADMM-CSNet: A} deep learning
  approach for image compressive sensing,'' \emph{IEEE Transactions on Pattern
  Analysis and Machine Intelligence}, vol.~42, no.~3, pp. 521--538, 2020.

\bibitem{Hammernik2017}
K.~Hammernik, T.~Klatzer, E.~Kobler, M.~P. Recht, D.~K. Sodickson, T.~Pock, and
  F.~Knoll, ``Learning a variational network for reconstruction of accelerated
  {MRI} data,'' \emph{Magnetic Resonance in Medicine}, vol.~79, no.~6, pp.
  3055--3071, 2017.

\bibitem{Mardani2019b}
M.~Mardani, E.~Gong, J.~Y. Cheng, S.~Vasanawala, G.~Zaharchuk, L.~Xing, and
  J.~M. Pauly, ``{Deep generative adversarial neural networks for compressive
  sensing MRI},'' \emph{IEEE Transactions on Medical Imaging}, vol.~38, no.~1,
  pp. 167--179, 2019.

\bibitem{Schlemper2017}
J.~Schlemper, J.~Caballero, J.~V. Hajnal, A.~Price, and D.~Rueckert, ``{A Deep
  Cascade of Convolutional Neural Networks for {MR} Image Reconstruction},'' in
  \emph{{Proceedings of IPMI}}, 2017, pp. 647--658.

\bibitem{MoDl}
H.~K. {Aggarwal}, M.~P. {Mani}, and M.~{Jacob}, ``{MoDL: Model-Based} deep
  learning architecture for inverse problems,'' \emph{IEEE Transactions on
  Medical Imaging}, vol.~38, no.~2, pp. 394--405, 2019.

\bibitem{Quan2018c}
T.~M. Quan, T.~Nguyen-Duc, and W.-K. Jeong, ``{Compressed sensing MRI
  reconstruction with cyclic loss in generative adversarial networks},''
  \emph{IEEE Transactions on Medical Imaging}, vol.~37, no.~6, pp. 1488--1497,
  2018.

\bibitem{Variatonal_end2end}
A.~Sriram, J.~Zbontar, T.~Murrell, A.~Defazio, C.~L. Zitnick, N.~Yakubova,
  F.~Knoll, and P.~Johnson, ``End-to-end variational networks for accelerated
  {MRI} reconstruction,'' in \emph{Proceedings of MICCAI}, 2020, pp. 64--73.

\bibitem{Biswas2019}
S.~Biswas, H.~K. Aggarwal, and M.~Jacob, ``Dynamic {MRI} using model-based deep
  learning and {SToRM }priors: {MoDL-SToRM},'' \emph{Magnetic resonance in
  medicine}, vol.~82, no.~1, pp. 485--494, 2019.

\bibitem{KnollGeneralization}
F.~Knoll, K.~Hammernik, E.~Kobler, T.~Pock, M.~P. Recht, and D.~K. Sodickson,
  ``Assessment of the generalization of learned image reconstruction and the
  potential for transfer learning,'' \emph{Magnetic Resonance in Medicine},
  vol.~81, no.~1, pp. 116--128, 2019.

\bibitem{Dar2017}
S.~U.~H. Dar, M.~{\"{O}}zbey, A.~B. {\c{C}}atl{\i}, and T.~{\c{C}}ukur, ``A
  transfer-learning approach for accelerated {MRI} using deep neural
  networks,'' \emph{Magnetic Resonance in Medicine}, vol.~84, no.~2, pp.
  663--685, 2020.

\bibitem{Zhu2018}
B.~Zhu, J.~Z. Liu, B.~R. Rosen, and M.~S. Rosen, ``{Image reconstruction by
  domain transform manifold learning},'' \emph{Nature}, vol. 555, no. 7697, pp.
  487--492, 2018.

\bibitem{oh2020}
G.~Oh, B.~Sim, H.~Chung, L.~Sunwoo, and J.~C. Ye, ``Unpaired deep learning for
  accelerated {MRI} using optimal transport driven {cycleGAN},'' \emph{IEEE
  Transactions on Computational Imaging}, vol.~6, pp. 1285--1296, 2020.

\bibitem{Eun2020tt}
D.-i. Eun, R.~Jang, W.~S. Ha, H.~Lee, S.~C. Jung, and N.~Kim,
  ``{Deep-learning-based image quality enhancement of compressed sensing
  magnetic resonance imaging of vessel wall: comparison of self-supervised and
  unsupervised approaches},'' \emph{Scientific Reports}, vol.~10, no.~1, p.
  13950, 2020.

\bibitem{chung2020progressive}
H.~Chung, E.~Cha, L.~Sunwoo, and J.~C. Ye, ``Two-stage deep learning for
  accelerated {3D} time-of-flight {MRA} without matched training data,''
  \emph{Medical Image Analysis}, vol.~71, p. 102047, 2021.

\bibitem{Tamir2019}
J.~I. Tamir, S.~X. Yu, and M.~Lustig, ``Unsupervised deep basis pursuit:
  {Learning} reconstruction without ground-truth data,'' in \emph{Proceedings
  of ISMRM}, 2019, p. 0660.

\bibitem{Wang2020self}
A.~Q. Wang, A.~V. Dalca, and M.~R. Sabuncu, ``Neural network-based
  reconstruction in compressed sensing {MRI} without fully-sampled training
  data,'' in \emph{{MLMIR}}, 2020, pp. 27--37.

\bibitem{Cole2020}
E.~K. Cole, J.~M. Pauly, S.~S. Vasanawala, and F.~Ong, ``Unsupervised {MRI}
  reconstruction with generative adversarial networks,''
  \emph{arXiv:2008.13065}, 2020.

\bibitem{yaman2020}
B.~Yaman, S.~A.~H. Hosseini, S.~Moeller, J.~Ellermann, K.~U{\u{g}}urbil, and
  M.~Ak{\c{c}}akaya, ``Self-supervised learning of physics-guided
  reconstruction neural networks without fully sampled reference data,''
  \emph{Magnetic resonance in medicine}, vol.~84, no.~6, pp. 3172--3191, 2020.

\bibitem{Huang2019self}
P.~Huang, C.~H. Li, S.~K. Gaire, R.~Liu, X.~Zhang, X.~Li, and L.~Ying, ``Deep
  {MRI} reconstruction without ground truth for training,'' in
  \emph{Proceedings of ISMRM}, 2019, p. 4668.

\bibitem{Liu2020}
J.~{Liu}, Y.~{Sun}, C.~{Eldeniz}, W.~{Gan}, H.~{An}, and U.~S. {Kamilov},
  ``{RARE}: Image reconstruction using deep priors learned without
  groundtruth,'' \emph{IEEE Journal of Selected Topics in Signal Processing},
  vol.~14, no.~6, pp. 1088--1099, 2020.

\bibitem{aggarwal2020}
H.~K. Aggarwal and M.~Jacob, ``{ENSURE: Ensemble Stein's} unbiased risk
  estimator for unsupervised learning,'' \emph{arXiv:2010.10631}, 2020.

\bibitem{Jin2019}
K.~H. Jin, H.~Gupta, J.~Yerly, M.~Stuber, and M.~Unser, ``Time-dependent deep
  image prior for dynamic {MRI},'' \emph{arXiv:1910.01684}, 2019.

\bibitem{Arora2020ismrm}
S.~Arora, V.~Roeloffs, and M.~Lustig, ``Untrained modified deep decoder for
  joint denoising and parallel imaging reconstruction,'' in \emph{Proceedings
  of ISMRM}, 2020, p. 3585.

\bibitem{Zou2021}
Q.~Zou, A.~H. Ahmed, P.~Nagpal, S.~Kruger, and M.~Jacob, ``Deep generative
  {SToRM} model for dynamic imaging,'' \emph{arXiv:2101.12366}, 2021.

\bibitem{Darestani2021}
M.~Z. Darestani and R.~Heckel, ``Accelerated mri with un-trained neural
  networks,'' \emph{IEEE Transactions on Computational Imaging}, vol.~7, pp.
  724--733, 2021.

\bibitem{Knoll2019inverseGANs}
D.~Narnhofer, K.~Hammernik, F.~Knoll, and T.~Pock, ``{Inverse GANs for
  accelerated MRI reconstruction},'' in \emph{Proceedings of SPIE}, vol. 11138,
  2019, pp. 381 -- 392.

\bibitem{Konukoglu2019}
K.~C. {Tezcan}, C.~F. {Baumgartner}, R.~{Luechinger}, K.~P. {Pruessmann}, and
  E.~{Konukoglu}, ``{MR} image reconstruction using deep density priors,''
  \emph{IEEE Transactions on Medical Imaging}, vol.~38, no.~7, pp. 1633--1642,
  2019.

\bibitem{Liu2020mrm}
Q.~Liu, Q.~Yang, H.~Cheng, S.~Wang, M.~Zhang, and D.~Liang, ``Highly
  undersampled magnetic resonance imaging reconstruction using autoencoding
  priors,'' \emph{Magnetic Resonance in Medicine}, vol.~83, no.~1, pp.
  322--336, 2020.

\bibitem{Zhang2019}
H.~Zhang, I.~Goodfellow, D.~Metaxas, and A.~Odena, ``Self-attention generative
  adversarial networks,'' in \emph{Proceedings of ICML}, 2019, pp. 7354--7363.

\bibitem{chen2021transunet}
J.~Chen, Y.~Lu, Q.~Yu, X.~Luo, E.~Adeli, Y.~Wang, L.~Lu, A.~L. Yuille, and
  Y.~Zhou, ``{TransUNet: T}ransformers make strong encoders for medical image
  segmentation,'' \emph{arXiv:2102.04306}, 2021.

\bibitem{dalmaz2021resvit}
O.~Dalmaz, M.~Yurt, and T.~\c{C}ukur, ``{ResViT: R}esidual vision transformers
  for multi-modal medical image synthesis,'' \emph{arXiv:2106.16031}, 2021.

\bibitem{Yilmaz2021ismrm}
Y.~Korkmaz, S.~U. Dar, M.~Yurt, M.~Ozbey, and T.~\c{C}ukur, ``Zero-shot
  learning for unsupervised reconstruction of accelerated {MRI} acquisitions,''
  in \emph{Proceedings of ISMRM}, 2021.

\bibitem{Hyun2018}
C.~M. Hyun, H.~P. Kim, S.~M. Lee, S.~Lee, and J.~K. Seo, ``{Deep learning for
  undersampled MRI reconstruction},'' \emph{Physics in Medicine and Biology},
  vol.~63, no.~13, p. 135007, 2018.

\bibitem{Yoon2018}
J.~Yoon, E.~Gong, I.~Chatnuntawech, B.~Bilgic, J.~Lee, W.~Jung, J.~Ko, H.~Jung,
  K.~Setsompop, G.~Zaharchuk, E.~Y. Kim, J.~Pauly, and J.~Lee, ``Quantitative
  susceptibility mapping using deep neural network: {QSMnet},''
  \emph{NeuroImage}, vol. 179, pp. 199--206, 2018.

\bibitem{ChulYe2018}
J.~C. Ye, Y.~Han, and E.~Cha, ``Deep convolutional framelets: {A} general deep
  learning framework for inverse problems,'' \emph{SIAM Journal on Imaging
  Sciences}, vol.~11, no.~2, pp. 991--1048, 2018.

\bibitem{lee2018deep}
D.~Lee, J.~Yoo, S.~Tak, and J.~C. Ye, ``Deep residual learning for accelerated
  {MRI} using magnitude and phase networks,'' \emph{IEEE Transactions on
  Biomedical Engineering}, vol.~65, no.~9, pp. 1985--1995, 2018.

\bibitem{Hauptmann2018}
A.~Hauptmann, S.~Arridge, F.~Lucka, V.~Muthurangu, and J.~A. Steeden,
  ``Real-time cardiovascular {MR} with spatio-temporal artifact suppression
  using deep learning-a proof of concept in congenital heart disease,''
  \emph{Magnetic Resonance in Medicine}, vol.~81, no.~2, pp. 1143--1156, 2019.

\bibitem{Kwon2017}
K.~Kwon, D.~Kim, and H.~Park, ``{A parallel MR imaging method using multilayer
  perceptron},'' \emph{Medical Physics}, vol.~44, no.~12, pp. 6209--6224, 2017.

\bibitem{Yang2016}
Y.~Yang, J.~Sun, H.~Li, and Z.~Xu, ``Deep {ADMM-Net} for compressive sensing
  {MRI},'' in \emph{Advances in Neural Information Processing Systems},
  vol.~29, 2016.

\bibitem{KikiNet}
T.~Eo, Y.~Jun, T.~Kim, J.~Jang, H.-J. Lee, and D.~Hwang, ``{KIKI-net:}
  cross-domain convolutional neural networks for reconstructing undersampled
  magnetic resonance images,'' \emph{Magnetic Resonance in Medicine}, vol.~80,
  no.~5, pp. 2188--2201, 2018.

\bibitem{Primal_dual}
J.~Cheng, H.~Wang, L.~Ying, and D.~Liang, ``Model learning: {Primal} dual
  networks for fast {MR} imaging,'' in \emph{Proceedings of MICCAI}, 2019, pp.
  21--29.

\bibitem{Adler2018}
J.~{Adler} and O.~{Oktem}, ``Learned primal-dual reconstruction,'' \emph{IEEE
  Transactions on Medical Imaging}, vol.~37, no.~6, pp. 1322--1332, 2018.

\bibitem{Wang2019}
S.~Wang, Z.~Ke, H.~Cheng, S.~Jia, L.~Ying, H.~Zheng, and D.~Liang,
  ``{DIMENSION: Dynamic MR imaging with both k-space and spatial prior
  knowledge obtained via multi-supervised network training},'' \emph{NMR in
  Biomedicine}, p. e4131, 2019.

\bibitem{Conv_recur}
C.~{Qin}, J.~{Schlemper}, J.~{Caballero}, A.~N. {Price}, J.~V. {Hajnal}, and
  D.~{Rueckert}, ``Convolutional recurrent neural networks for dynamic {MR}
  image reconstruction,'' \emph{IEEE Transactions on Medical Imaging}, vol.~38,
  no.~1, pp. 280--290, 2019.

\bibitem{Hosseini2020b}
S.~A.~H. {Hosseini}, B.~{Yaman}, S.~{Moeller}, M.~{Hong}, and M.~{Akcakaya},
  ``Dense recurrent neural networks for accelerated {MRI: History}-cognizant
  unrolling of optimization algorithms,'' \emph{IEEE Journal of Selected Topics
  in Signal Processing}, vol.~14, no.~6, pp. 1280--1291, 2020.

\bibitem{Chen2021}
Y.~Chen, D.~Firmin, and G.~Yang, ``Wavelet improved {GAN for MRI}
  reconstruction,'' in \emph{Proceedings of SPIE}, vol. 11595, 2021, p.
  1159513.

\bibitem{Yu2018c}
S.~Yu, H.~Dong, G.~Yang, G.~Slabaugh, P.~L. Dragotti, X.~Ye, F.~Liu,
  S.~Arridge, J.~Keegan, D.~Firmin, and Y.~Guo, ``{DAGAN: Deep de-aliasing
  generative adversarial networks for fast compressed sensing MRI
  reconstruction},'' \emph{IEEE Transactions on Medical Imaging}, vol.~37,
  no.~6, pp. 1310--1321, 2018.

\bibitem{Mardani2017}
M.~Mardani, H.~Monajemi, V.~Papyan, S.~Vasanawala, D.~Donoho, and J.~Pauly,
  ``Recurrent generative adversarial networks for proximal learning and
  automated compressive image recovery,'' \emph{arXiv:1711.10046}, 2017.

\bibitem{rgan}
S.~U. Dar, M.~Yurt, M.~Shahdloo, M.~E. Ild{\i}z, B.~T{\i}naz, and
  T.~{\c{C}}ukur, ``Prior-guided image reconstruction for accelerated
  multi-contrast {MRI} via generative adversarial networks,'' \emph{IEEE
  Journal of Selected Topics in Signal Processing}, vol.~14, no.~6, pp.
  1072--1087, 2020.

\bibitem{Polakjointvvn2020}
D.~Polak, S.~Cauley, B.~Bilgic, E.~Gong, P.~Bachert, E.~Adalsteinsson, and
  K.~Setsompop, ``Joint multi-contrast variational network reconstruction
  {(jVN)} with application to rapid {2D} and {3D} imaging,'' \emph{Magnetic
  Resonance in Medicine}, vol.~84, no.~3, pp. 1456--1469, 2020.

\bibitem{data_diff}
Q.~{Chang}, H.~{Qu}, Y.~{Zhang}, M.~{Sabuncu}, C.~{Chen}, T.~{Zhang}, and D.~N.
  {Metaxas}, ``Synthetic learning: {Learn} from distributed asynchronized
  discriminator {GAN} without sharing medical image data,'' in \emph{{IEEE
  CVPR}}, 2020, pp. 13\,853--13\,863.

\bibitem{Han2018a}
Y.~Han, J.~Yoo, H.~H. Kim, H.~J. Shin, K.~Sung, and J.~C. Ye, ``{Deep learning
  with domain adaptation for accelerated projection-reconstruction MR},''
  \emph{Magnetic Resonance in Medicine}, vol.~80, no.~3, pp. 1189--1205, 2018.

\bibitem{Dar2021ismrm}
S.~U. Dar, M.~Yurt, and T.~\c{C}ukur, ``A few-shot learning approach for
  accelerated {MRI} via fusion of data-driven and subject-driven priors,'' in
  \emph{Proceedings of ISMRM}, 2021, p. 1949.

\bibitem{Beker2019}
Y.~Arefeen, O.~Beker, H.~Yu, E.~Adalsteinsson, and B.~Bilgic, ``Scan-specific,
  parameter-free artifact reduction in k-space ({SPARK}),''
  \emph{arXiv:2104.01188}, 2021.

\bibitem{lei2020}
K.~Lei, M.~Mardani, J.~M. Pauly, and S.~S. Vasanawala, ``Wasserstein gans for
  mr imaging: From paired to unpaired training,'' \emph{IEEE Transactions on
  Medical Imaging}, vol.~40, no.~1, pp. 105--115, 2021.

\bibitem{raki}
M.~Akcakaya, S.~Moeller, S.~Weingartner, and K.~Ugurbil, ``Scan-specific robust
  artificial-neural-networks for k-space interpolation {(RAKI)} reconstruction:
  {Database}-free deep learning for fast imaging,'' \emph{Magnetic Resonance in
  Medicine}, vol.~81, no.~1, pp. 439--453, 2019.

\bibitem{sraki}
S.~A.~H. Hosseini, C.~Zhang, S.~Weingartner, S.~Moeller, M.~Stuber, K.~Ugurbil,
  and M.~Akcakaya, ``Accelerated coronary {MRI} with {sRAKI: A} database-free
  self-consistent neural network k-space reconstruction for arbitrary
  undersampling,'' \emph{PLOS ONE}, vol.~15, no.~2, pp. 1--13, 2020.

\bibitem{loraki}
T.~H. Kim, P.~Garg, and J.~P. Haldar, ``{LORAKI}: Autocalibrated recurrent
  neural networks for autoregressive {MRI} reconstruction in k-space,''
  \emph{arXiv:1904.09390}, 2019.

\bibitem{Huang2020ismrm}
P.~Huang, C.~Zhang, H.~Li, S.~K. Gaire, R.~Liu, X.~Zhang, X.~Li, L.~Dong, and
  L.~Ying, ``Unsupervised deep learning reconstruction using the {MR} imaging
  model,'' in \emph{Proceedings of ISMRM}, 2020, p. 3617.

\bibitem{Eldeniz2020ismrm}
C.~Eldeniz, W.~Gan, S.~Chen, J.~Liu, U.~S. Kamilov, and H.~An, ``{Phase2Phase:
  Reconstruction of free-breathing MRI into multiple respiratory phases using
  deep learning without a ground truth},'' in \emph{Proceedings of ISMRM},
  2020, p. 0807.

\bibitem{Liu2020ismrm}
J.~Liu, C.~Eldeniz, W.~Sun, Yu~Gan, S.~Chen, H.~An, and U.~S. Kamilov,
  ``{RED-N2N: Image reconstruction for MRI using deep CNN priors trained
  without ground truth},'' in \emph{Proceedings of ISMRM}, 2020, p. 0993.

\bibitem{Hosseini2020}
S.~A. {Hossein Hosseini}, B.~{Yaman}, S.~{Moeller}, and M.~{Akcakaya},
  ``High-fidelity accelerated {MRI} reconstruction by scan-specific fine-tuning
  of physics-based neural networks,'' in \emph{{IEEE EMBC}}, 2020, pp.
  1481--1484.

\bibitem{aggarwal2021}
H.~K. Aggarwal and M.~Jacob, ``Model adaptation for image reconstruction using
  generalized {Stein's} unbiased risk estimator,'' \emph{arXiv:2102.00047},
  2021.

\bibitem{Ke2020ismrm}
Z.~Ke, Y.~Zhu, J.~Cheng, L.~Ying, X.~Liu, H.~Zheng, and D.~Liang, ``Assessment
  of the generalization of learned unsupervised deep learning method,'' in
  \emph{Proceedings of ISMRM}, 2020, p. 3630.

\bibitem{Jin2016}
K.~H. Jin, D.~Lee, and J.~C. Ye, ``A general framework for compressed sensing
  and parallel mri using annihilating filter based low-rank hankel matrix,''
  \emph{IEEE Transactions on Computational Imaging}, vol.~2, no.~4, pp.
  480--495, 2016.

\bibitem{Zhang2020}
X.~Zhang, D.~Guo, Y.~Huang, Y.~Chen, L.~Wang, F.~Huang, Q.~Xu, and X.~Qu,
  ``Image reconstruction with low-rankness and self-consistency of k-space data
  in parallel mri,'' \emph{Medical Image Analysis}, vol.~63, p. 101687, 2020.

\bibitem{DIP}
D.~Ulyanov, A.~Vedaldi, and V.~Lempitsky, ``Deep image prior,'' in \emph{{IEEE
  CVPR}}, 2018, pp. 9446--9454.

\bibitem{StyleGAN2}
T.~Karras, S.~Laine, M.~Aittala, J.~Hellsten, J.~Lehtinen, and T.~Aila,
  ``Analyzing and improving the image quality of {StyleGAN},'' in \emph{{IEEE
  CVPR}}, 2020, pp. 8107--8116.

\bibitem{hudson2021generative}
D.~A. Hudson and C.~L. Zitnick, ``Generative adversarial transformers,''
  \emph{arXiv:2103.01209}, 2021.

\bibitem{yang2021focal}
J.~Yang, C.~Li, P.~Zhang, X.~Dai, B.~Xiao, L.~Yuan, and J.~Gao, ``Focal
  self-attention for local-global interactions in vision transformers,''
  \emph{arXiv:2107.00641}, 2021.

\bibitem{liu2021Swin}
Z.~Liu, Y.~Lin, Y.~Cao, H.~Hu, Y.~Wei, Z.~Zhang, S.~Lin, and B.~Guo, ``Swin
  transformer: Hierarchical vision transformer using shifted windows,''
  \emph{{IEEE ICCV}}, 2021.

\bibitem{adain}
X.~Huang and S.~J. Belongie, ``Arbitrary style transfer in real-time with
  adaptive instance normalization,'' \emph{arXiv:1703.06868}, 2017.

\bibitem{Goodfellow2014a}
I.~J. Goodfellow, J.~Pouget-Abadie, M.~Mirza, B.~Xu, D.~Warde-Farley, S.~Ozair,
  A.~Courville, and Y.~Bengio, ``{Generative adversarial networks},'' in
  \emph{Advances in Neural Information Processing Systems}, 2014, pp.
  2672--2680.

\bibitem{transformer1}
R.~Child, S.~Gray, A.~Radford, and I.~Sutskever, ``Generating long sequences
  with sparse transformers,'' \emph{arXiv:1904.10509}, 2019.

\bibitem{Bishop2006}
C.~M. Bishop, \emph{\BIBforeignlanguage{English}{{Pattern recognition and
  machine learning}}}.\hskip 1em plus 0.5em minus 0.4em\relax Springer Verlag,
  2006.

\bibitem{pmlr-v80-mescheder18a}
L.~Mescheder, A.~Geiger, and S.~Nowozin, ``Which training methods for {GAN}s do
  actually converge?'' in \emph{Proceedings of ICML}, vol.~80, 2018, pp.
  3481--3490.

\bibitem{XLSA18}
Y.~Xian, C.~H. Lampert, B.~Schiele, and Z.~Akata, ``Zero-shot learning—a
  comprehensive evaluation of the good, the bad and the ugly,'' \emph{IEEE
  Transactions on Pattern Analysis and Machine Intelligence}, vol.~41, no.~9,
  pp. 2251--2265, 2019.

\bibitem{Uecker2014}
M.~Uecker, P.~Lai, M.~J. Murphy, P.~Virtue, M.~Elad, J.~M. Pauly, S.~S.
  Vasanawala, and M.~Lustig, ``{ESPIRiT-an eigenvalue approach to
  autocalibrating parallel MRI: Where SENSE meets GRAPPA},'' \emph{Magnetic
  Resonance in Medicine}, vol.~71, no.~3, pp. 990--1001, 2014.

\bibitem{Bydder2002}
M.~Bydder, D.~J. Larkman, and J.~V. Hajnal, ``{Combination of signals from
  array coils using image-based estimation of coil sensitivity profiles},''
  \emph{Magnetic Resonance in Medicine}, vol.~47, no.~3, pp. 539--548, 2002.

\bibitem{Haldar2016}
J.~P. Haldar and J.~Zhuo, ``{P-LORAKS: Low-Rank Modeling of Local k-Space
  Neighborhoods with Parallel Imaging Data},'' \emph{Magnetic resonance in
  medicine}, vol.~75, no.~4, p. 1499, 2016.

\bibitem{lorakstoolbox}
T.~Kim and J.~Haldar, ``Loraks software version 2.0: Faster implementation and
  enhanced capabilities,'' \emph{University of Southern California, Los
  Angeles, CA, Tech. Rep. USC-SIPI-443}, 2018.

\bibitem{fastmri}
F.~Knoll, J.~Zbontar, A.~Sriram, M.~J. Muckley, M.~Bruno, A.~Defazio,
  M.~Parente, K.~J. Geras, J.~Katsnelson, H.~Chandarana, Z.~Zhang,
  M.~Drozdzalv, A.~Romero, M.~Rabbat, P.~Vincent, J.~Pinkerton, D.~Wang,
  N.~Yakubova, E.~Owens, C.~L. Zitnick, M.~P. Recht, D.~K. Sodickson, and Y.~W.
  Lui, ``{fastMRI: A} publicly available raw k-space and {DICOM} dataset of
  knee images for accelerated {MR} image reconstruction using machine
  learning,'' \emph{Radiology: Artificial Intelligence}, vol.~2, no.~1, p.
  e190007, 2020.

\bibitem{Zhang2013}
T.~Zhang, J.~M. Pauly, S.~S. Vasanawala, and M.~Lustig, ``{Coil compression for
  accelerated imaging with Cartesian sampling.}'' \emph{Magnetic Resonance in
  Medicine}, vol.~69, no.~2, pp. 571--82, 2013.

\bibitem{Wu2019}
Y.~Wu, Y.~Ma, J.~Liu, J.~Du, and L.~Xing, ``Self-attention convolutional neural
  network for improved {MR} image reconstruction,'' \emph{Information
  Sciences}, vol. 490, pp. 317--328, 2019.

\bibitem{Lan2020}
H.~Lan, , A.~W. Toga, and F.~Sepehrband, ``{SC-GAN}: {3D} self-attention
  conditional {GAN} with spectral normalization for multi-modal neuroimaging
  synthesis,'' \emph{bioRxiv:2020.06.09.143297}, 2020.

\bibitem{Oktay2018}
O.~Oktay, J.~Schlemper, L.~L. Folgoc, M.~Lee, M.~Heinrich, K.~Misawa, K.~Mori,
  S.~McDonagh, N.~Y. Hammerla, B.~Kainz, B.~Glocker, and D.~Rueckert,
  ``Attention {U-Net}: {Learning} where to look for the pancreas,''
  \emph{arXiv:1804.03999}, 2018.

\bibitem{Yuan2020}
Z.~Yuan, M.~Jiang, Y.~Wang, B.~Wei, Y.~Li, P.~Wang, W.~Menpes-Smith, Z.~Niu,
  and G.~Yang, ``{SARA-GAN: S}elf-attention and relative average discriminator
  based generative adversarial networks for fast compressed sensing {MRI}
  reconstruction,'' \emph{Frontiers in Neuroinformatics}, vol.~14, p.~58, 2020.

\bibitem{optimizersoverview}
S.~Ruder, ``An overview of gradient descent optimization algorithms,''
  \emph{arXiv:1609.04747}, 2016.

\bibitem{Zhang2021}
X.~Zhang, H.~Lu, D.~Guo, L.~Bao, F.~Huang, Q.~Xu, and X.~Qu, ``A guaranteed
  convergence analysis for the projected fast iterative soft-thresholding
  algorithm in parallel mri,'' \emph{Medical Image Analysis}, vol.~69, p.
  101987, 2021.

\bibitem{Huang2021}
W.~Huang, P.~Hand, R.~Heckel, and V.~Voroninski, ``{A Provably Convergent
  Scheme for Compressive Sensing Under Random Generative Priors},'' \emph{J
  Fourier Anal Appl}, vol.~27, no.~2, p.~19, 2021.

\bibitem{Heckel2020}
R.~Heckel and M.~Soltanolkotabi, ``Compressive sensing with un-trained neural
  networks: Gradient descent finds the smoothest approximation,''
  \emph{arXiv:2005.03991}, 2020.

\bibitem{Asim2020}
M.~Asim, M.~Daniels, O.~Leong, A.~Ahmed, and P.~Hand, ``Invertible generative
  models for inverse problems: mitigating representation error and dataset
  bias,'' in \emph{Proceedings of ICML}, vol. 119, 2020, pp. 399--409.

\bibitem{flowgan}
A.~Grover, M.~Dhar, and S.~Ermon, ``Flow-gan: Bridging implicit and prescribed
  learning in generative models,'' \emph{arXiv:1705.08868}, 2017.

\end{thebibliography}

\newpage
\begin{minipage}{1\textwidth}
\setcounter{section}{0}
\section{Supplementary Text}

\renewcommand{\thesupptab}{\Roman{supptab}}

\renewcommand{\labelitemii}{$\dot$}

\subsection{Positional encoding variables}
\label{sec:pos_enc}
For input feature maps $X\in \mathbb{R}^{h_1 \times h_2 \times u}$, sinusoidal position encoding variables $PE \in \mathbb{R}^{h_1 \times h_2 \times u}$ are set at location ($loc_{h_1}$, $loc_{h_2}$, $loc_{u}$) as [74]:
\begin{align}
PE[loc_{h_1}, loc_{h_2},loc_{u}]=
\begin{cases}
sin \left(\frac{loc_{h_2}}{10000^{4(\frac{loc_{u}}{u})}}\right) & \quad  0 < loc_{u} \leq \frac{u}{4} \notag \\
cos \left(\frac{loc_{h_2}}{10000^{4(\frac{loc_{u}}{u}-\frac{1}{4})}}\right) & \quad  \frac{u}{4} \leq loc_{u} \leq \frac{u}{2} \notag\\
sin \left(\frac{loc_{h_1}}{10000^{4(\frac{loc_{u}}{u}-\frac{2}{4})}}\right) & \quad  \frac{u}{2} \leq loc_{u} \leq \frac{u}{3}\notag \\
cos \left(\frac{loc_{h_1}}{10000^{4(\frac{loc_{u}}{u}-\frac{3}{4})}}\right) & \quad  \frac{u}{3} \leq loc_{u} \leq u \notag \\
\end{cases}
\label{positional embedding}
\end{align} 
where $loc_{h_1}$ and $loc_{h_2}$ lie in range [-1, 1], covering complete field of view along the first two spatial dimensions, and $loc_u$ is the channel index.
\end{minipage}
\clearpage
\begin{minipage}{1\textwidth}
\subsection{Architectural Details}
\label{sec:architecture}
\subsubsection{Synthesizer}

\begin{itemize}
\item Layer 1 (4x4): Input(Constant) $\,\to\,$ Cross-Attention Transformer Block $\,\to\,$ Output
\item Layer 2 (8x8): Input $\,\to\,$ Upsample$\,\to\,$ Cross-Attention Transformer Block +  Upsample(Input) $\,\to\,$ Output
\item Layer 3 (16x16): Input $\,\to\,$ Upsample$\,\to\,$ Cross-Attention Transformer Block +  Upsample(Input) $\,\to\,$ Output
\item Layer 4 (32x32): Input $\,\to\,$ Upsample $\,\to\,$ Cross-Attention Transformer Block +  Upsample(Input) $\,\to\,$ Output
\item Layer 5 (64x64): Input $\,\to\,$ Upsample$\,\to\,$ Cross-Attention Transformer Block +  Upsample(Input) $\,\to\,$ Output
\item Layer 6 (128x128): Input $\,\to\,$ Upsample$\,\to\,$ Cross-Attention Transformer Block +  Upsample(Input) $\,\to\,$ Output
\item Layer 7 (256x256): Input $\,\to\,$ Upsample$\,\to\,$ Modulated Convolution $\,\to\,$ Output
\item Cross-attention Transformer Block: Input $\,\to\,$ Cross-Attention + Noise $\,\to\,$ Modulated Convolution $\,\to\,$ Cross-Attention + Noise $\,\to\,$ Output \\

\end{itemize}
\subsubsection{Mapper}
\paragraph{\underline{\textbf{Local Stream}}}
\begin{itemize}
\item Layer 1: Input  $\,\to\,$ Self-Attention Block $\,\to\,$ Output
\item Layer 2: Input  $\,\to\,$ Self-Attention Block $\,\to\,$ Output
\item Layer 3: Input  $\,\to\,$ Self-Attention Block $\,\to\,$ Output
\item Layer 4: Input  $\,\to\,$ Self-Attention Block $\,\to\,$ Output
\item Layer 5: Input  $\,\to\,$ Fully-connected$\,\to\,$ Output
\item Self-Attention Block: Input $\,\to\,$ Self-Attention $\,\to\,$ Fully-connected $\,\to\,$ Fully-connected + Input $\,\to\,$Output 
\\
\end{itemize}
\paragraph{\underline{\textbf{Global Stream}}}
\begin{itemize}
\item Layer 1: Input  $\,\to\,$ Fully-connected $\,\to\,$ Output
\item Layer 2: Input  $\,\to\,$ Fully-connected $\,\to\,$ Output
\item Layer 3: Input  $\,\to\,$ Fully-connected $\,\to\,$ Output
\item Layer 4: Input  $\,\to\,$ Fully-connected $\,\to\,$ Output
\item Layer 5: Input  $\,\to\,$ Fully-connected $\,\to\,$ Output
\item Layer 6: Input  $\,\to\,$ Fully-connected $\,\to\,$ Output
\item Layer 7: Input  $\,\to\,$ Fully-connected $\,\to\,$ Output
\item Layer 8: Input  $\,\to\,$ Fully-connected $\,\to\,$ Output
\item Layer 9: Input  $\,\to\,$ Fully-connected $\,\to\,$ Output
\\
\end{itemize}
\subsubsection{Discriminator}

\begin{itemize}
\item Layer 1 (256x256): Input $\,\to\,$ Convolution $\,\to\,$ Downsample +  Downsample(Input) $\,\to\,$  Output
\item Layer 2 (128x128): Input $\,\to\,$ Convolution $\,\to\,$ Downsample +  Downsample(Input) $\,\to\,$  Output
\item Layer 3 (64x64): Input $\,\to\,$ Convolution $\,\to\,$ Downsample +  Downsample(Input) $\,\to\,$  Output
\item Layer 4 (32x32): Input $\,\to\,$ Convolution $\,\to\,$ Downsample +  Downsample(Input) $\,\to\,$  Output
\item Layer 5 (16x16): Input $\,\to\,$ Convolution $\,\to\,$ Downsample +  Downsample(Input) $\,\to\,$  Output
\item Layer 6 (8x8): Input $\,\to\,$ Convolution $\,\to\,$ Downsample +  Downsample(Input) $\,\to\,$  Output
\item Layer 7 (4x4): Input $\,\to\,$ Convolution $\,\to\,$ Downsample +  Downsample(Input) $\,\to\,$  Output
\end{itemize}
\clearpage
\end{minipage}

\newpage
\clearpage
\begin{minipage}{1\textwidth}
\section{Supplementary Tables} 
\end{minipage}
\null
\vfill
\begin{supptab}[ht]
\centering
\begin{minipage}{1\textwidth}
\captionsetup{justification=justified,singlelinecheck=false}
\caption{Within-domain reconstruction performance for T\SB{1}- and T\SB{2}-weighted acquisitions in the IXI dataset at R=4 and 8.}
\label{supptab:IXI_IN_DOMAIN}
\end{minipage}
\centering
\resizebox{1\textwidth}{!}{%
\begin{tabular}{|l|l|l|l|l|l|l|l|l|l|l|l|l|}
\hline
 &
 \multicolumn{2}{c|}{\textbf{LORAKS}} &
  \multicolumn{2}{c|}{\textbf{GAN\SB{sup}}} &
  \multicolumn{2}{c|}{\textbf{SSDU}} &
  \multicolumn{2}{c|}{\textbf{GAN\SB{prior}}} &
  \multicolumn{2}{c|}{\textbf{SAGAN}} &
  \multicolumn{2}{c|}{\textbf{SLATER}} \\ \hline
 &
   \multicolumn{1}{c|}{PSNR} &
  \multicolumn{1}{c|}{SSIM(\%)} &
  \multicolumn{1}{c|}{PSNR} &
  \multicolumn{1}{c|}{SSIM(\%)} &
  \multicolumn{1}{c|}{PSNR} &
  \multicolumn{1}{c|}{SSIM(\%)} &
  \multicolumn{1}{c|}{PSNR} &
  \multicolumn{1}{c|}{SSIM(\%)} &
  \multicolumn{1}{c|}{PSNR} &
  \multicolumn{1}{c|}{SSIM(\%)} &
  \multicolumn{1}{c|}{PSNR} &
  \multicolumn{1}{c|}{SSIM(\%)} \\ \hline
T\SB{1}, R=4 & 30.7$\pm$1.2 & 91.7$\pm$1.0
 &37.5$\pm$0.5 & 97.8$\pm$0.2 
 & 37.9$\pm$0.6  & 97.8$\pm$0.2 
 &  34.4$\pm$0.8& 94.4$\pm$0.7 
&       32.1$\pm$0.9     &      92.1$\pm$0.7
 &38.8$\pm$0.8 &97.9$\pm$0.5  \\ \hline
T\SB{1}, R=8 & 26.8$\pm$0.9 & 87.3$\pm$1.1 
&33.3$\pm$0.6 & 95.7$\pm$0.3 
& 33.1$\pm$0.7 & 93.9$\pm$0.7 
&29.3$\pm$1.2  &  89.7$\pm$1.4
&   28.6$\pm$0.9         &      88.3$\pm$1.2  
&33.2$\pm$0.9 &95.2$\pm$0.9\\ \hline
T\SB{2}, R=4 & 35.4$\pm$0.5 & 92.3$\pm$1.2 
&38.7$\pm$0.8 & 96.8$\pm$0.3 
& 38.9$\pm$0.7 & 96.3$\pm$0.4 
&33.4$\pm$0.9  & 87.5$\pm$1.0
&     34.9$\pm$0.6       &      91.6$\pm$1.1  
& 40.0$\pm$0.8&97.7$\pm$0.5 \\ \hline
T\SB{2}, R=8 & 31.4$\pm$0.4 & 88.2$\pm$1.3 
&34.2$\pm$0.8 &94.3$\pm$0.6  
& 33.7$\pm$0.9 & 91.6$\pm$1.1 
& 31.2$\pm$0.7 &85.3$\pm$1.0  
&      30.7$\pm$0.5      &           86.4$\pm$1.4
& 34.1$\pm$0.8&94.8$\pm$0.7\\ \hline

\end{tabular}%
}
\end{supptab}
\vfill
\clearpage

\null
\vfill
\begin{supptab}[ht]
\centering
\begin{minipage}{1\textwidth}
\captionsetup{justification   = justified,singlelinecheck = false}
\caption{Across-domain reconstruction performance for T\SB{1}- and T\SB{2}-weighted acquisitions in the IXI and fastMRI datasets. In A-\textgreater{}B, A and B denote the acceleration rates in training versus test domains. Because LORAKS is untrained, and GAN\SB{prior}, SAGAN and SLATER do not make any assumptions regarding the imaging operator during training, their across-domain reconstruction performance is equivalent to the within-domain performance for the target acceleration rate.}
\label{supptab:CROSS_DOMAIN_R}
\end{minipage}
\resizebox{1\textwidth}{!}{%
\begin{tabular}{|l|l|l|l|l|l|l|l|l|l|l|l|l|}
\hline
 &
  \multicolumn{2}{c|}{\textbf{LORAKS}} &
  \multicolumn{2}{c|}{\textbf{GAN\SB{sup}}} &
  \multicolumn{2}{c|}{\textbf{SSDU}} &
  \multicolumn{2}{c|}{\textbf{GAN\SB{prior}}} &
  \multicolumn{2}{c|}{\textbf{SAGAN}} &
  \multicolumn{2}{c|}{\textbf{SLATER}} \\ \hline
 &
  \multicolumn{1}{c|}{PSNR} &
  \multicolumn{1}{c|}{SSIM(\%)} &
  \multicolumn{1}{c|}{PSNR} &
  \multicolumn{1}{c|}{SSIM(\%)} &
  \multicolumn{1}{c|}{PSNR} &
  \multicolumn{1}{c|}{SSIM(\%)} &
  \multicolumn{1}{c|}{PSNR} &
  \multicolumn{1}{c|}{SSIM(\%)} &
  \multicolumn{1}{c|}{PSNR} &
  \multicolumn{1}{c|}{SSIM(\%)} &
  \multicolumn{1}{c|}{PSNR} &
  \multicolumn{1}{c|}{SSIM(\%)} \\ \hline

\textbf{IXI}  T\SB{1}, 8-\textgreater{}4 
& 30.7$\pm$1.2 & 91.7$\pm$1.0
&32.8$\pm$0.9 & 96.6$\pm$0.3 
& 33.1$\pm$1.2  & 94.5$\pm$0.9 
&  34.4$\pm$0.8& 94.4$\pm$0.7 
&       32.1$\pm$0.9     &      92.1$\pm$0.7
&38.8$\pm$0.8 &97.9$\pm$0.5  \\ \hline
\textbf{IXI}  T\SB{2}, 8-\textgreater{}4 
& 35.4$\pm$0.5 & 92.3$\pm$1.2 
&33.7$\pm$0.5 & 93.8$\pm$0.4 
& 34.8$\pm$0.8 & 92.4$\pm$1.0 
&33.4$\pm$0.9  & 87.5$\pm$1.0
&     34.9$\pm$0.6       &      91.6$\pm$1.1  
& 40.0$\pm$0.8&97.7$\pm$0.5 \\ \hline

\textbf{fastMRI}  T\SB{1}, 8-\textgreater{}4  &33.4$\pm$2.7 & 82.2$\pm$7.7 &  34.8$\pm$2.0& 93.7$\pm$5.7 & 35.0$\pm$2.5 & 92.1$\pm$7.4 &  32.8$\pm$2.0& 92.5$\pm$5.2 & 36.1$\pm$2.6 &  94.1$\pm$5.1 &37.6$\pm$3.2 & 93.9$\pm$9.5 \\ \hline
\textbf{fastMRI}  T\SB{2}, 8-\textgreater{}4 &34.3$\pm$1.0 &90.8$\pm$1.6 &  33.3$\pm$1.0& 94.8$\pm$0.6 & 32.0$\pm$1.9 & 92.5$\pm$1.6 &  33.5$\pm$1.1&  91.5$\pm$1.8& 33.5$\pm$1.3 & 94.1$\pm$0.8 &36.3$\pm$1.2 &95.5$\pm$0.7 \\ \hline
\end{tabular}%
}
\end{supptab}
\vfill
\clearpage

\null
\vfill
\begin{supptab}[ht]
\centering
\begin{minipage}{1\textwidth}
\captionsetup{justification   = justified,singlelinecheck = false}
\caption{Reconstruction performance in ablation experiments for SLATER. Metrics are reported for T\SB{1}- and T\SB{2}- weighted acquisitions in the IXI dataset at R=4.}
\label{supptab:ablation}
\centering
\resizebox{.8\textwidth}{!}{%
\begin{tabular}{|l|l|l|l|l|l|l|l|l|}
\hline
   & \multicolumn{2}{c|}{\textbf{None}} & \multicolumn{2}{c|}{\textbf{Latent}} & \multicolumn{2}{c|}{\textbf{Latent+Noise}} & \multicolumn{2}{c|}{\textbf{Latent+Noise+Weight}} \\ \hline
 &
  \multicolumn{1}{c|}{PSNR} &
  \multicolumn{1}{c|}{SSIM(\%)} &
  \multicolumn{1}{c|}{PSNR} &
  \multicolumn{1}{c|}{SSIM(\%)} &
  \multicolumn{1}{c|}{PSNR} &
  \multicolumn{1}{c|}{SSIM(\%)} &
  \multicolumn{1}{c|}{PSNR} &
  \multicolumn{1}{c|}{SSIM(\%)} \\ \hline
T\SB{1} &   26.7$\pm$1.3      &   87.1$\pm$1.2      & 32.4$\pm$1.1 &  94.1$\pm$0.7          & 34.0$\pm$1.2         & 96.6$\pm$0.4        & 38.8$\pm$0.8          & 97.9$\pm$0.5        \\ \hline
T\SB{2} &  30.4$\pm$0.7          &  80.5$\pm$1.4     &     32.9$\pm$0.8       &     88.0$\pm$0.9      
&      36.2$\pm$0.8        &     94.3$\pm$0.6   
&        40.0$\pm$0.8&97.7$\pm$0.5          \\ \hline
\end{tabular}%
}
\end{minipage}
\end{supptab}
\vfill
\clearpage

\null
\vfill
\begin{supptab}[ht]
\centering
\begin{minipage}{1\textwidth}
\captionsetup{justification= justified,singlelinecheck = false}
\caption{Average training time of models in min:sec format per epoch in the IXI dataset. Note that LORAKS does not perform any training.}
\label{supptab:training}
\centering
\resizebox{.6\textwidth}{!}{%
\begin{tabular}{|l|c|c|c|c|c|c|}
\hline
  & {\textbf{LORAKS}}& {\textbf{GAN\SB{sup}}} & {\textbf{SSDU}} &  {\textbf{GAN\SB{prior}}} & {\textbf{SAGAN}} & {\textbf{SLATER}} \\ \hline
Time (min:sec)  & -- & 6:49 & 1:49 &  6:22 &  6:49  & 8:10  \\ \hline
\end{tabular}%
}
\end{minipage}
\end{supptab}

\vfill
\clearpage

\null
\vfill
\begin{supptab}[ht]
\centering
\begin{minipage}{1\textwidth}
\caption{Reconstruction performance for T\SB{1}- and T\SB{2}-weighted acquisitions in the IXI dataset at R=4 and 8 based on the weight propagation procedure. Note that weight propagation only affects the performance of GAN\SB{prior} and SLATER for which weight optimization is performed during inference.}
\label{supptab:WP_performance}
\captionsetup{justification= justified,singlelinecheck = false}

\resizebox{\textwidth}{!}{%
\begin{tabular}{|l|ll|ll|ll|ll|ll|ll|}
\hline
 &
  \multicolumn{2}{c|}{\textbf{LORAKS}} &
  \multicolumn{2}{c|}{\textbf{GAN\SB{sup}}} &
  \multicolumn{2}{c|}{\textbf{SSDU}} &
  \multicolumn{2}{c|}{\textbf{GAN\SB{prior}}} &
  \multicolumn{2}{c|}{\textbf{SAGAN}} &
  \multicolumn{2}{c|}{\textbf{SLATER}} \\ \hline
 &
  \multicolumn{1}{c|}{PSNR} &
  SSIM(\%) &
  \multicolumn{1}{c|}{PSNR} &
  SSIM(\%) &
  \multicolumn{1}{c|}{PSNR} &
  SSIM(\%) &
  \multicolumn{1}{c|}{PSNR} &
  \multicolumn{1}{c|}{SSIM(\%)} &
  \multicolumn{1}{c|}{PSNR} &
  SSIM(\%) &
  \multicolumn{1}{c|}{PSNR} &
  \multicolumn{1}{c|}{SSIM(\%)} \\ \hline
T1, R=4 &
  \multicolumn{1}{l|}{30.7$\pm$1.2} &
  91.7$\pm$1.0 &
  \multicolumn{1}{l|}{37.5$\pm$0.5} &
  97.8$\pm$0.2 &
  \multicolumn{1}{l|}{37.9$\pm$0.6} &
  97.8$\pm$0.2 &
  \multicolumn{1}{l|}{34.15$\pm$0.93} &
  95.06$\pm$0.55 &
  \multicolumn{1}{l|}{32.1$\pm$0.9} &
  92.1$\pm$0.7 &
  \multicolumn{1}{l|}{38.63$\pm$0.88} &
  98.17$\pm$0.24 \\ \hline
T1, R=8 &
  \multicolumn{1}{l|}{26.8$\pm$0.9} &
  87.3$\pm$1.1 &
  \multicolumn{1}{l|}{33.3$\pm$0.6} &
  95.7$\pm$0.3 &
  \multicolumn{1}{l|}{33.1$\pm$0.7} &
  93.9$\pm$0.7 &
  \multicolumn{1}{l|}{29.02$\pm$1.11} &
  88.91$\pm$1.41 &
  \multicolumn{1}{l|}{28.6$\pm$0.9} &
  88.3$\pm$1.2 &
  \multicolumn{1}{l|}{33.04$\pm$1.05} &
  96.06$\pm$0.52 \\ \hline
T2, R=4 &
  \multicolumn{1}{l|}{35.4$\pm$0.5} &
  92.3$\pm$1.2 &
  \multicolumn{1}{l|}{38.7$\pm$0.8} &
  96.8$\pm$0.3 &
  \multicolumn{1}{l|}{38.9$\pm$0.7} &
  96.3$\pm$0.4 &
  \multicolumn{1}{l|}{33.04$\pm$0.84} &
  88.39$\pm$1.23 &
  \multicolumn{1}{l|}{34.9$\pm$0.6} &
  91.6$\pm$1.1 &
  \multicolumn{1}{l|}{39.80$\pm$0.80} &
  97.77$\pm$0.27 \\ \hline
T2, R=8 &
  \multicolumn{1}{l|}{31.4$\pm$0.4} &
  88.2$\pm$1.3 &
  \multicolumn{1}{l|}{34.2$\pm$0.8} &
  94.3$\pm$0.6 &
  \multicolumn{1}{l|}{33.7$\pm$0.9} &
  91.6$\pm$1.1 &
  \multicolumn{1}{l|}{30.83$\pm$0.64} &
  87.30$\pm$1.17 &
  \multicolumn{1}{l|}{30.7$\pm$0.5} &
  86.4$\pm$1.4 &
  \multicolumn{1}{l|}{33.96$\pm$0.77} &
  94.10$\pm$0.46 \\ \hline
\end{tabular}%
}

\end{minipage}
\end{supptab}
\vfill
\clearpage

\begin{minipage}{1\textwidth}
\section{Supplementary Figures}   
\end{minipage}

\null
\vfill
\begin{minipage}{1\textwidth}
\begin{suppfig}[H]
\centering
\includegraphics[width=1\textwidth]{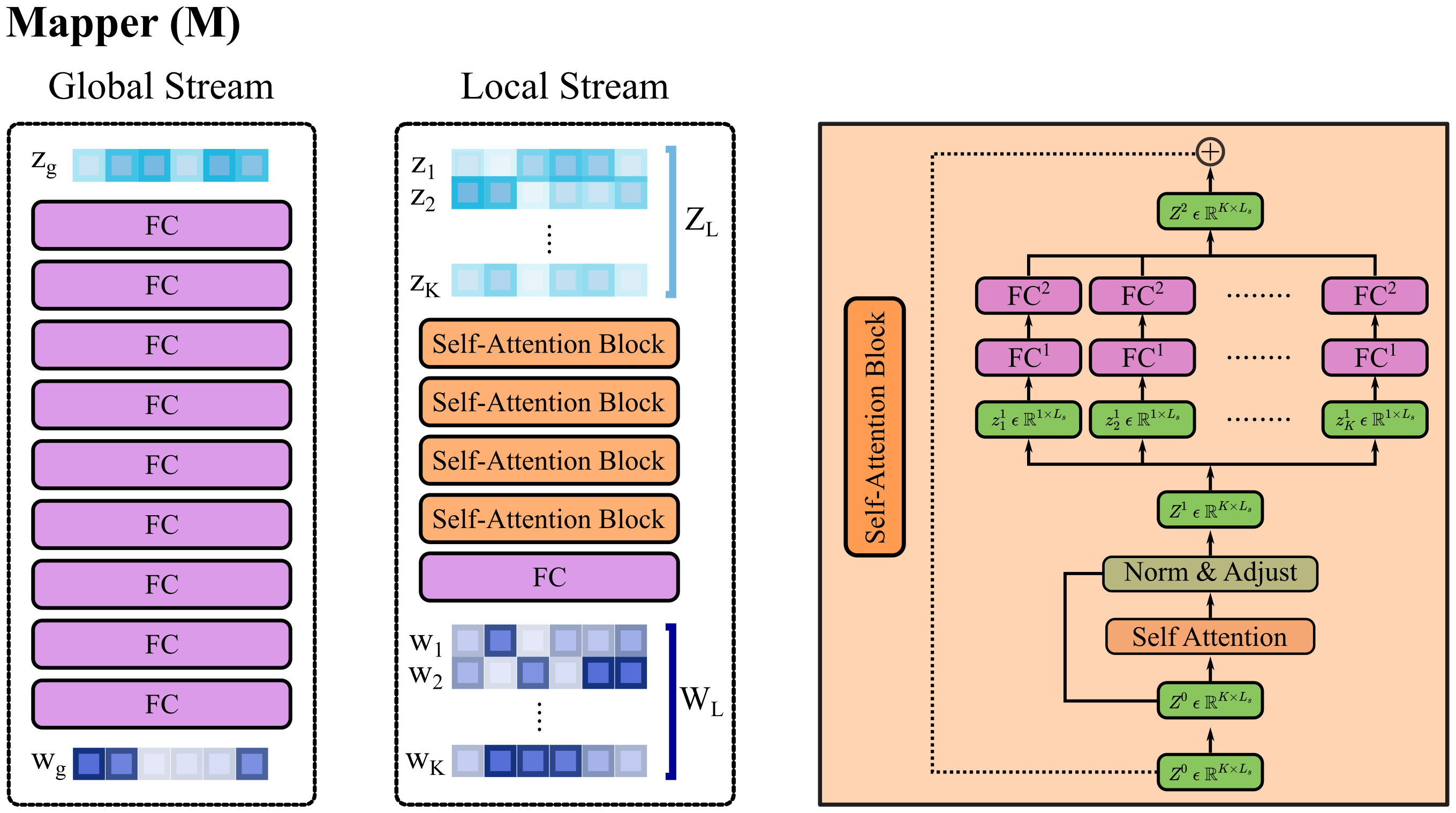}
\captionsetup{justification=justified,singlelinecheck=false}
\caption{Mapper is a multi-layered architecture comprising two separate processing streams: a global stream dedicated to the global latent variable $w_g$, and a local stream dedicated to the local latent variables $W_l = \{w_1, w_2, ..., w_K\}$. The global stream contains a cascade of fully-connected sub-blocks. Meanwhile, the local stream is a cascade of self-attention sub-blocks followed by a fully-connected sub-block (see rightmost panel for the architecture of the self-attention sub-block). Self-attention sub-blocks enable interactions among individual local latents.}
\label{suppfig:mapper}
\end{suppfig} 
\end{minipage}
\vfill
\clearpage

\null
\vfill
\begin{minipage}{1\textwidth}
\begin{suppfig}[H]
\centering
\includegraphics[width=0.65\textwidth]{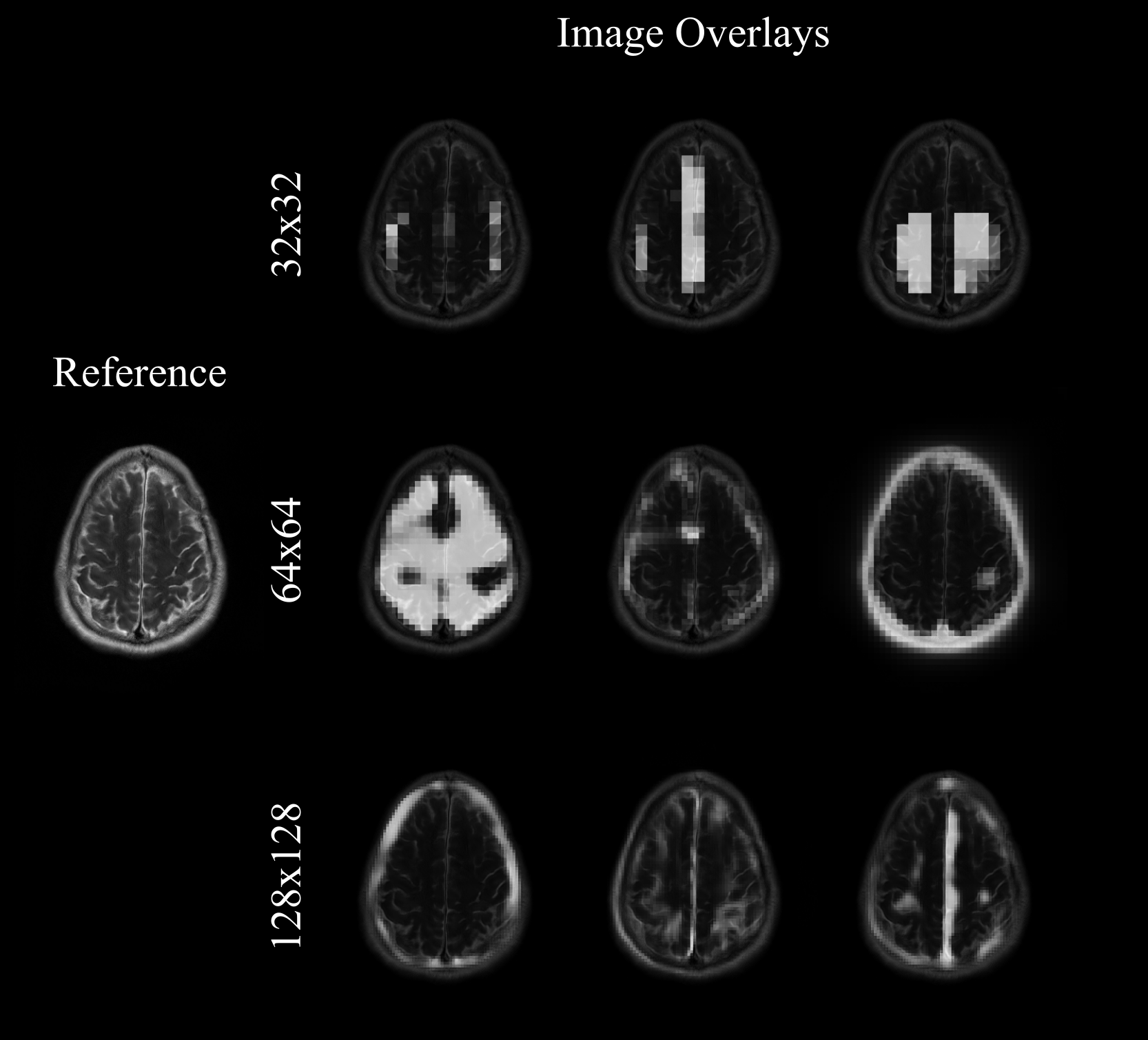}
\captionsetup{justification=justified,singlelinecheck=false}
\caption{Cross-attention maps in SLATER for a T\SB{2}-weighted acquisition. Sample attention maps from the first cross-attention transformer block are displayed across three resolutions (i.e., 32x32, 64x64, 128x128 at network layers 4-6). At each resolution, respective maps are displayed in overlaid format onto the MR image, and the reference MR image is also shown. Attention maps for separate latents show segregated spatial distribution. They also tend to group tissue clusters with similar signal intensity and texture, where the clusters are broadly distributed across the image and they are often spatially noncontiguous.}
\label{suppfig:attmap2}
\end{suppfig} 
\end{minipage}
\vfill
\clearpage

\null
\vfill
\begin{minipage}{1\textwidth}
\begin{suppfig}[H]
\includegraphics[width=1\textwidth]{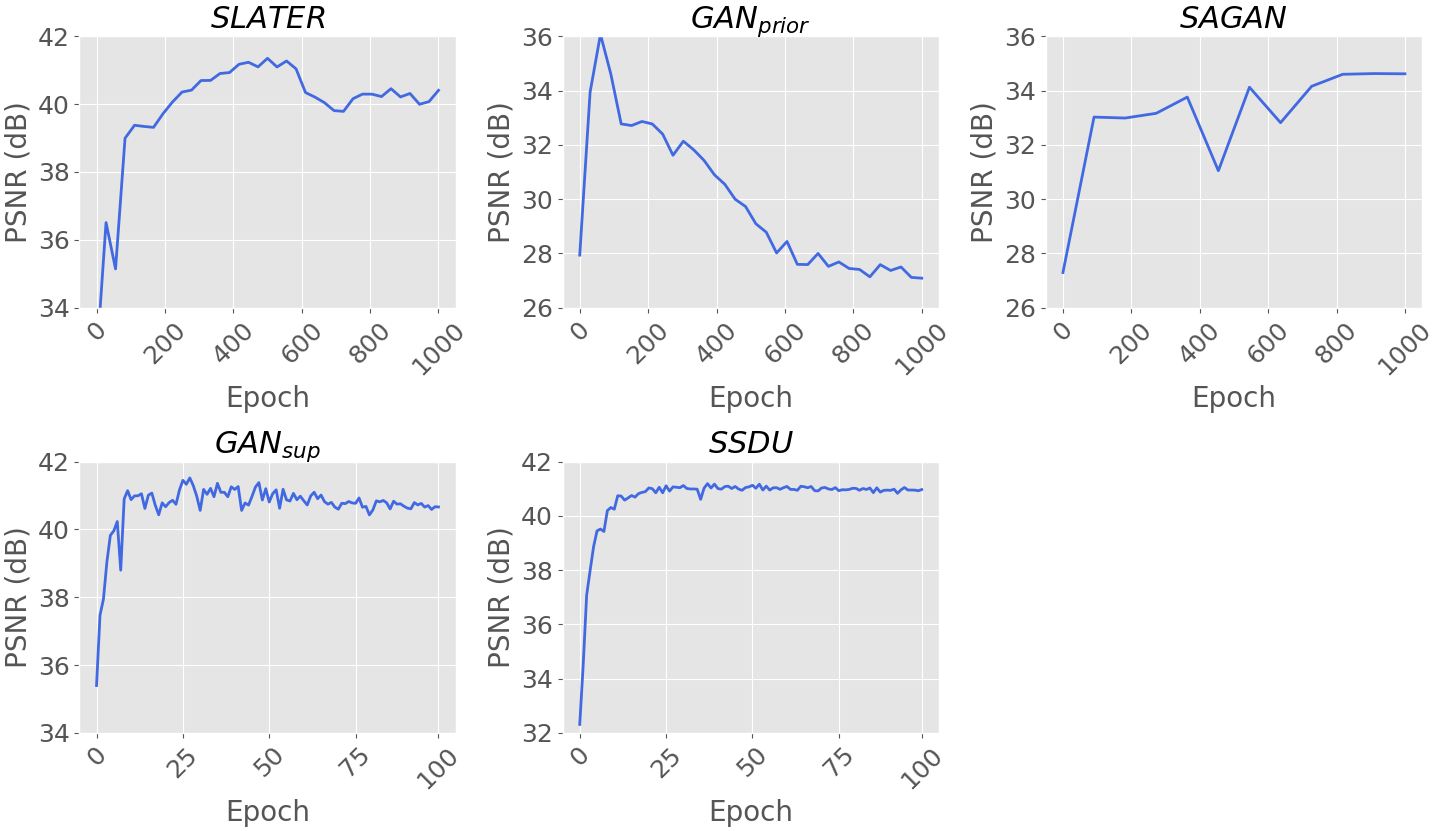}
\captionsetup{justification=justified,singlelinecheck=false}
\caption{Reconstruction performance in the validation set as a function of number of training epochs. Results from supervised (GAN\SB{sup}) and unsupervised models (SSDU, GAN\SB{prior}, SAGAN and SLATER) are shown for T\SB{1}-weighted acquisitions in IXI at R=4. For unsupervised models, hyperparameter selection in the validation set was actually performed based on the difference between recovered and acquired k-space samples in undersampled data. However, to facilitate interpretation, here performance for all methods is displayed as PSNR between reconstructed and ground-truth images. }
\label{suppfig:hyper}
\end{suppfig} 
\end{minipage}
\vfill
\clearpage

\null
\vfill
\begin{minipage}{1\textwidth}
\begin{suppfig}[H]
\begin{center}
\includegraphics[width=.40\textwidth]{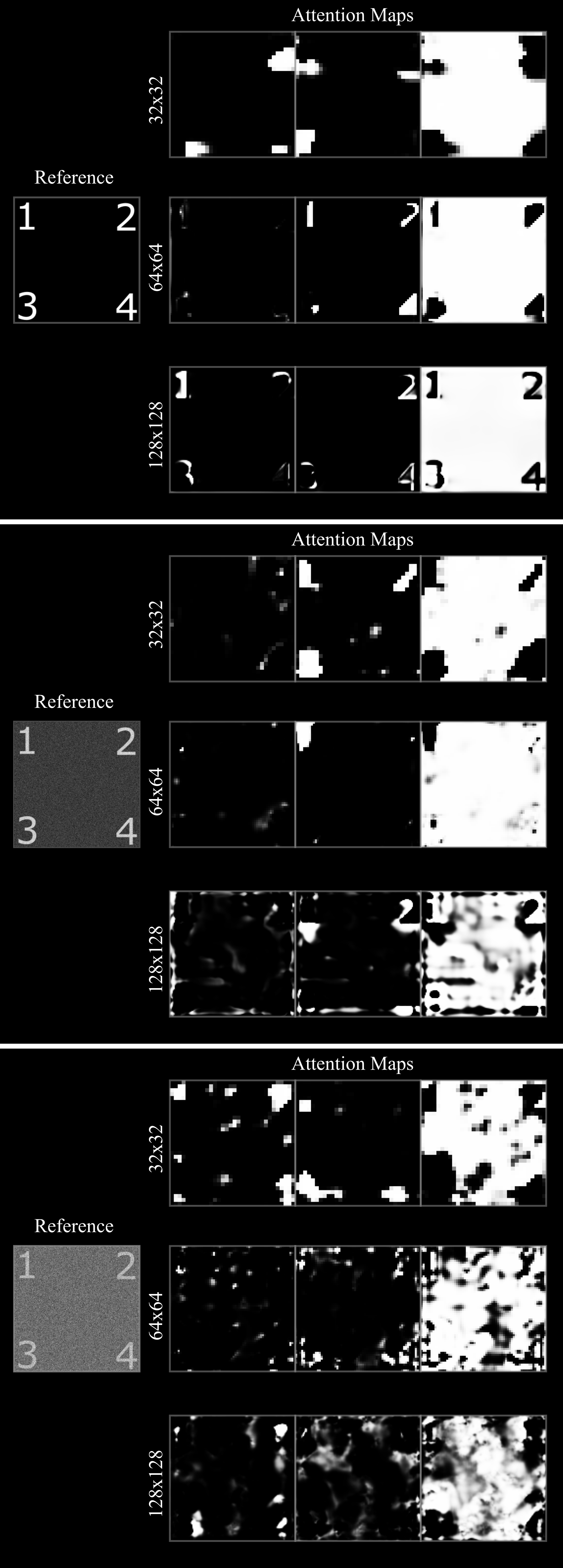}
\end{center}
\captionsetup{justification=justified,singlelinecheck=false}
\caption{Cross-attention maps in SLATER for a simulated digit phantom with varying levels of noise. Relative  to  a  peak signal intensity of 1, top, middle and bottom panels display sample attention maps for no noise, noise variance of 0.01, and noise variance of 0.1, respectively. Within each panel, maps from the first cross-attention sub-block are shown at three resolutions (i.e., 32x32, 64x64, 128x128 at network layers 4-6), along with the reference phantom images.}
\label{suppfig:att_maps_for_all_phantoms}
\end{suppfig}
\end{minipage}
\vfill
\clearpage

\null
\vfill
\begin{minipage}{1\textwidth}
\begin{suppfig}[H]
\includegraphics[width=1\textwidth]{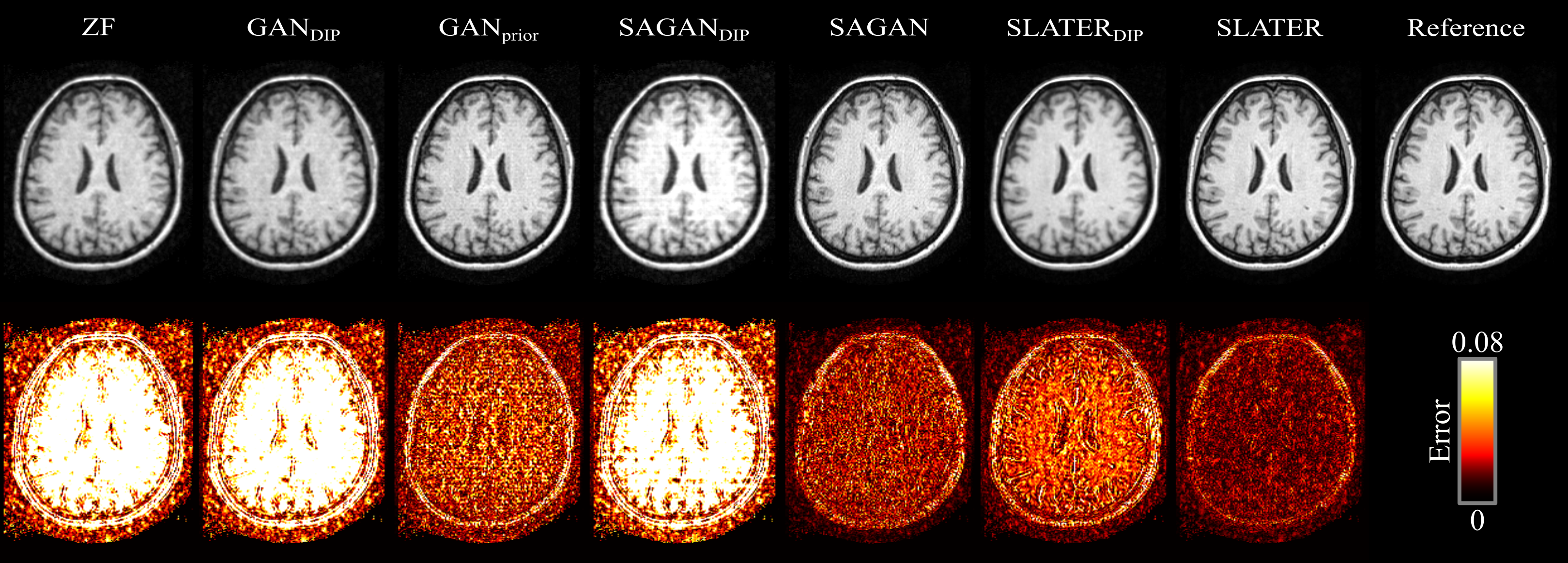}
\captionsetup{justification=justified,singlelinecheck=false}
\caption{Reconstructions of a representative T\SB{1}-weighted acquisition at R=4 are shown for the Fourier method (ZF), DIP methods (GAN\SB{DIP}, SAGAN\SB{DIP}, SLATER\SB{DIP}) and zero-shot reconstructions (GAN\SB{prior}, SAGAN, SLATER) along with the reference image. Corresponding error maps are underneath the images for each method.}
\label{suppfig:T1_IXI_DIP}
\end{suppfig} 
\end{minipage}
\vfill
\clearpage

\null
\vfill
\begin{minipage}{1\textwidth}
\begin{suppfig}[H]
\includegraphics[width=1\textwidth]{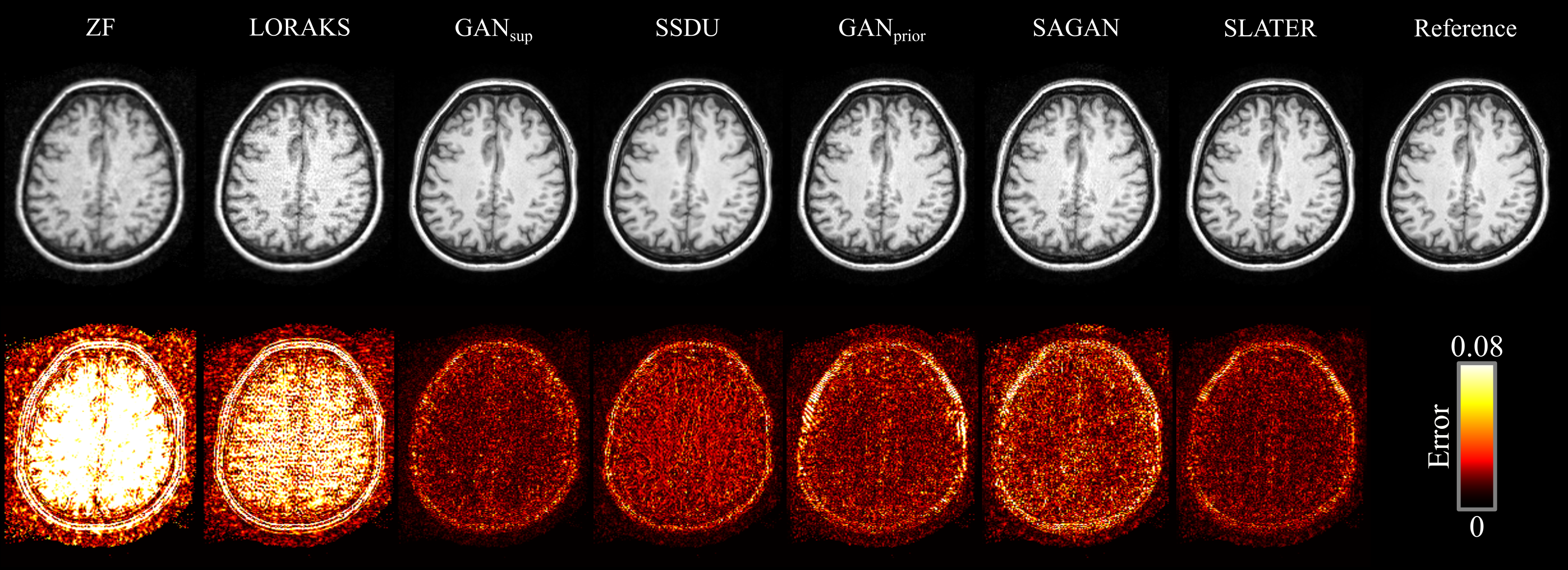}
\captionsetup{justification=justified,singlelinecheck=false}
\caption{Within-domain reconstructions of a T\SB{1}-weighted acquisition in the IXI dataset at R=4 are shown for the Fourier method (ZF), a traditional low-rank method (LORAKS), a supervised baseline (GAN\SB{sup}), unsupervised baselines (SSDU, GAN\SB{prior}, SAGAN) and the proposed method (SLATER) along with the reference image. Corresponding error maps are underneath the images for each method. }
\label{suppfig:T1_IXI_IN}
\end{suppfig} 
\end{minipage}
\vfill
\clearpage

\null
\vfill
\begin{minipage}{1\textwidth}
\begin{suppfig}[H]
\includegraphics[width=1\textwidth]{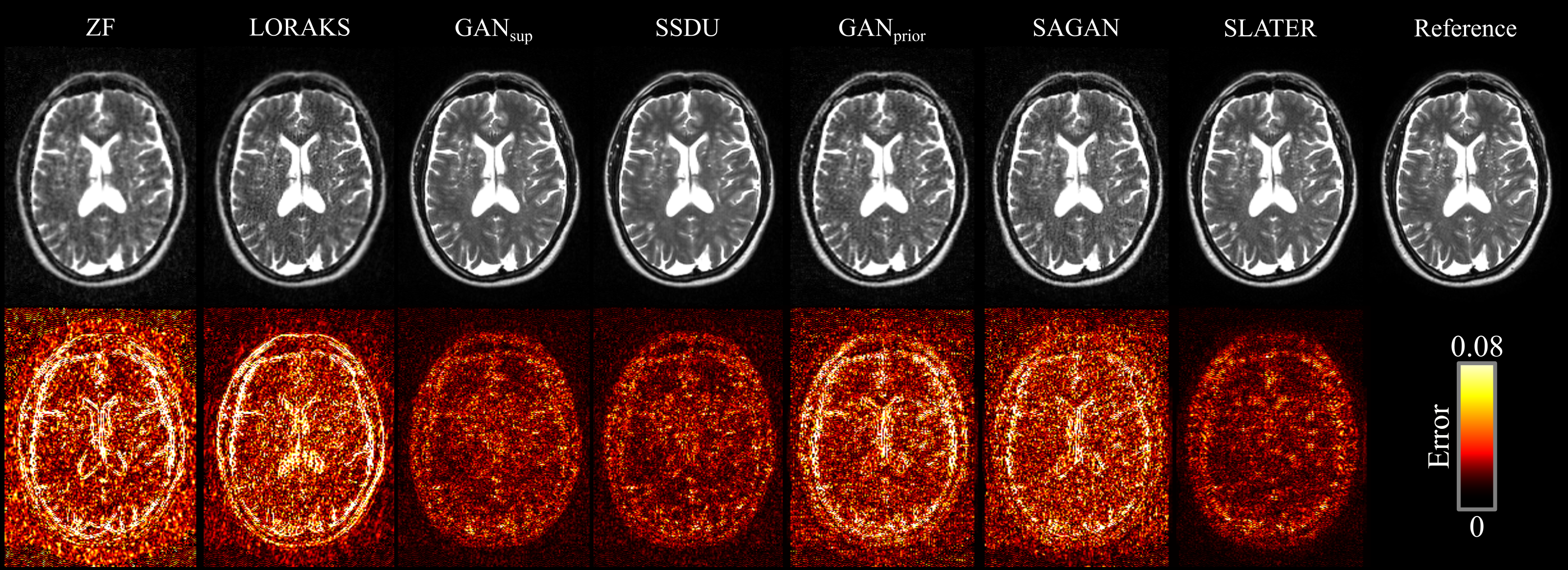}
\captionsetup{justification=justified,singlelinecheck=false}
\caption{Within-domain reconstructions of a T\SB{2}-weighted acquisition in the IXI dataset at R=4. Results are shown for ZF, LORAKS, GAN\SB{sup}, SSDU, GAN\SB{prior}, SAGAN and SLATER along with the reference image. Corresponding error maps are underneath the images for each method. }
\label{suppfig:T2_IXI_IN}
\end{suppfig} 
\end{minipage}
\vfill
\clearpage

\null
\vfill
\begin{minipage}{1\textwidth}
\begin{suppfig}[H]
\includegraphics[width=1\textwidth]{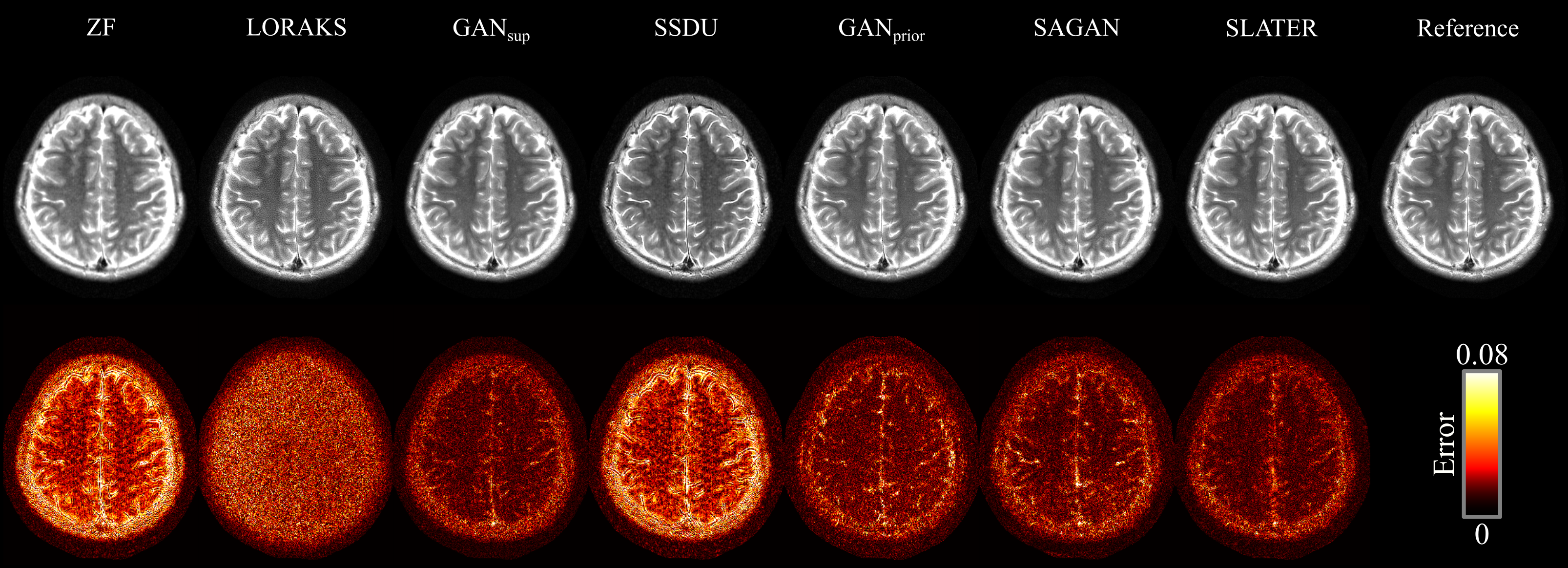}
\captionsetup{justification=justified,singlelinecheck=false}
\caption{Within-domain reconstructions of a T\SB{2}-weighted acquisition in the fastMRI dataset at R=4. Results are shown for ZF, LORAKS, GAN\SB{sup}, SSDU, GAN\SB{prior}, SAGAN and SLATER along with the reference image. Corresponding error maps are underneath the images for each method.}
\label{suppfig:T2_FASTMRI_IN}
\end{suppfig}
\end{minipage}
\vfill
\clearpage

\null
\vfill
\begin{minipage}{1\textwidth}
\begin{suppfig}[H]
\includegraphics[width=1\textwidth]{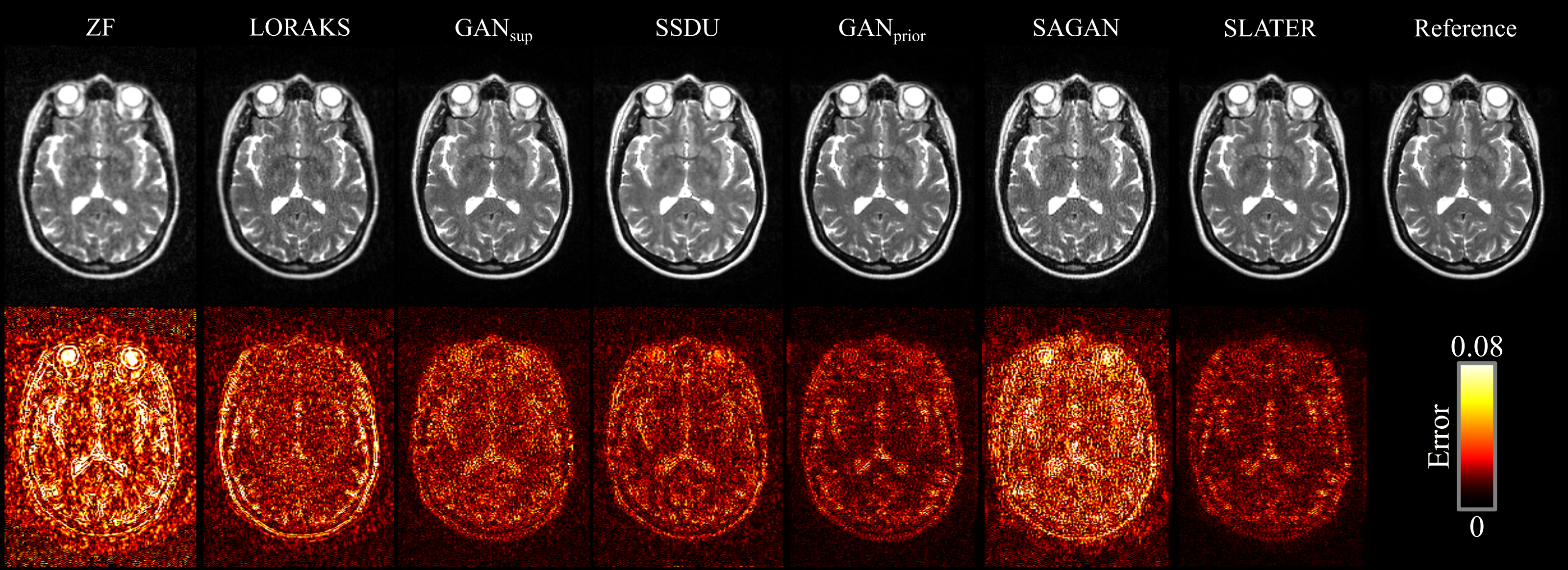}
\captionsetup{justification=justified,singlelinecheck=false}
\caption{Across-domain reconstructions of a T\SB{2}-weighted acquisition in the IXI dataset  at R=4. Results are shown for ZF, LORAKS, GAN\SB{sup}, SSDU, GAN\SB{prior}, SAGAN and SLATER along with the reference image. Corresponding error maps are underneath the images for each method.}
\label{suppfig:T2_IXI_CD}
\end{suppfig}
\end{minipage}
\vfill
\clearpage

\null
\vfill
\begin{minipage}{1\textwidth}
\begin{suppfig}[H]
\includegraphics[width=1\textwidth]{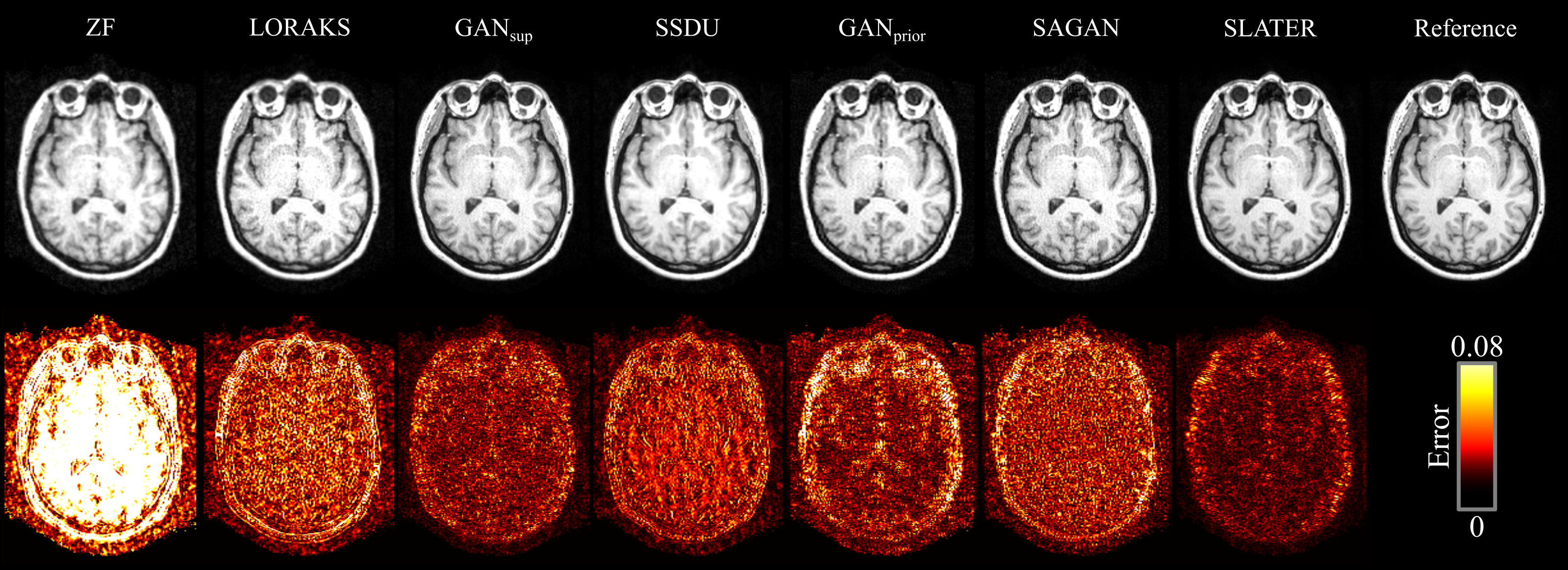}
\captionsetup{justification=justified,singlelinecheck=false}
\caption{Across-domain reconstructions of a T\SB{1}-weighted acquisition in the IXI dataset  at R=4. Results are shown for ZF, LORAKS, GAN\SB{sup}, SSDU, GAN\SB{prior}, SAGAN and SLATER along with the reference image. Corresponding error maps are underneath the images for each method.}
\label{suppfig:T1_IXI_CD}
\end{suppfig}
\end{minipage}
\vfill
\clearpage

\null
\vfill
\begin{minipage}{1\textwidth}
\begin{suppfig}[H]
\includegraphics[width=1\textwidth]{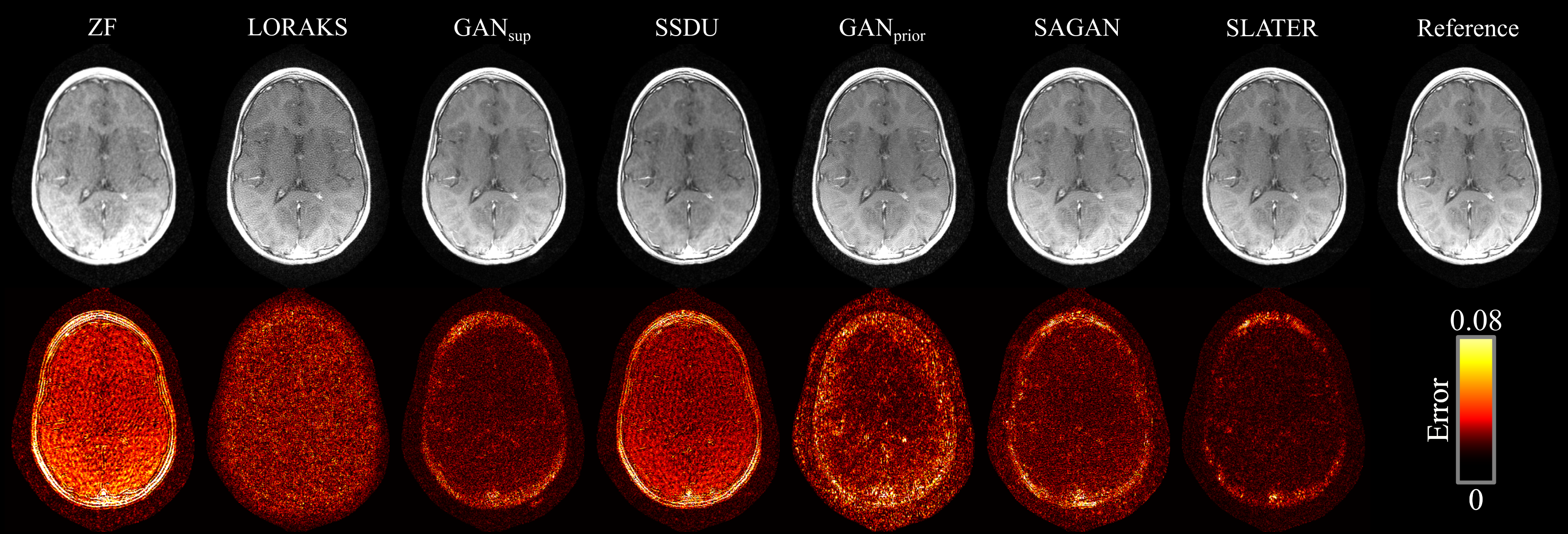}
\captionsetup{justification=justified,singlelinecheck=false}
\caption{Across-domain reconstructions of a T\SB{1}-weighted acquisition in the fastMRI dataset  at R=4. Results are shown for ZF, LORAKS, GAN\SB{sup}, SSDU, GAN\SB{prior}, SAGAN and SLATER along with the reference image. Corresponding error maps are underneath the images for each method.}
\label{suppfig:T1_FASTMRI_CD}
\end{suppfig}
\end{minipage}
\vfill
\clearpage

\end{document}